\documentclass[a4]{spie}

\usepackage{amsmath,amsfonts,amssymb}
\usepackage{graphicx}
\graphicspath{ {./images/} }
\usepackage[colorlinks=true, allcolors=blue]{hyperref}
\usepackage{multirow}
\usepackage{wasysym}
\usepackage[section]{placeins}
\usepackage{siunitx}
\usepackage{gensymb}
\usepackage{subcaption}
\usepackage{booktabs}
\usepackage{array}
\usepackage{textcomp}
\usepackage{threeparttable}

\title{Design, manufacture and metrology of additively manufactured, metal and ceramic lightweight circular mirror prototypes}

\author[a]{Greg Lister}
\author[a]{Rhys Tuck}
\author[a]{Younes Chahid}
\author[a]{Katherine Morris}
\author[a]{Richard Kotlewski}
\author[a]{Scott McPhee}
\author[b]{Cyril Bourgenot}
\author[b]{Ken Parkin}
\author[c]{Mat Beardsley}
\author[d]{Marta Civitani}
\author[d]{Gabriele Vecchi}

\author[a]{Carolyn Atkins}
\affil[a]{UK Astronomy Technology Centre, Royal Observatory Edinburgh, EH9 3HJ, Edinburgh, UK}
\affil[b]{Durham University, NETPark Research Institute, Sedgefield,  S1 3JD, UK}
\affil[c]{RAL Space, Harwell Science and Innovation Campus, OX11 0QX, UK}
\affil[d]{INAF Astronomical Observatory of Brera, Via E. Bianchi 46, 23807, Merate(LC), Italy}

\authorinfo{Further author information:\\ E-mail: carolyn.atkins@stfc.ac.uk}

\begin{document} 
\maketitle

\begin{abstract}
Spaced-based mirrors are a developing use-case for Additive Manufacturing (AM), the process that builds a part layer-by-layer. The increased geometric freedom results in novel and advantageous designs previously unachievable. Conventionally, mirror fabrication uses subtractive (milling \& turning), formative (casting) and fabricative (bonding) manufacturing methods; however, an additive method can simplify an assembly by consolidating individual components into one, and incorporating lattice structures and function optimised geometries to reduce the mass of components, which are beneficial to space-based instrumentation as mass and volume are constrained. Attention must be given to the printability of the design - build orientation and powder/resin removal from lattices and internal cavities are challenges when designing for AM.

This paper will describe the design, manufacture and metrology of mirror prototypes from the Active Deployable Optical Telescope (ADOT) 6U CubeSat project. The AM mirror is 52mm in diameter, 10mm deep, with a convex 100mm radius of curvature reflective surface and deploys telescopically on three booms. The objectives of the designs were to combine the boom mounting features into the mirror and to lightweight both prototypes by 50\% and 70\% using internal, thin-walled lattices. Four final lattice designs were downselected through simulation and prototype validation. Prototypes were printed in the aluminium alloy AlSi10Mg using powder bed fusion and fused silica using stereolithography. Aluminium mirrors were single point diamond turned and had surface roughness measurements taken. Fused silica designs were adapted from the aluminium designs and have completed printing. 

\end{abstract}


\section{INTRODUCTION}
\label{introduction} 

The uptake of Additive Manufacturing (AM) across the engineering environment has been rapidly increasing over recent years. At its core, AM, often called 3D printing, is a process where parts are built in layers, it exists in its own category of manufacturing process due to the distinct differences over subtractive (milling, turning), formative (casting, injection moulding) and fabricative (bonding) processes which are more commonly used in industry. AM opens up the available design space and allows for unique and challenging designs to solve engineering problems. Features such as complex internal geometries, large overhangs, lattice structures and optimised free-form shapes are traditionally difficult or impractical to manufacture in a cost and time effective manner. With AM, these features can be more easily incorporated into designs without conventional manufacturing constraints. The primary advantage of including these AM-enabled features is reducing the mass of components while maintaining structural integrity under specified loads, which is frequently sought after in engineering, but designs can also benefit from rapid prototyping, minimal waste material, and part consolidation.

Typically, AM is associated with desktop fused deposition modelling (FDM) printers that extrude melted polymer filament, due to increasing popularity among hobbyists as cost and barriers to entry drop; however, the term AM is much broader and encompasses several technologies depending on the type of material that is being printed, but all printers function on the same layer-by-layer build principle. For engineering applications, where higher strength materials are needed, it is common to use powder bed fusion (PBF) or stereolithography (SLA) to process metals, polymers and ceramics. Across all of the AM technologies, the range and type of metal, polymer and ceramic materials that can be printed commercially through bureaux or independently with stand-alone systems is also increasing, further promoting the use of AM as a viable and cost-effective alternative manufacturing method. 

There are a growing number of uses of AM in the aerospace, automotive and medical sector, where the increasing complexity of designs can surpass the capabilities of conventional manufacturing methods. End-use examples such as porous titanium orthopedic implants \cite{Hipreplacement} or topology optimised pistons for high performance cars \cite{Porschepistons} show the clear benefits of AM. They are important to demonstrate the viability and increase confidence in this relatively new manufacturing method. Within astronomy however, the adoption of AM is not as mature, but it is emerging as a potentially more cost effective solution to hardware manufacture by producing lighter, optimised, one-off components. 

\begin{figure}[!h]
  \centering
    \includegraphics[width=\textwidth]{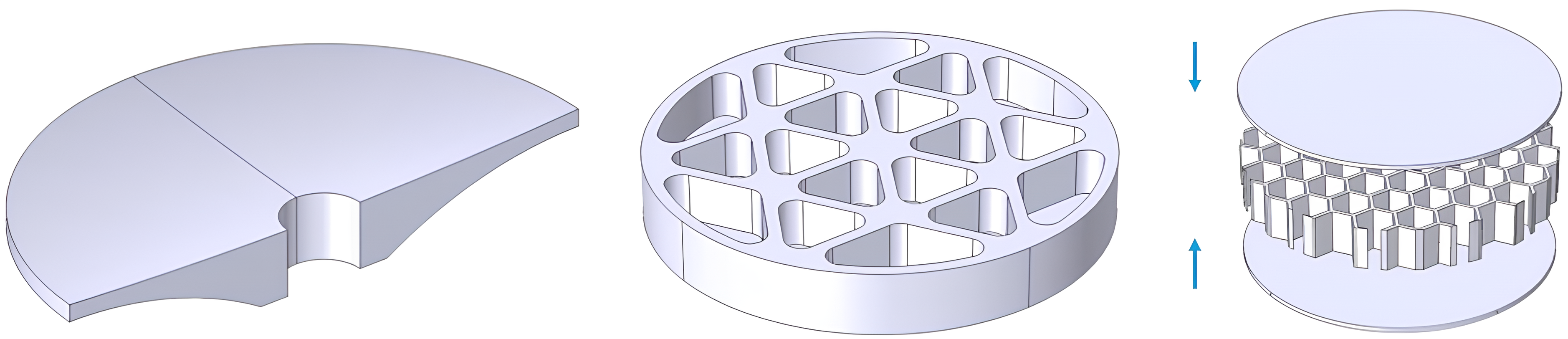}
    \caption{Examples of contoured back, open back and sandwich mirror designs. Image credit: \textit{Atkins et al.(2022)\cite{CookBook}}}
    \label{fig:3LWmirrordesigns}
\end{figure}

Launch vehicles have limited payload capacity, which ultimately constrains the mass and volume of hardware, this drives the demand for designers to reduce the mass of components destined for space to minimise launch costs, or to increase the mass budget for other components. Mass reduction will also reduce the structural demands on supporting systems. Telescope mirrors are one such component where lightweighting has been effectively used. Lightweighting of mirrors can be achieved by adopting a contoured back, open back or sandwich design, as shown in Figure \ref{fig:3LWmirrordesigns}. The three lightweight mirror variants each have their own advantages and disadvantages with respect to cost, stiffness and manufacturing complexity when using convention manufacturing techniques. Open back and sandwich mirrors use a honeycomb, or 2.5D lattice structure which can be manufactured by milling or waterjet cutting, these subtractive techniques limit the type of honeycomb structures that can be used. 

AM can  improve on both conventional open back mirrors by using more complex, thin-walled ($<\SI{1}{mm}$) lattice structures to reduce print-through effects\cite{XRayOpticsprintthru} and increase stiffness, and sandwich mirrors by printing all parts of the mirror at once in addition to using stiffer lattices. All while potentially reducing the cost and time to manufacture. Previous metal lightweight mirrors have been printed in the aluminium alloy AlSi10Mg\cite{XRayOpticsprintthru, Woodard}, the quality of the reflective surface is variable, with porosity and scratch marks degrading the otherwise good surface, separate efforts to investigate scratches and reduce roughness aim to investigate this\cite{RSnellDefects,CAtkinsHIP}. The lattice structures used in these studies were generated from non-specialised computer aided design (CAD) packages, which do not have the functionality to model intricate geometry and handle the large file sizes associated with them, as such the lattices were not optimised for the mirror geometry and anticipated loads. 

To target short wavelength applications (visible - ultraviolet), mirrors with low surface roughness values are desired ($RMS \leq \SI{1}{nm}$). The achievable surface roughness with aluminium is limited (4-5 nm RMS\cite{CAtkinsHIP}) because it is a relatively soft metal. 3D printed ceramics such as fused silica using stereolithography and silicon carbide using binder jetting can be a suitable option for short wavelength mirrors, as they are mechanically harder and have lower coefficients of thermal expansion (CTE). 

There are several challenges associated with lightweight AM mirrors, for example:

\begin{itemize}
    \item \textbf{Porosity} - For PBF (laser and electron beam), porosity (voids in the material) is practically unavoidable, they can be caused by uneven powder distribution, gas bubbles, lack of fusion, too much fusion etc. Porosity reduces the mechanical properties\cite{PorosityMech} and fatigue life of a part\cite{PorosityonMechandFatigue}, and for mirror applications, they can degrade the reflective surface quality. 
    \item \textbf{Surface Roughness} - The surface roughness of AM parts is generally much larger than machined or cast parts, and it varies with the chosen printing technology, layer height and overhang angle to the build plate. Mirror surfaces will need additional diamond turning or polishing steps from any as-manufactured part, but the high surface roughness will necessitate machining mounting interfaces and features. Roughness can also impact fatigue life\cite{SurfRoughFatigueLife} and outgassing properties\cite{CBreenOutGassing}. 
    \item \textbf{Lack of Standardisation} - Previous AM mirror projects have used external manufacturing bureaux, who will use different machines/powders/settings and will have used different mixtures of virgin and recycled powder. The variations in the reflective surface quality of mirrors from different suppliers therefore, cannot necessarily be attributed solely to part geometry. There are factors outside of the designer's control and currently no widely adopted methods of material traceability.
\end{itemize}
    
The goal of this study is to explore the design freedoms and material choices that AM offers for mirror fabrication. The three primary objectives were to utilise AM's ability to mass reduce and part consolidate, as well as exploring metal and ceramic materials. This will be realised through the redesign of a mirror from a nanosat/CubeSat technology demonstrator. The AM mirror prototypes will aim to have mass reduction targets of 50\% and 70\% from an equivalent solid mirror, this will be achieved by using an internal lattice structure. The attachment points for the mirror to the CubeSat will also be consolidated into the part. Prototypes will be printed in both metal and ceramic to assess their suitability for future mirror fabrication after producing a reflective surface. AM designs will need to meet manufacturing requirements, both from the printing process and subtractive machining. Metrology will be performed on the reflective surface, however pursuing a perfect reflective surface was not in scope for this project, instead it will focus primarily on the design and manufacture of the prototypes.

The CubeSat application, along with the optical, mechanical and AM requirements are outlined in Section \ref{Sec:Mirror Specs and Reqs}. Section \ref{sec:Mirror Design} will explain the process of lattice design and downselection, in addition to the fixture point designs. To compare surface deformations between lattice designs, Zernike analysis will be performed on the final parts in Section \ref{sec:Zernikes}. Section \ref{sec:manufacturing} will cover the manufacturing, this includes printing, rough machining and single point diamond turning, metrology results will be presented in Section \ref{sec:metrology}. Sections \ref{Sec:Mirror Specs and Reqs}-\ref{sec:metrology} focus on the metal prototypes, progress on the fused silica prototypes will be reported in Section \ref{sec:Fused silica}. The paper concludes with a summary and future works.

\section{Mirror Specifications and AM Constraints} \label{Sec:Mirror Specs and Reqs}
\subsection{ADOT 6U CubeSat} \label{ADOT6UCubesat}

\begin{figure}[h]
    \centering
    \includegraphics[width=1\linewidth]{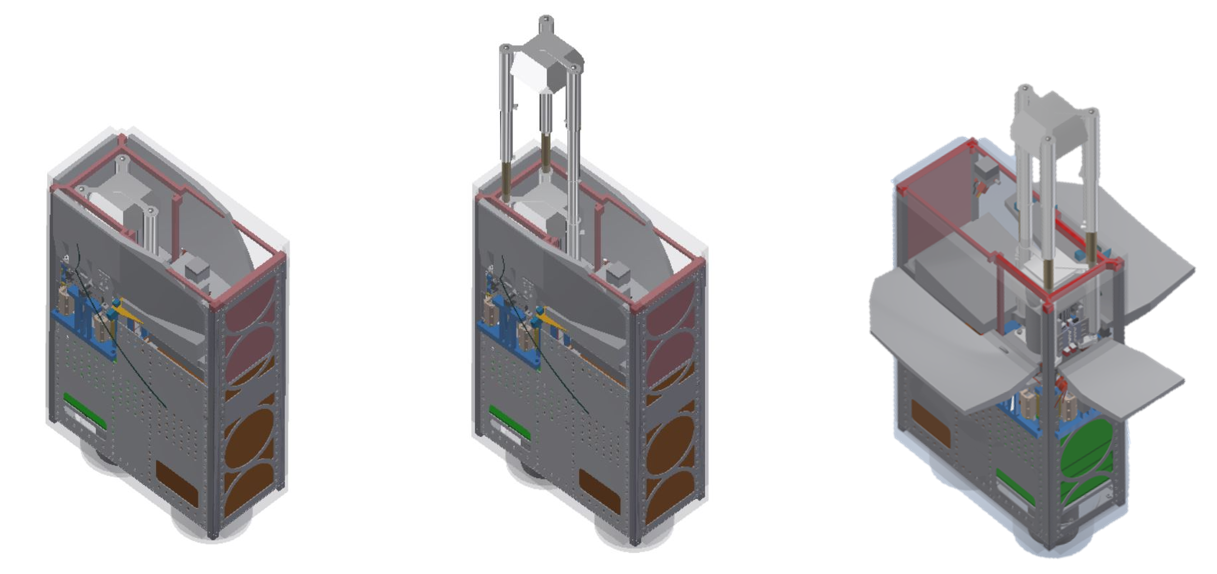}
    \caption{ADOT configurations, left to right: Fully stowed. Only secondary mirror deployed. All mirrors deployed.}
    \label{fig:ADOTs}
\end{figure}

The mechanical and optical requirements for the prototype mirror come from the Active Deployable Optical Telescope (ADOT) 6U CubeSat project (1U=10 cm\textsuperscript{3}), currently being undertaken in house at the UK Astronomy Technology Centre (UKATC)\cite{Cubesatpaper}. Since spatial resolution is proportional to aperture size; deployable mirrors are used to achieve a higher spatial resolution than a fixed aperture satellite with the same payload volume. The \SI{30}{cm} primary mirror is segmented into four petals, one of them is fixed within the CubeSat chassis, the remaining three rotate outwards and the secondary mirror deploys telescopically on three booms. Figure \ref{fig:ADOTs} shows how the primary petals fold out and the deployment of the secondary mirror. The proposed low cost, high resolution Earth/Astronomy observation satellite is intended to operate in the visible to near infrared range.

\subsection{Mirror Specifications} \label{Mirror Specifications}

\begin{figure}[h]
    \centering
    \includegraphics[width=0.5\linewidth]{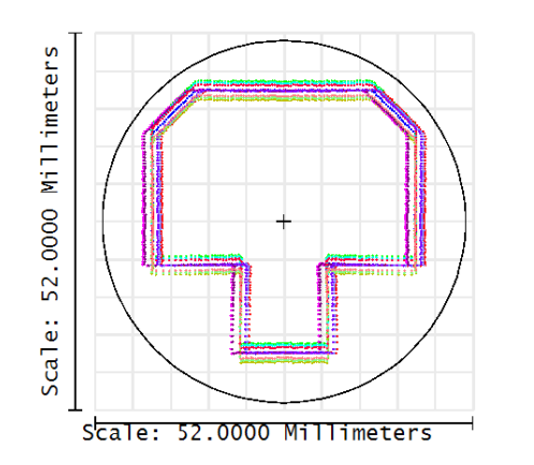}
    \caption{Circular AM approximation to M2. Black circle encompasses the reflection of the primary mirror (M1) at M2.}
    \label{fig:M2 approx}
\end{figure}

\begin{figure}[h]
    \centering
    \includegraphics[width=0.4\linewidth]{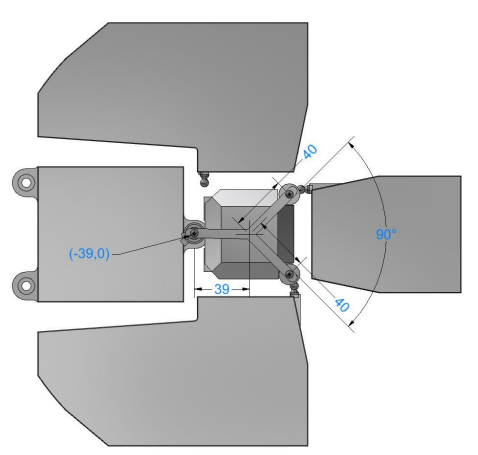}
    \caption{Top-down view of the mirror arrangement, with M2 mounting feature dimensions}
    \label{fig:Top Down Mirrors}
\end{figure}

The secondary mirror (M2) design, used in this study, is an approximation to the actual M2 mirror of ADOT (see Figure \ref{fig:M2 approx}). This was done to explore a circular geometry, more specifically, to explore how lattices, which usually consist of repeating cubic structures, could fit into a non-cubic geometry. The AM approximation to M2 is circular, convex, \SI{100}{mm} spherical radius of curvature (ROC), \SI{52}{mm} in diameter with a maximum height of \SI{10}{mm}. The thickness of the mirror substrate and walls is \SI{1}{mm}. Key specifications are summarised in Table \ref{tab:m2specs}. The mirror mounts on the three telescopic booms using M3 bolts, the bolt thread interfaces with the booms and are only a clearance fit with the mounts themselves, this will help to alleviate some of the issue of screw pressure distorting the mirror surface\cite{Westik}. Dimensions for the mounting locations can be found in Figure \ref{fig:Top Down Mirrors}. Considerations for the optical baffle are out of scope for this project. While there are no well defined volumetric constraints on the mounting features, their mass should be kept to a minimum while meeting structural requirements. 

\begin{table}[!h]
\caption{Summary table of key M2 specifications.} 
\label{tab:m2specs}
\begin{center}       
\begin{tabular}{|l|l|} 
\hline
     \textbf{Parameter} & \textbf{Specification} \\
\hline
\rule[-1ex]{0pt}{3.5ex}Material & Aluminium (AlSi10Mg), fused silica  \\
\hline
\rule[-1ex]{0pt}{3.5ex}Mass reduction & 50\% and 70\%  \\
\hline
\rule[-1ex]{0pt}{3.5ex}Mechanical aperture & \SI{52}{mm} \diameter  \\
\hline
\rule[-1ex]{0pt}{3.5ex}Clear aperture & \SI{50}{\mm} \diameter   \\
\hline
\rule[-1ex]{0pt}{3.5ex}Height & \SI{10}{mm} (at peak)   \\
\hline
\rule[-1ex]{0pt}{3.5ex}Mirror wall thickness & \SI{1}{mm}   \\
\hline
\rule[-1ex]{0pt}{3.5ex}Mounting solution & 3 x M3 bolts   \\
\hline
\rule[-1ex]{0pt}{3.5ex}Optical prescription & Spherical, convex  \\
\hline
\rule[-1ex]{0pt}{3.5ex}Spherical ROC & \SI{100}{mm}  \\
\hline
\rule[-1ex]{0pt}{3.5ex}Extra material on machined surfaces & \SI{1}{mm} \\
\hline
\rule[-1ex]{0pt}{3.5ex}Edge chamfer & \SI{0.5}{mm} x 45\degree \\
\hline
\rule[-1ex]{0pt}{3.5ex}Extra feature & 6 x Fiducial markers on sides \\
\hline
\end{tabular}
\end{center}
\end{table} 

\subsection{AM Design Considerations}
\label{sec:amdesigncons}

Design for AM introduces different constraints and freedoms compared to conventional manufacturing techniques, the distinctly different design rules that need to be adhered to and the overall change in design mindset can be a barrier to designing parts that make use of the advantages that AM has to offer. 

Print orientation is a significant challenge and is likely to dictate much of the geometry of a part. Most 3D printing processes cannot print large overhanging features, due to the nature of the machine building a part in layers, successive layers need to be placed on existing material, if the machine attempts to create overhanging features, there is increased chance of the material sagging down or in the case of PBF, warping up. Maximum overhang angles and unsupported horizontal distances vary depending on the AM process, material and machine settings, a general set of design parameters are presented by \textit{Atkins et al. (2019)}\cite{CookBook}  in the \textit{Opticon A2IM Cookbook} for several AM processes. If overhangs cannot be avoided by redesign or alternative print orientation, sacrificial support material can be used to ensure a successful build, this is not ideal however, as it requires additional time and effort to remove the supports, increases the print time and cost due to the extra material being processed, and leaves small defects on the part where the supports connected to the part.

Lattices are also affected by the constraints of print orientation, any internal lattices need to be self supporting during the build, since removing support material from within them will be impractical, if not impossible. The choice of unit cells, which is the single repeating cuboid in the structure, will determine if a lattice is self supporting, the printability of the lattice in all build directions should also be considered in the design phase. The appeal of using lattice structures is to reduce mass while maintaining stiffness, but not all unit cells have the same mechanical properties. Lightweight mirrors that use lattices under the substrate have previously suffered from print through effects after machining\cite{XRayOpticsprintthru}, this is a result of material deflecting more in between where lattice members support the substrate, attempts should be made to even out a surface deflection by first choosing appropriate unit cells, then modifying the thickness of the walls if needed.

After printing, parts usually require some form of post-processing, this can differ with the process used, for the prototype secondary mirrors, which will be printed using PBF and SLA, the immediate step after printing is to remove the excess powder or resin. For intricate internal geometries, there needs to be some escape holes to let the excess material flow out, if there is still trapped powder/resin, this could effectively increase the mass of the part, and contaminate future machining operations, there are also health and safety risks associated with fine powder. 

Printed parts do not necessarily conform to the original digital design file, powder based printing processes especially can see large geometric deviations. Rough machining is needed to transform the coarse and uneven surface from the printer to a uniform and flat one prior to SPDT. As such this will require additional material to be added to any surfaces that need to be machined. This stage will also require some work holding solution, given the geometric freedoms of AM, it is possible to design complex parts that are difficult or impossible to hold for machining. Additional features can be created on the part itself to assist with this, or custom one-off fixtures can be printed to stabilise and locate complex parts. 

\section{Mirror Design} \label{sec:Mirror Design}

\subsection{Lattice Design} \label{Lattice Design}

\begin{figure}[!h]
    \centering
    \includegraphics[width=0.4\linewidth]{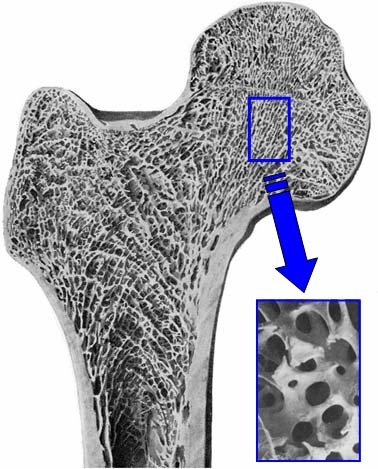}
    \caption{Cancellous bone structure. Image credit: \textit{Kunio et al.(2010)\cite{BonePaper}}}
    \label{fig:bonestructure}
\end{figure}

Though various lattices are seen across engineering, parallels can be drawn between them and natural structures. Stochastic (randomly distributed) lattices (Figure \ref{fig:UnitCells}) for example, share similarities to the porous structure found in bones (Figure \ref{fig:bonestructure}). To mimic these natural designs (biomimicry) and harness their increased specific stiffness to the fullest extent, a new manufacturing technique, like AM, is needed. Section \ref{introduction} highlighted the difficulty of traditional CAD to create and manipulate lattices in the 3D modelling environment, to overcome this challenge, an AM specific modelling package will be used. The software uses implicit modelling\cite{ImplicitModelling}, unlike traditional CAD, which not only allows for easy and fast modelling of large or intricate lattices, but also makes it possible to manipulate and optimise lattices and bodies with tools such as field-drive design\cite{FieldDrivenDesign}.

\subsubsection{Downselection}
\label{sec:downselection}

\begin{figure}[!h]
  \centering
  \begin{minipage}[!]{0.2\textwidth}
    \includegraphics[width=\textwidth]{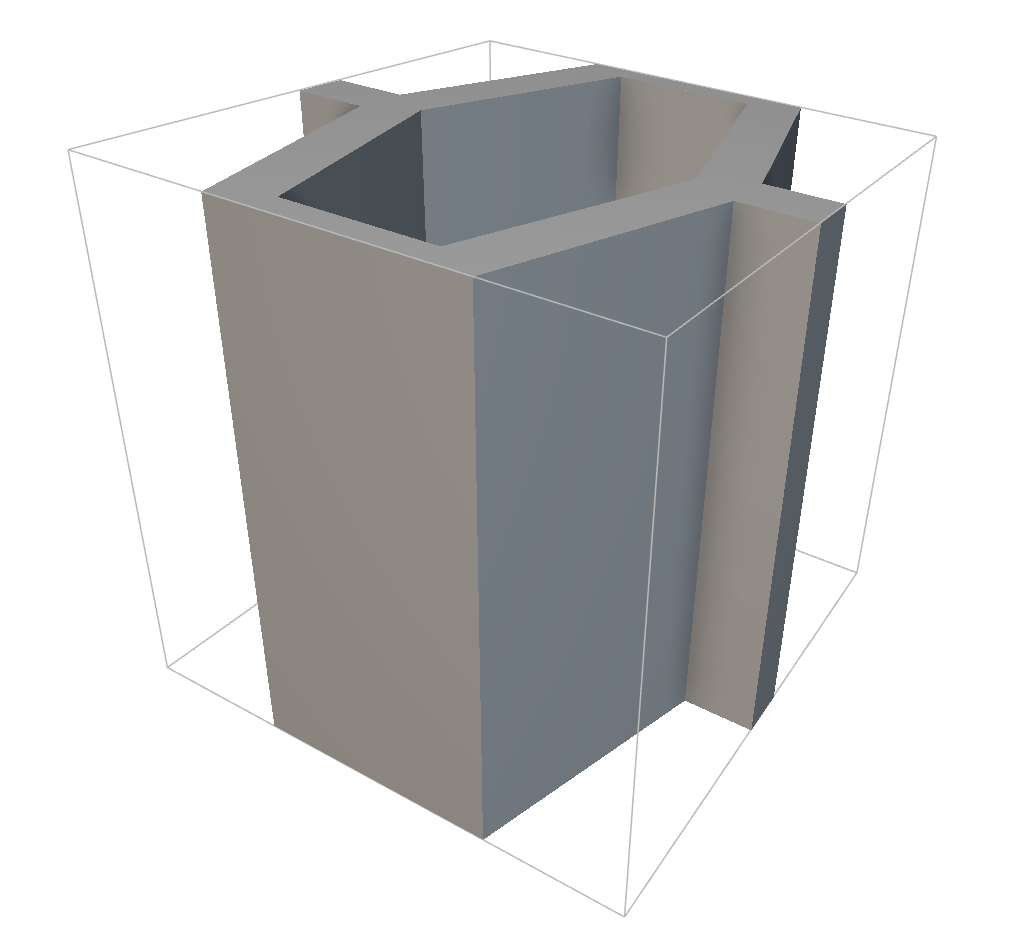}
    \label{fig:Honeycombunitcell}
  \end{minipage}
  \hfill
  \begin{minipage}[!]{0.2\textwidth}
    \includegraphics[width=\textwidth]{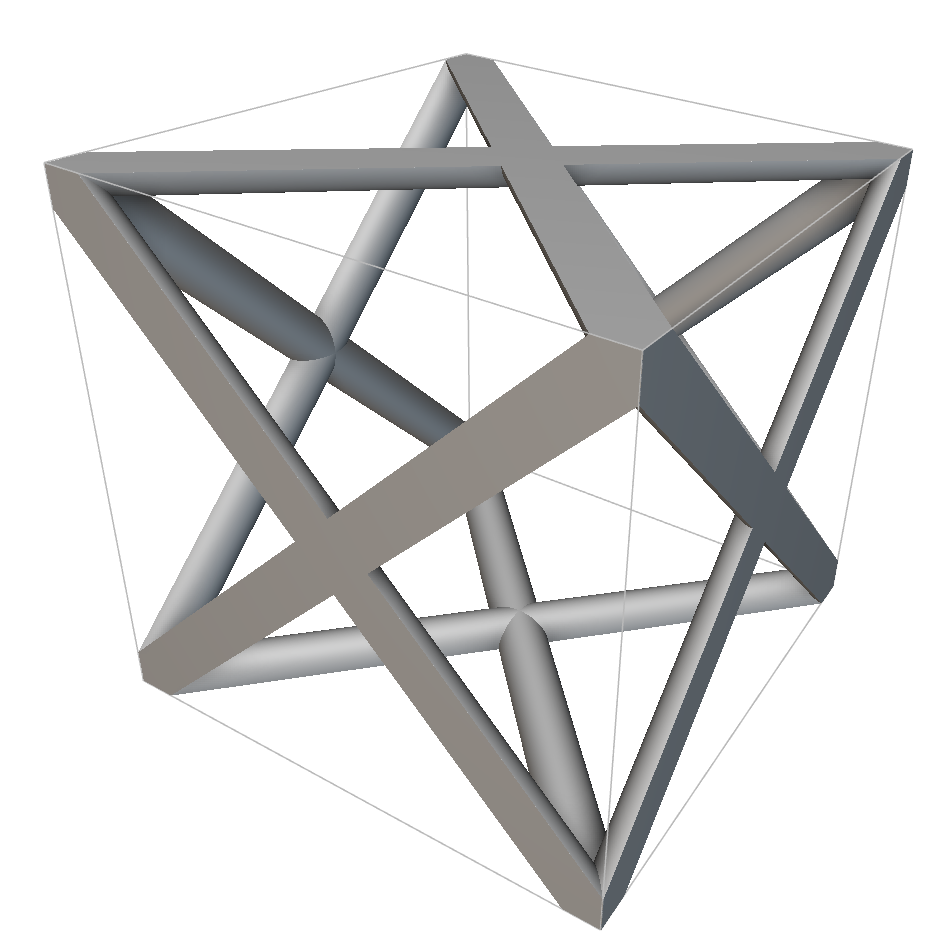}
    \label{fig:FluoriteUnitCell}
  \end{minipage}
  \hfill
  \begin{minipage}{0.2\textwidth}
    \includegraphics[width=\textwidth]{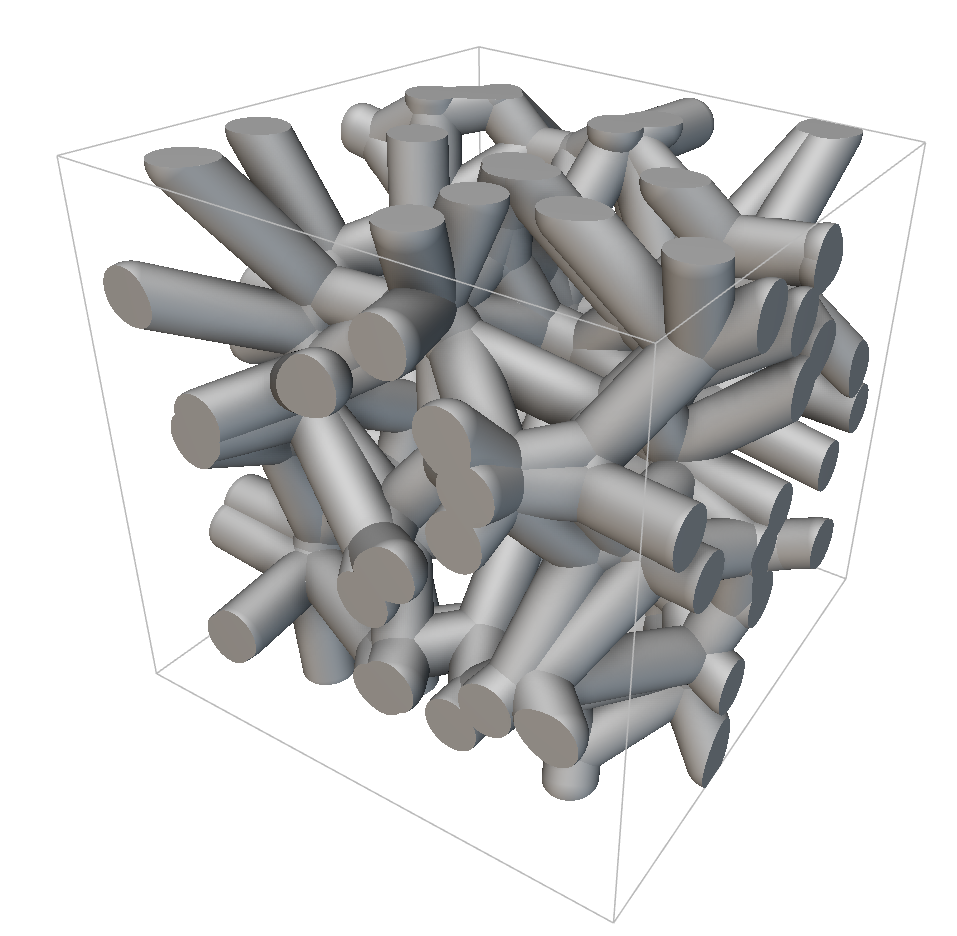}
    \label{fig:GyroidUnitCell}
  \end{minipage}
  \hfill
  \begin{minipage}{0.2\textwidth}
    \includegraphics[width=\textwidth]{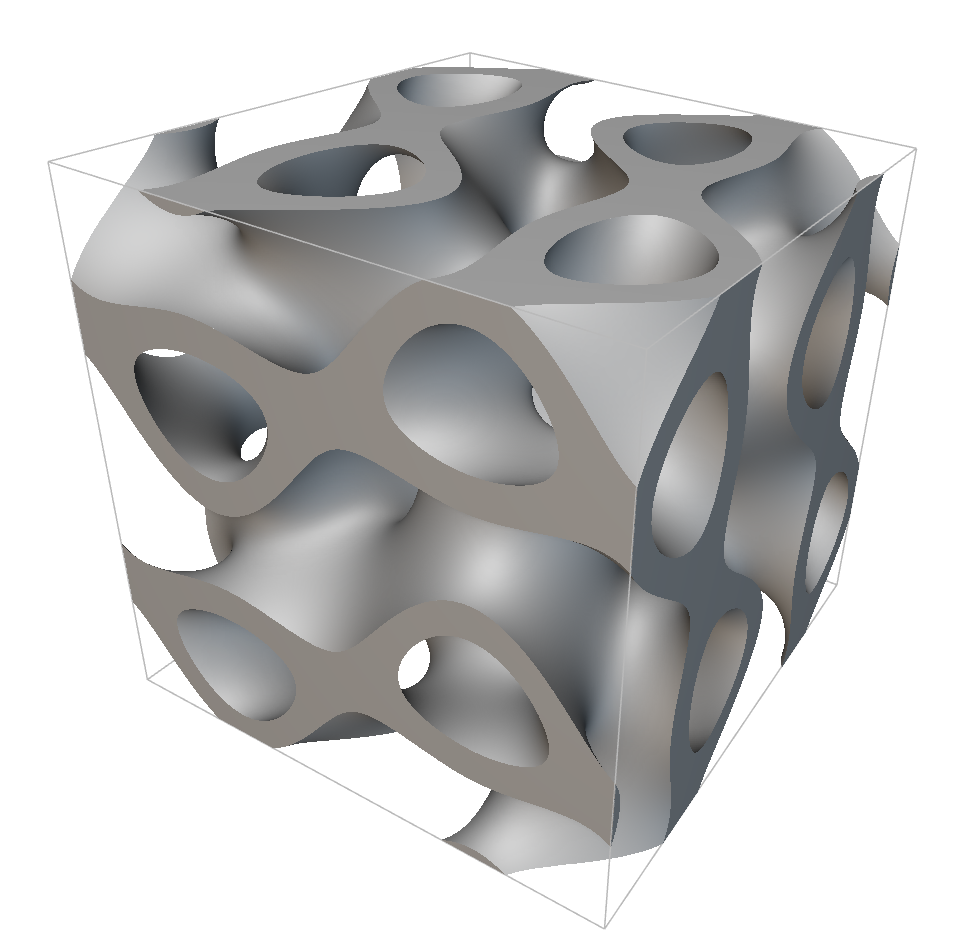}
    \label{fig:Vornoi}
  \end{minipage}
  \caption{Left to right: Honeycomb unit cell (Hexagonal), strut-based unit cell (Face centred cubic), TPMS unit cell (SplitP), stochastic unit cell (Voronoi)}
  \label{fig:UnitCells}
\end{figure}

There are 33 unit cells to choose from across several categories; honeycomb, strut/graph, stochastic and triply periodic minimal surface (TPMS), examples of each can be seen in Figure \ref{fig:UnitCells}. Honeycomb unit cells have constant cross sections along one direction and are typically composed of sheets or surfaces. These are seen used in conventionally lightweighted mirror structures. Strut unit cells are comprised of arrangements of uniform cylindrical beams, they are the simplest type of 3D lattice and the least computationally demanding, which makes them ideal for modelling in traditional CAD. Stochastic lattices also consist of beams, but are arranged in a random order, they often have uneven mechanical properties due to the variation in structure density and introduce too much uncertainty into the design process, therefore they are not suitable for a lightweight mirror design. TPMS unit cells are defined parametrically by a surface stretched over a set of points with the smallest surface area, they can more efficiently distribute material for a given volume and reduce local stress concentrations present in strut lattices. Strut unit cells have been previously shown to have worse mechanical properties compared to TPMS unit cells\cite{UnitCellComparison,TPMSSuperior}.

\begin{figure}[!h]
  \centering
  \hspace*{\fill}%
  \begin{minipage}{0.3\textwidth}
    \includegraphics[width=\textwidth]{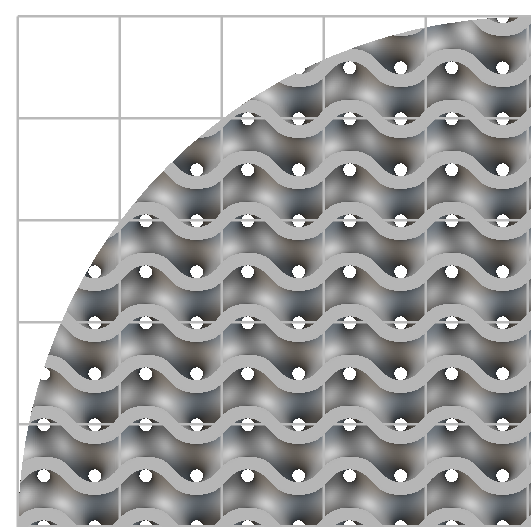}
    \caption{Cubic cell map overlaid onto a cylindrical volume with Gyroid unit cells}
    \label{fig:CubicGyroid}
  \end{minipage}%
  \hspace*{\fill}%
  \begin{minipage}{0.3\textwidth}
    \includegraphics[width=\textwidth]{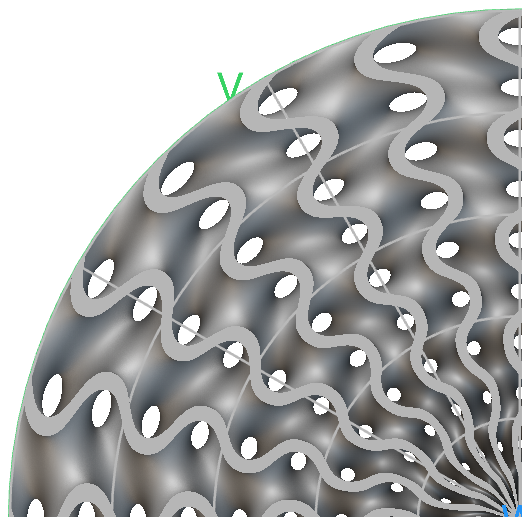}
    \caption{Cylindrical coordinate cell map overlaid onto a cylindrical volume with Gyroid unit cells}
    \label{fig:ConformalGyroid}
  \end{minipage}%
  \hspace*{\fill}%
\end{figure}

Conformal lattices were also investigated to assess their structural performance and suitability for a circular mirror. Typically, lattices are cubic structures, like those in Figure \ref{fig:UnitCells}, that repeat uniformly in the three axes of a Cartesian coordinate system. This is implemented in the AM software by creating a `cell map', which is a representation of the volumes that the unit cells will populate. An issue rises when the volume of a part intended for a lattice is not rectangular, in the case of circular or ellipsoidal mirrors for example. The outer cells are frequently truncated, see Figure \ref{fig:CubicGyroid}, leaving behind unsupported or unconnected geometry, this is a particular issue for strut lattices. A conformal lattice, however, does not use fixed shape cells in a uniform arrangement, instead the cells deform and stretch to match the boundary of the volume, leaving behind no incomplete cells, this increases the potential solution space and should be investigated for potentially superior designs. 
 
 Conformal lattices have been discussed in literature in terms of design approach and methodology\cite{ConformalTheory, RobSconformal} but printed examples are limited\cite{ConformalActual, Conformalntop}. For the case of a circular conformal lattice, a cylindrical coordinate system is used to map the unit cells into, resulting in a polar grid structure, which can be seen in Figure \ref{fig:ConformalGyroid}. If the radius of the cells is a multiple of the volume radius, the cell map will terminate at the edge of the volume, as in Figure \ref{fig:ConformalGyroid}, if not, then the lattice will need to be trimmed down, but the continuous nature of TPMS unit cells is not as affected as having truncated cells. One apparent issue is the cells becoming more spaced out along the radial axis, for supporting a mirror surface, this uneven distribution in lattice density is undesirable as the edges will deflect more than the centre under a polishing/turning load, increasing the surface form error. 

Several stages of downselection were implemented to converge on a set of prototype designs, this involved comparing different unit cells and lattice arrangements (cubic vs conformal) based on several criteria. The first stage eliminated unit cells based on general printability constraints set out in Section \ref{sec:amdesigncons}. This removed a number of strut-based unit cells that cannot be confidently printed in any orientation without support. Although some of these unit cells have been shown to print without failure at a small scale\cite{overhangunitcellexample}, the prototype with 70\% mass removed will have a more widely spaced lattice, making unsupported bridges pose a greater threat. Additionally, the TPMS cell Lidinoid was removed as it has enclosed volumes that would trap powder/resin. 

\begin{figure}[!h]
  \begin{subfigure}{0.31\textwidth}
    \includegraphics[width=\linewidth]{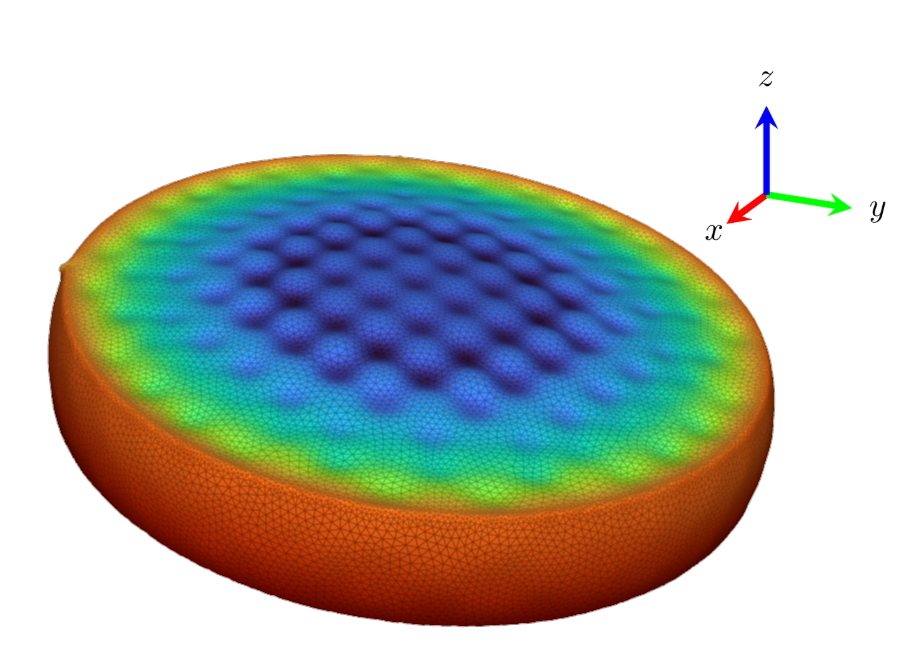}
 \label{fig:press}
  \end{subfigure}%
  \hfill
  \begin{subfigure}{0.31\textwidth}
    \includegraphics[width=\linewidth]{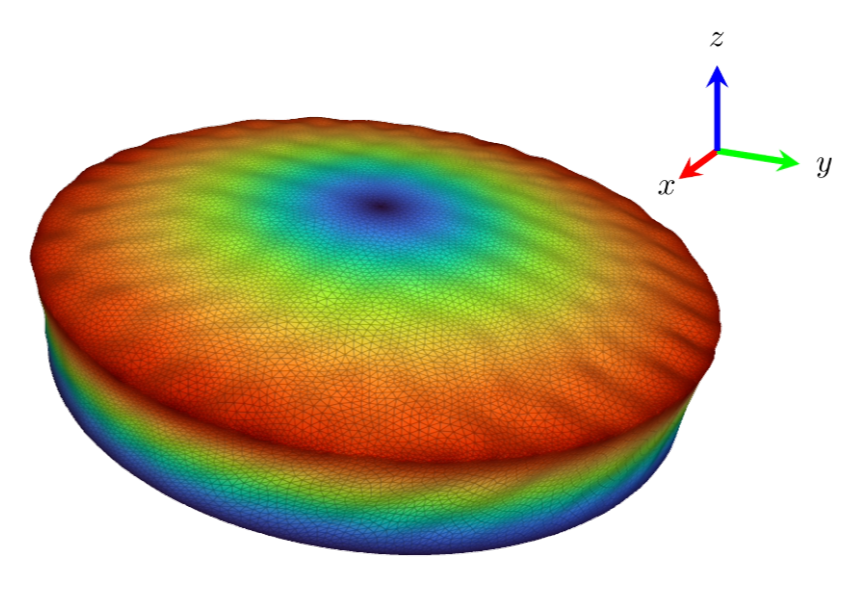}
    \label{fig:torque}
  \end{subfigure}%
  \hfill   
  \begin{subfigure}{0.31\textwidth}
    \includegraphics[width=\linewidth]{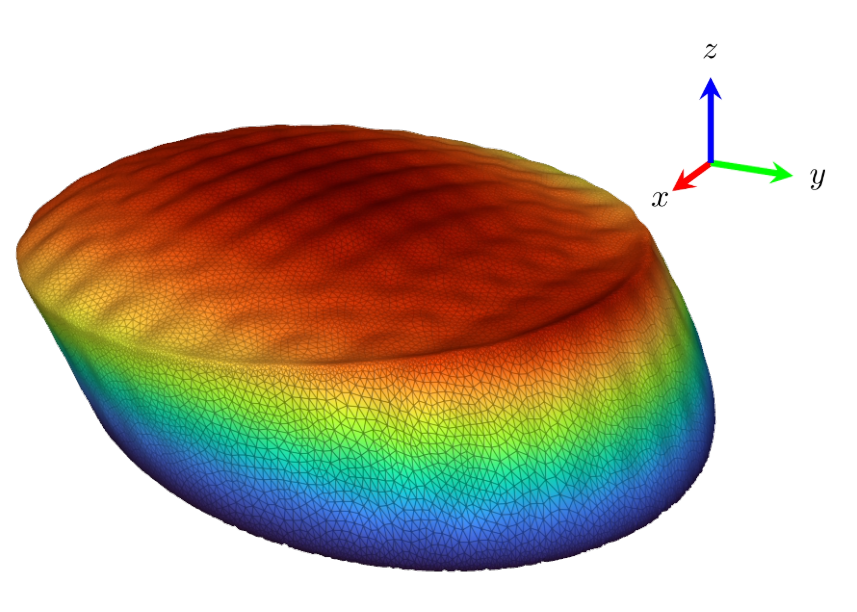}
     \label{fig:lat}
  \end{subfigure}
  \caption{Scaled FEA deformations from uniform pressure (left), torque (middle), and lateral force (right)}
  \label{fig:boundary conditions}
  \end{figure}

Numerical simulation was then used to compare remaining options based on mechanical properties. Unit cell simulations have been performed in literature \cite{UnitCellComparison} and there are online databases such as \textit{LatticeRobot} that document the structural responses of a range of common unit cells. This data can be very useful in performing initial downselection, removing the need for time-consuming and computationally demanding simulations. However, given the inclusion of conformal lattices in the study and nature of a circular lattice volume, custom simulations were performed. The main loading condition that influenced downselection was a compressive turning or polishing pressure acting evenly on the mirror surface, an approximate value for this pressure is \SI{3500}{Pa}\cite{polishingpressure} since the exact boundary condition load for this type of machining process is very difficult to replicate in a finite element analysis (FEA) model. This load is the primary cause for print-through, or quilting \cite{XRayOpticsprintthru}, the other loads that were considered are a torque around the centre of the mirror, as well as two lateral shearing forces at 45\degree apart, Figure \ref{fig:boundary conditions} demonstrates these loads.

The model was restrained by fixing the bottom flat surface in \textit{z} only and fixing two vertical edges in \textit{x} or \textit{y} to prevent rotation\cite{polishingpressure}. Due to the unit cells each having a different volume fraction, i.e the ratio of solid lattice volume to the volume of the unit cube, some lattice parameter will need to change for each analysis to maintain the target mass reduction. The chosen method was to vary the wall/strut thickness for each unit cell and keep the cell size constant. The test model has the same lattice volume that the prototype mirror will have, initial test results were skewed by edge effects (very high displacements), this was solved by adding the \SI{1}{mm} wall around the edge.

A \SI{0.4}{mm} edge length was chosen for the FEA mesh, this value was a balance between capturing the intricate geometry accurately enough and minimising the computing power and time required to simulate each lattice. Generating meshes for additive manufacture or simulation is challenging for complex lattice structures, a relatively coarse mesh will not accurately model the lattice and is likely to incur errors, but the mesh size can quickly become impractical if a relatively fine setting is used. It is common practice to perform a mesh convergence study prior to FEA, however, given the challenges faced, this was not performed and the chosen edge length and mesh settings were kept constant across all simulations. Even with relatively fine mesh, some unit cells failed to produce results. 

There are other options for simulating lattices, such as meshing strut-based lattices using single beam elements, or homogenising the lattice by using a solid with the equivalent mechanical properties of the unit cell, though these may be quicker, they neglect edge effects, stress concentrations, are not applicable for all types of unit cell and lattice arrangement and are not as accurate as solid elements. 

\begin{figure}[!h]
  \begin{subfigure}[t]{0.5\textwidth}
    \includegraphics[width=\linewidth]{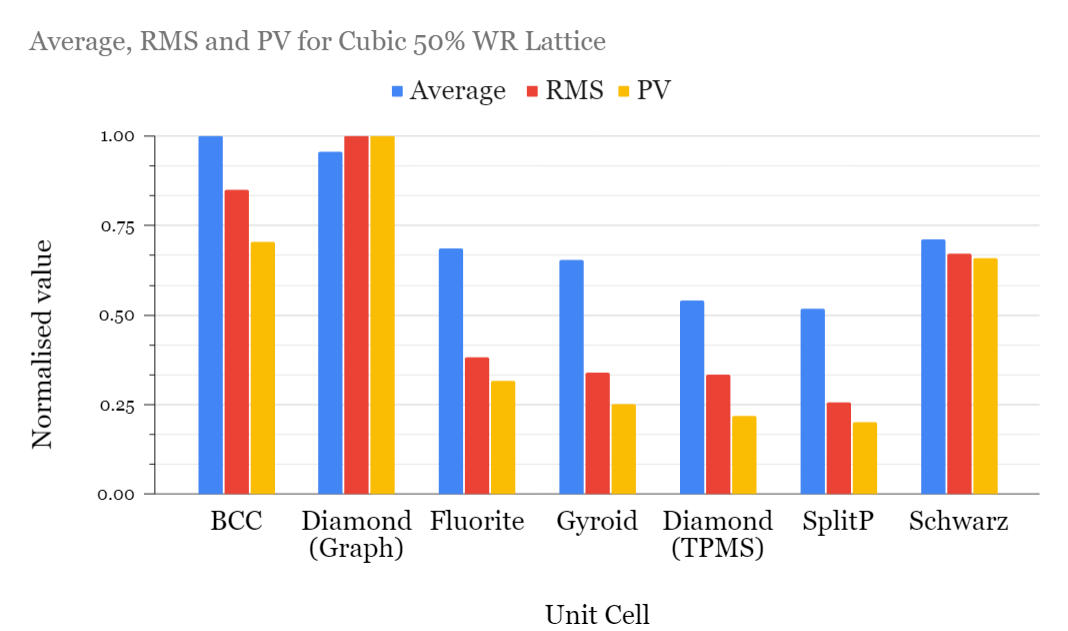}
  \end{subfigure}
  \hfill
  \begin{subfigure}[t]{0.5\textwidth}
    \includegraphics[width=\linewidth]{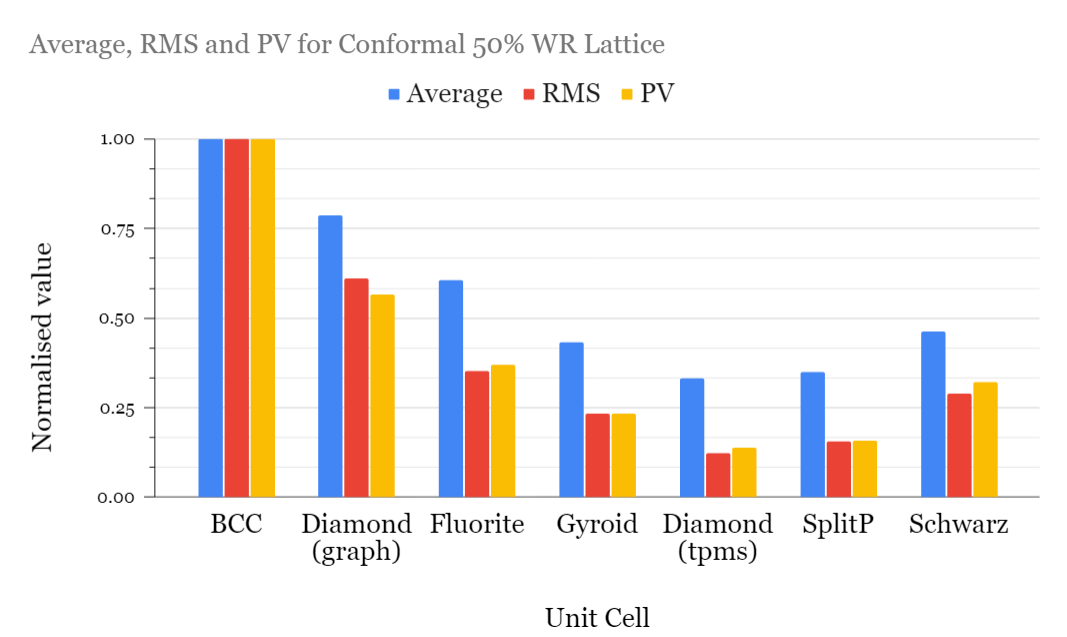}
  \end{subfigure}
  \caption{Average, RMS and PV results from uniform \SI{3500}{Pa} pressure for cubic lattice (left) and conformal lattice (right) at 50\% WR}
  \label{50Lattices}
  \end{figure}

\begin{figure}[!h]
  \begin{subfigure}[t]{0.5\textwidth}
    \includegraphics[width=\linewidth]{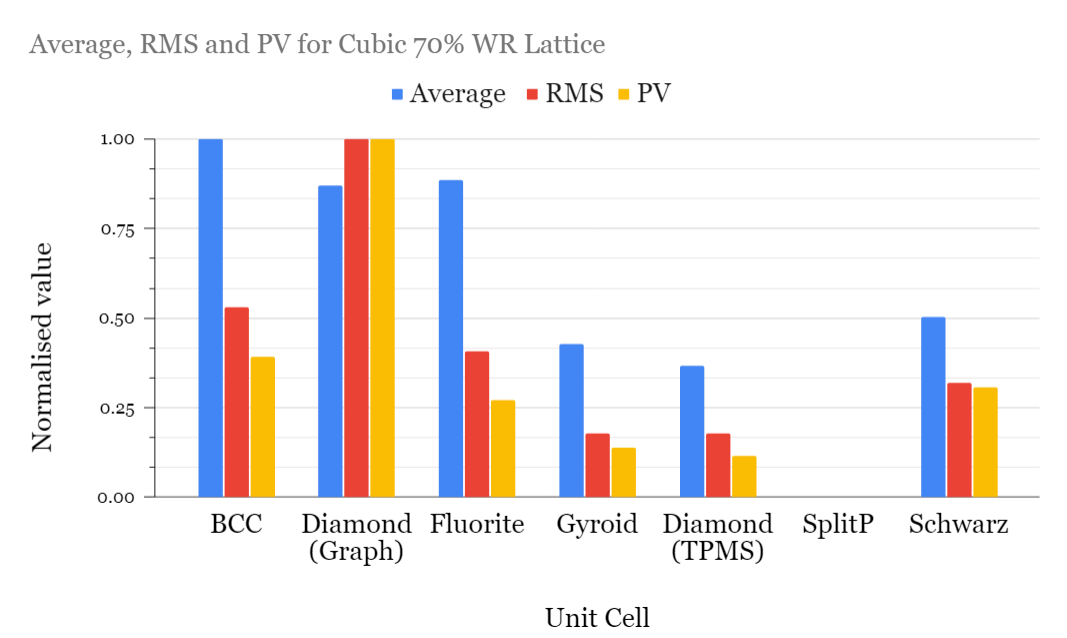}
  \end{subfigure}
  \hfill
  \begin{subfigure}[t]{0.5\textwidth}
    \includegraphics[width=\linewidth]{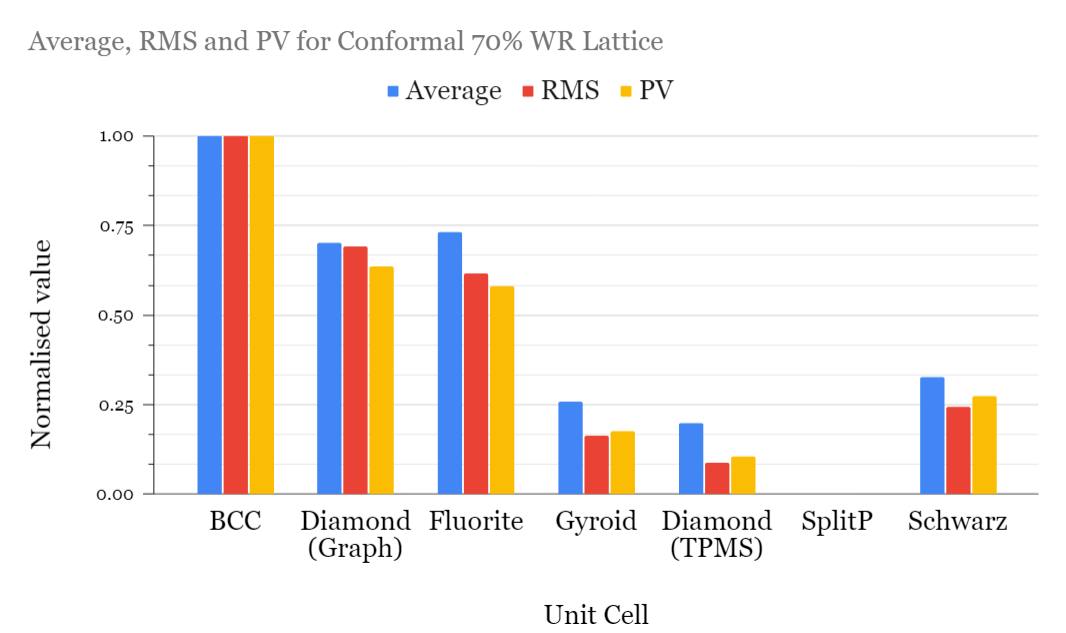}
 \end{subfigure}
  \caption{Average, RMS and PV results from uniform \SI{3500}{Pa} pressure for cubic lattice (left) and conformal lattice (right) at 70\% WR}
  \label{70Lattices}
 \end{figure}

For each load, the displacement map of the reflective surface was exported to MatLab for data analysis. Average, root mean square (RMS) and peak to value (PV) values were normalised and used for comparison. The results for the uniform pressure with 50\% and 70\% weight removed (WR) lattices are presented in Figures \ref{50Lattices} and \ref{70Lattices}. From the analysis, TPMS unit cells perform better than strut ones, which agrees with similar studies \cite{UnitCellComparison,TPMSSuperior}, both Diamond (TPMS) and SplitP share the lead for best performance, with BCC usually performing the worst out of the remaining options. The difficulty with simulating lattices is highlighted with SplitP at 70\% WR, which failed to mesh, even with fine mesh parameters, the required quality to get a successful mesh would increase the time of the other simulations beyond reason.

\begin{figure}[!h]
    \centering
   \includegraphics[width=0.75\linewidth]{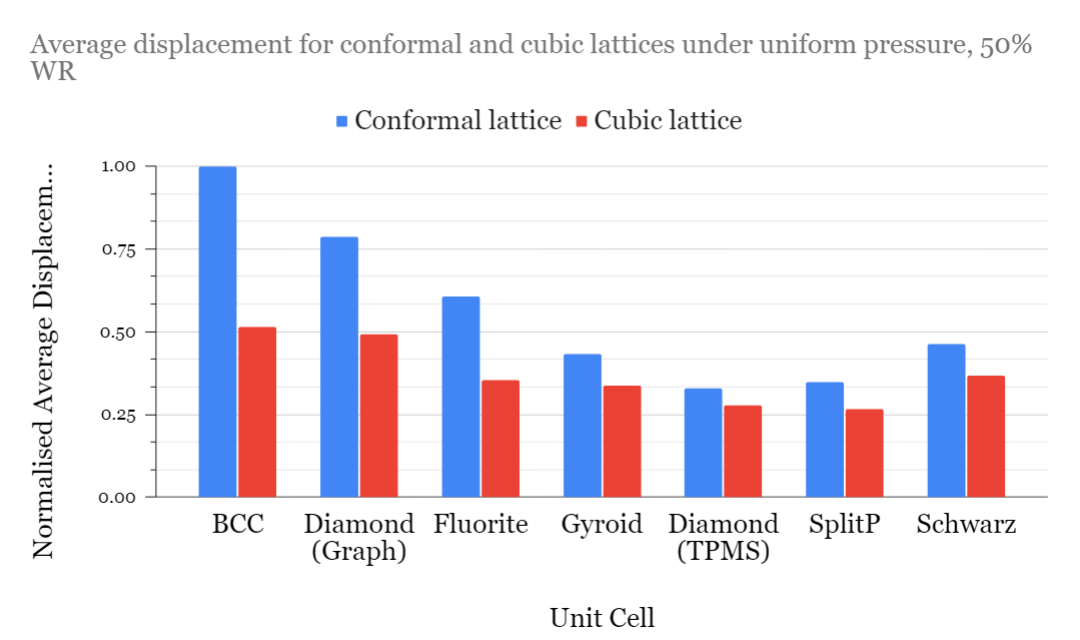}
   \caption{Conformal Vs cubic lattices under uniform pressure}
    \label{fig:conformalvscubic}
\end{figure}

Figure \ref{fig:conformalvscubic} highlights conformal lattices performing slightly worse than cubic ones, despite being a better fit within the circular lattice volume, the difference is larger for graph unit cells (BCC, Diamond, Fluorite), but TPMS unit cells have similar performance. Results from the torque and lateral forces also showed that Diamond (TPMS) and SplitP perform the best for all lattice arrangements and mass removed percentages, graphs for these were omitted as they are not the primary load to consider.

It was concluded from simulations that for the 50\% weight remaining prototype, both conformal and cubic lattices would be suitable. For the cubic lattices, Fluorite and Diamond were chosen as the graph and TPMS unit cells respectively, Schwarz and Gyroid were eliminated based on inferior performance to Diamond, SplitP was removed due to previously observed difficulties in powder removal via tests\cite{Westik}. For the conformal lattices, graph unit cells were shown not to be suitable due to how struts that support the reflective surface would be distributed in the polar grid arrangement, there would be large gaps between supporting members which could compromise printability, TPMS unit cells have more continuous support that lends them better to such a conformal lattice. For the 70\% WR prototype, conformal lattices did not perform as well as the cubic option, the decision was made to pursue the conformal lattice only for the 50\% WR prototype to allow for more flexibility in the design. 

\subsubsection{Conformal Lattice Design}
\label{sec:conformalatdesign}
\begin{figure}[!h]
 \centering
  \begin{minipage}{0.4\textwidth}
    \includegraphics[width=\textwidth]{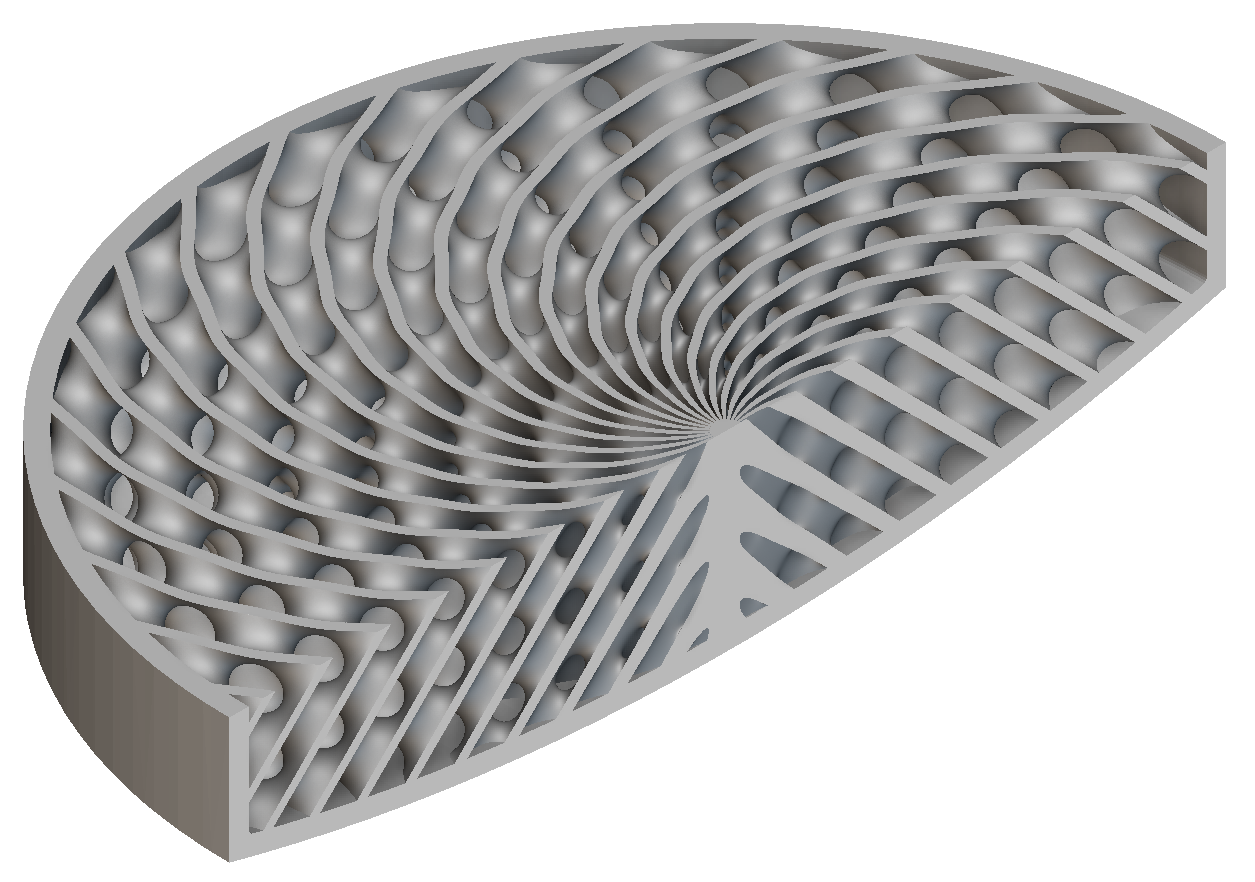}
    \caption{Sectioned CAD model of a conformal lattice using Diamond TPMS unit cell}
   \label{fig:ConformalExample}
  \end{minipage}
 \hfill
 \begin{minipage}{0.5\textwidth}
    \includegraphics[width=\textwidth]{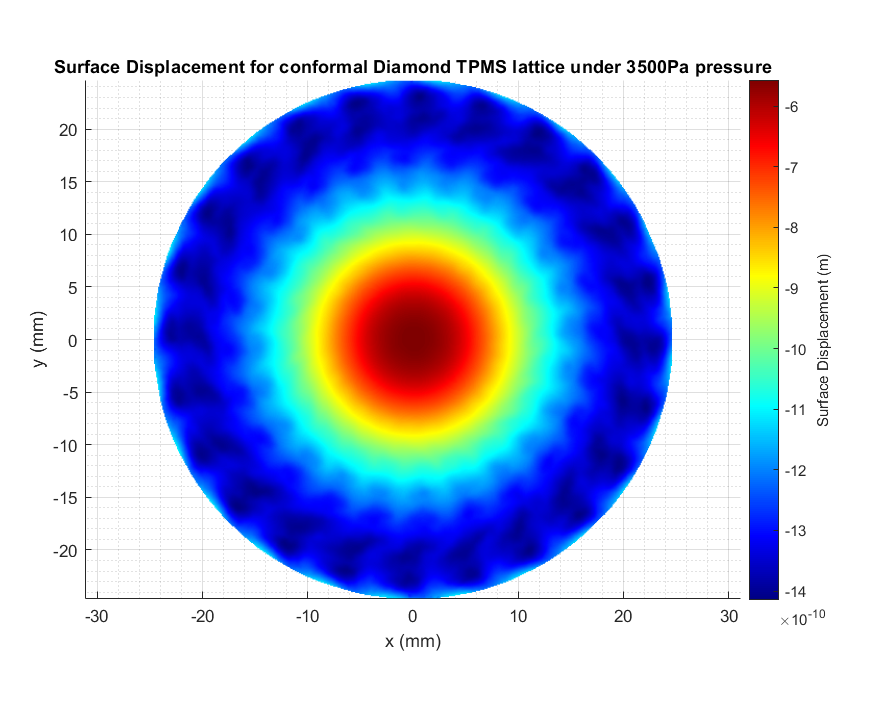}
    \caption{Surface displacement plot of conformal Diamond TPMS in MatLab}
    \label{fig:Matlabdisplconformal}
  \end{minipage}
\end{figure}

After deciding to take the conformal lattices forward, modifications were needed to optimise the lattice for a mirror. As shown in Figures \ref{fig:ConformalGyroid} and \ref{fig:ConformalExample}, the unmodified lattice becomes denser towards the centre and vice versa, to the point where there is a solid cylinder of material at the centre, this results in a very uneven displacement response, which is clearly shown in the displacement map of the surface (Figure \ref{fig:Matlabdisplconformal}). 

One solution to reduce this effect was to vary the wall thickness of the lattice radially, from thinner in the centre to thicker towards the edges. The displacement data from the unmodified lattice was used as the scale for the varying thickness, where higher displacements were observed in the original lattice, the thickness value of the new lattice will be higher and vice versa, the minimum and maximum values to vary the thickness over were chosen to meet the desired mass reduction target. It was found that after simulating the new lattice, the issue of the over supported centre was still present, though to a much lesser extent and the outer annulus of the surface had much more even support. 

\begin{figure}[!h]
 \centering
  \begin{subfigure}[t]{0.48\textwidth}
    \includegraphics[width=\textwidth]{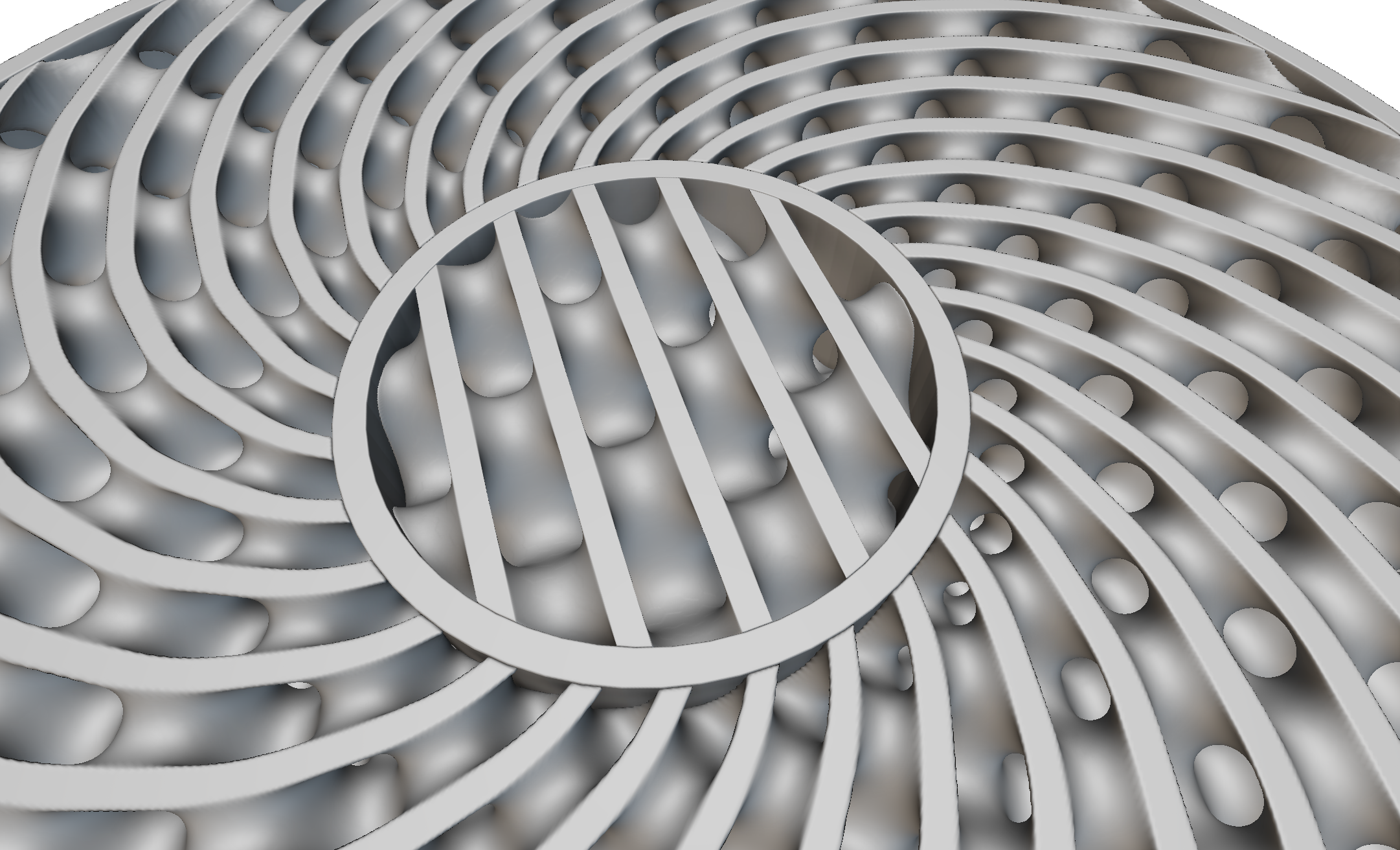}
    \caption{Centre of the conformal lattice replaced with a cubic lattice}
   \label{fig:innerwallsolid}
  \end{subfigure}
  \hfill
 \begin{subfigure}[t]{0.48\textwidth}
    \includegraphics[width=\textwidth]{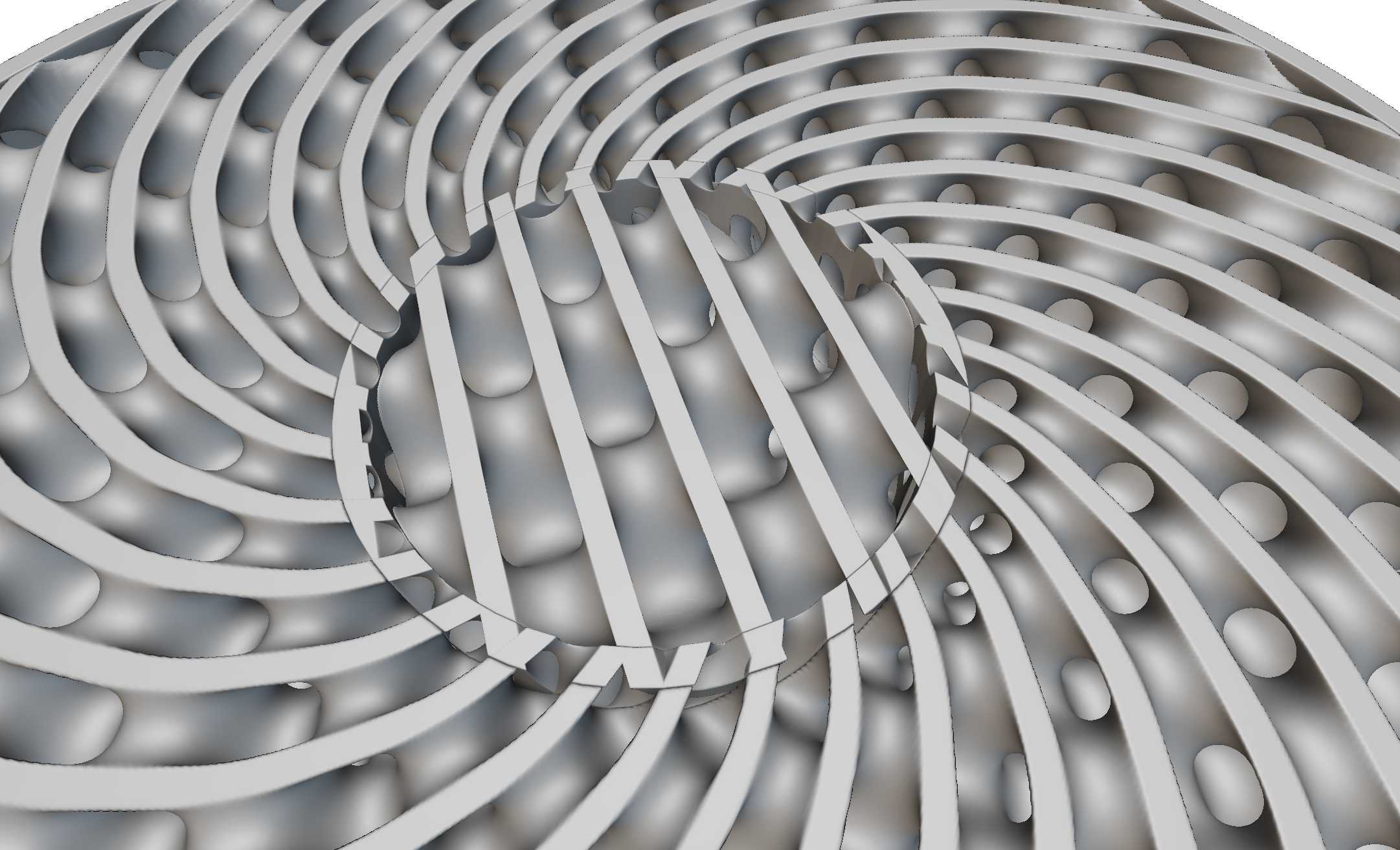}
    \caption{Separating solid wall changed for a latticed wall}
    \label{fig:innerwalllatticed}
  \end{subfigure}
  \caption{Conformal lattice design}
\end{figure}

Even with a very low minimum thickness value, the over supported column of material could not be avoided, the printability and powder removal from this central section becomes an increasing concern the smaller the minimum value goes. To overcome this, the decision was made to remove a portion of the centre and replace it with a cubic lattice, which has more even support. Figure \ref{fig:innerwallsolid} shows how this was implemented, with a solid \SI{0.8}{mm} cylindrical wall separating the two lattices. After simulating the same uniform pressure in Section \ref{sec:downselection}, it was found that the solid wall caused an over supported ring around the cubic lattice. The purpose of replacing the centre of the conformal lattice was to even out the support across the mirror surface, but the solid wall was presenting a similar problem, just in a different location. To fix this, several iterations of a lightweighted separating wall were tried, ultimately, both the cubic and conformal lattices were extended into the \SI{0.8}{mm} gap, then a Boolean union operation joined all of the lattices together, the result of this can be seen in Figure \ref{fig:innerwalllatticed}. Some of the resulting small features and gaps in the separating wall were below the minimum recommended sizes for most L-PDF printers, but it was not deemed a issue if they did not print successfully. 

\subsubsection{Open back Vs Closed back}

To increase bending stiffness, mirrors can use a sandwich design, this is ideal for AM since the internal structure can be printed with no additional bonding steps like conventional sandwich mirrors. Initially, a sandwich structure was used, but this limited the flexibility of the design since the lattice volume is quite small. This meant that in order to reduce the mass of the equivalent solid mirror by 70\%, the lattice density needed to be extremely sparse, the surface deflection with such a lattice was arguably worse than with no lattice at all, additionally, this became very challenging to print with respect to orientation. The decision was made to adopt an open back design for both mass reduction prototypes, this greatly increased flexibility in the lattice design and made powder removal much less challenging.

\subsubsection{Powder Removal Test}
\label{sec:powderremovaltest}

When iterating towards a final lattice design, there are several factors to consider when deciding on cell size and wall/strut thickness. Investigations into the effect of lattice density on structural performance found that in general, a denser lattice of the same mass performs better, this means decreasing both the wall thickness and cell size. However, there is a limit to how small both of these can go. The aluminium prototypes will be printed using a commercial bureau, who will have minimum recommended wall thicknesses, these typically range from \SI{0.8}{mm} down to \SI{0.5}{mm}, beyond which the thermal distortions are likely to affect the geometry significantly. If a lattice is made too dense in the design stage, there is a possibility that the excess powder cannot be easily removed. This is difficult to generalise since the ease that powder flows will depend on the part geometry and unit cell choice. 

There are general rules about minimum hole and feature sizes that can print successfully\cite{CookBook} available in literature, but these don't necessarily provide enough information to determine if a lattice is printable. \textit{Westsik et al. (2023)}\cite{Westik} printed a replica model of a lightweight mirror in a transparent polymer using SLA, then assessed the powder removal capabilities of several unit cells using red sand to simulate the metal powder. Given the relatively dense nature of the lattices for the M2 mirror, in addition to the novelty of a conformal lattice, a similar test was performed. Tests were not printed for the 70\% WR prototype since we were confident that trapped powder would not be a problem.

The polymer model was designed in such a way to assess powder removal through both an open back and side wall holes, as the print orientation was not yet decided. There were four lattice designs to test, two cubic: Diamond (TPMS) and Fluorite (Graph) and two conformal: Diamond (TPMS) and Gyroid (TPMS). These made up the `core' of the test model, which had a removable top and bottom plate. Holes were placed in the side walls to assess the powder removal capability if the mirror is printed convex surface up, for the conformal lattices, side holes were strategically placed where the radial `channels' meet the side walls. The holes were covered up when they were not used during the tests.  

  
\begin{figure}[!h]
 \centering
  \begin{minipage}{0.4\textwidth}
    \includegraphics[width=\textwidth]{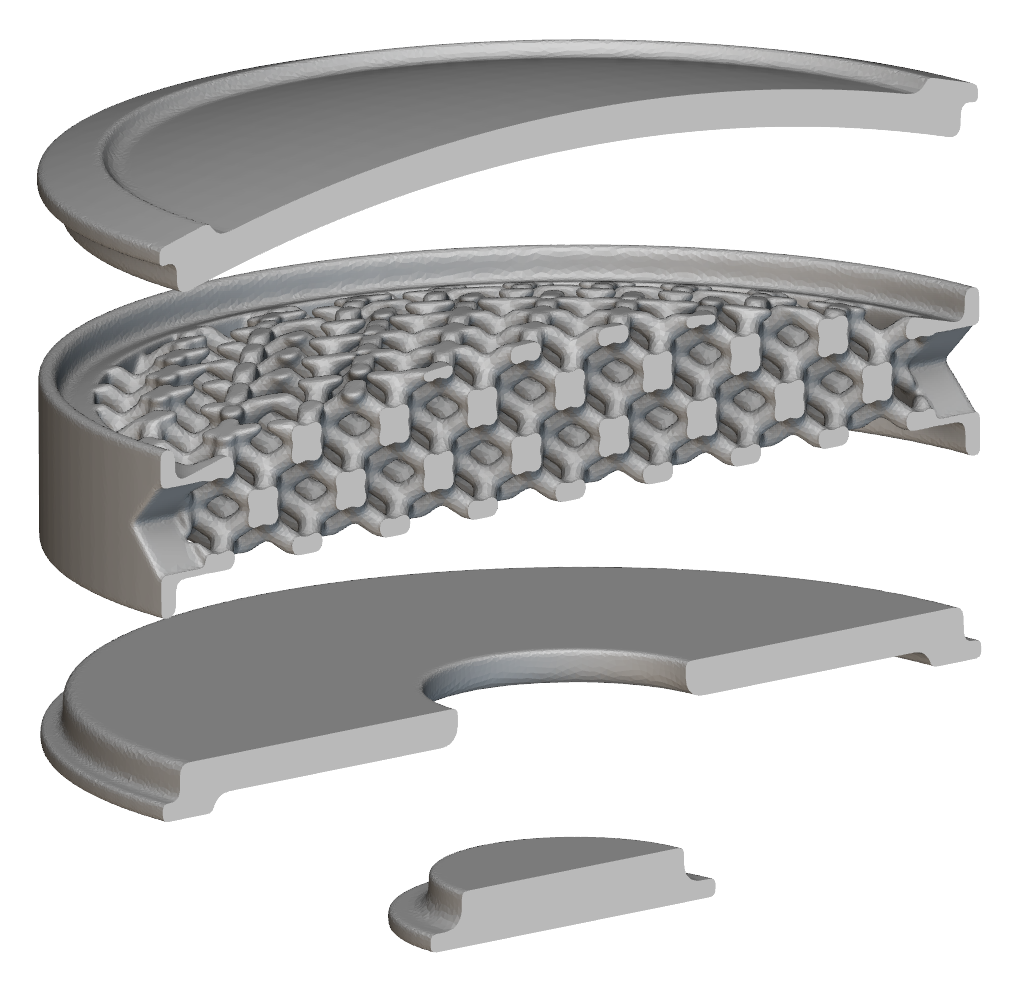}
    \caption{Exploded CAD view of the SLA test model, one of four lattice designs can be interchanged, cubic Fluorite is shown here}
    \label{fig:explodedcad}
  \end{minipage}
  \hfill
  \begin{minipage}{0.5\textwidth}
    \includegraphics[width=\textwidth]{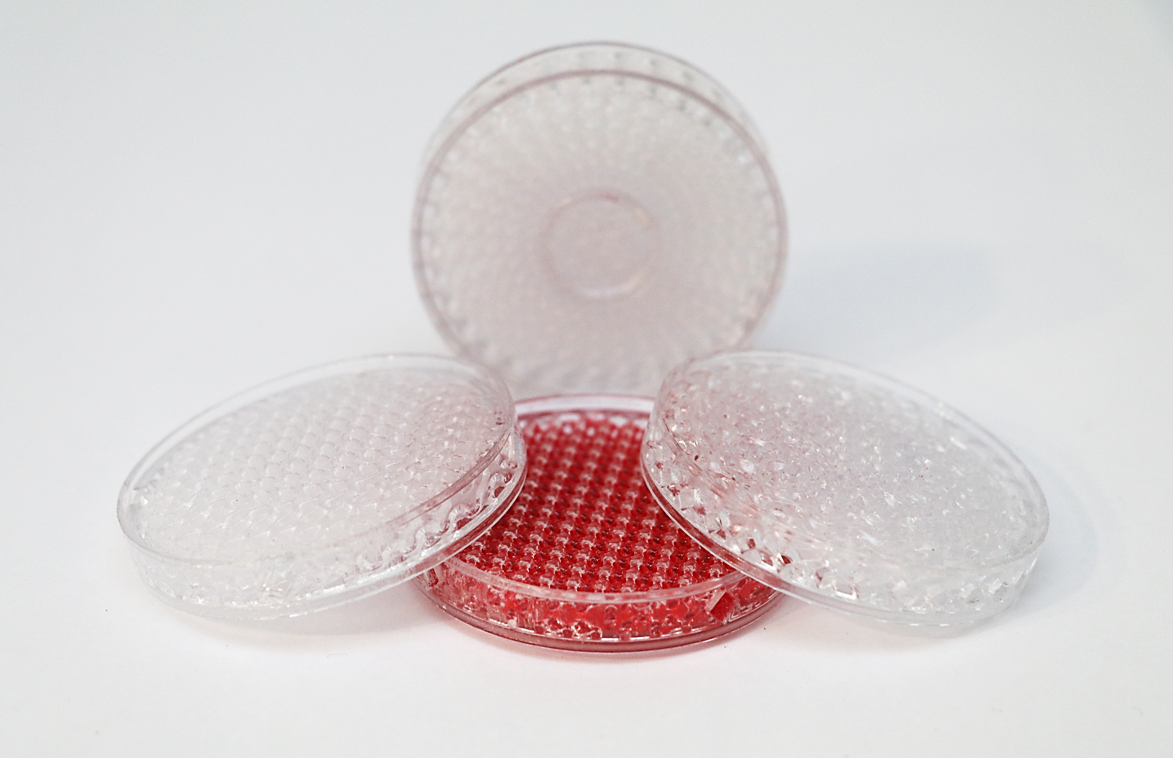}
    \caption{All four lattice tests printed in clear resin, red sand residue is clearly visible in the Fluorite lattice. Image credit: Jason Cowan}
    \label{fig:slaprints}
  \end{minipage}
\end{figure}

\begin{figure}[!h]
    \centering
    \includegraphics[width=1\linewidth]{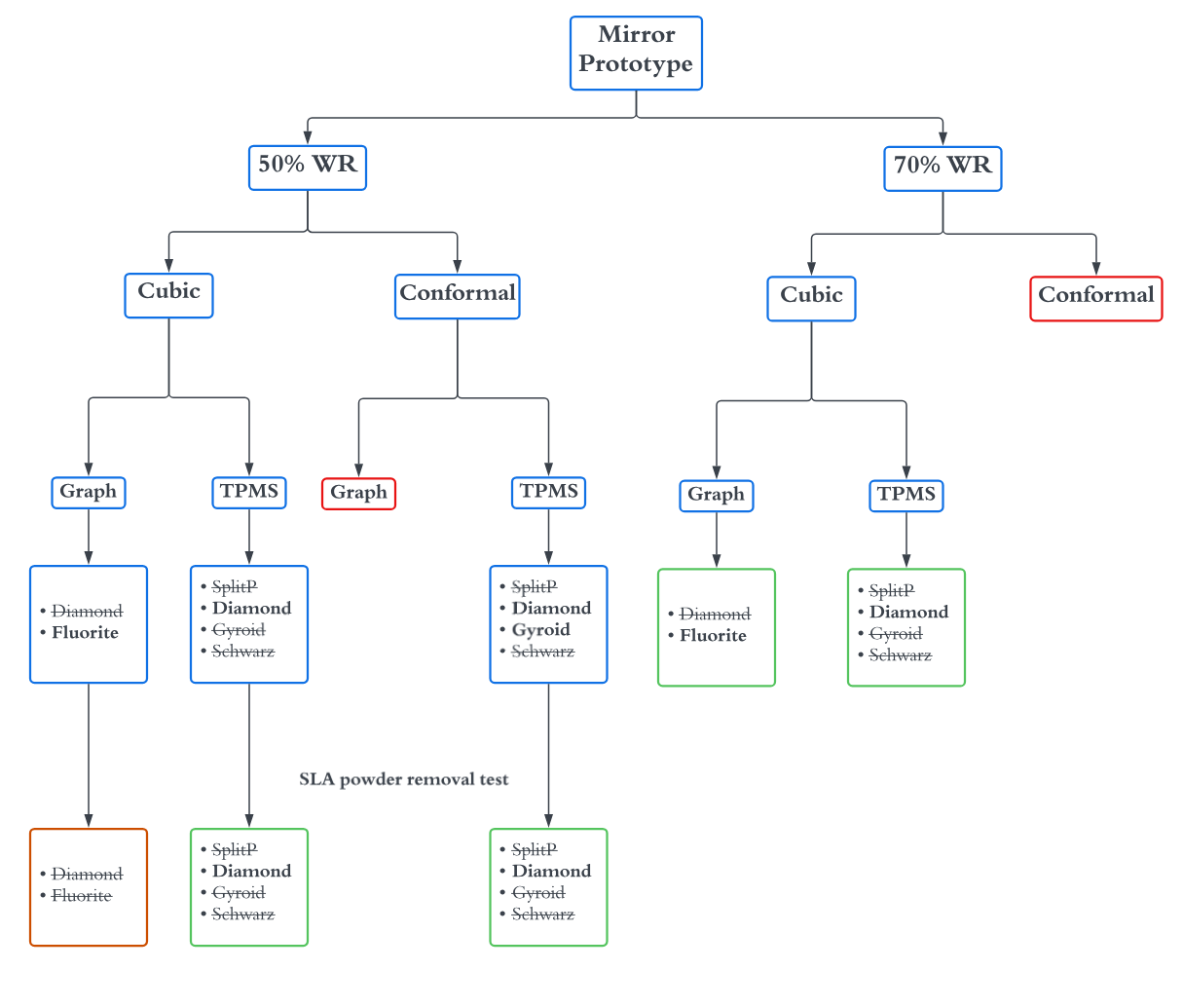}
    \caption{Downselection flow chart}
    \label{fig:downselectionflowchart}
\end{figure}

The models were printed in \textit{Accura ClearVue} resin, they were then lacquered and polished to produce a transparent surface. In hindsight, the lacquer should have been omitted, there is a chance that a relatively thick layer was deposited at sharp corners of the lattice, but the bigger issue was the `stickiness' of the lacquer, which likely didn't cure fully inside the lattice. This made the red sand adhere to the surface, it was then difficult to tell if the powder was stuck in the pockets of the lattice, or if it was just stuck to the surface. This problem primarily affected the cubic Fluorite lattice. Tests of each lattice showed that both cubic and conformal Diamond lattices were the easiest to remove powder from, the unit cell Diamond (TPMS) has excellent flow channels for powder in orthogonal directions. The Fluorite lattice was difficult to draw any conclusions from, other than it would probably be too dense for L-PBF. The Gyroid conformal lattice did not let powder flow out as easily as Diamond. Gyroid unit cells separate their unit cube into two independent channels of flow that never come into contact with one another, this could limit powder flow for 3D printing applications. Both Diamond lattices were taken forward for aluminium printing. Figure \ref{fig:downselectionflowchart} shows a down selection flow chart with all the starting lattice options and which ones were taken forward. 

The cell sizes and wall thicknesses used in the powder removal tests are summarised below:
\begin{itemize}
    \item Diamond cubic used 7 x 7 x 7 mm cells with \SI{1.1}{mm} wall thickness.
    \item Fluorite cubic used 6 x 6 x 6 mm cells with \SI{1.3}{mm} strut thickness
    \item Diamond conformal has two types of lattice. The inner cubic lattice used 7 x 7 x 7 mm cells with \SI{0.9}{mm} wall thickness. The outer conformal lattice is defined by a cell radius, height and arc count (number of segments in the circle) which were 6, 6 and 14 respectively.
    \item Gyroid conformal also has two types of lattice. The inner cubic lattice used 5 x 5 x 5 mm cells with \SI{0.9}{mm} wall thickness. The outer conformal lattice had a cell radius of \SI{5}{mm}, cell height of \SI{5}{mm} and an arc count of 16. 
\end{itemize}

\subsection{Mounting Feature Design}

Figure \ref{fig:Top Down Mirrors} showed the fixing point dimensions for M2. The fixture points that mount M2 to the telescopic booms were consolidated into the mirror. They needed to accommodate M3 bolts for mounting to the CubeSat, but they were also used for the single point diamond turning setup. The design approach for the mounts utilised topology optimisation (TO), which optimises geometry for a given objective and set of constraints. The mounts were optimised to minimise any distortion transfer to the reflective surface during single point diamond turning. Considerations for launch loads such as shock loading and vibrations and operational thermal loads were out of scope for this study,

The steps that were taken to perform the TO in this case are as follows: 
\begin{enumerate}
  \item Define a study space, this will constrain the volume within which the TO can add or remove material. Several study spaces were defined in CAD to explore the design freedom of TO. 
  \item Produce a solid volume mesh of the space. A finer mesh will result in a more refined solution, at the cost of higher computational power. If minimum feature size constraints are used, the quality of the mesh will need to be increased to suit this. 
  \item Define boundary conditions of the loading case and objectives of the study. TO iterates until the objective function, which could be structural compliance, displacement or stress for example, is minimised within some stopping threshold. Since thermal and vibrational loads are out of scope of this study, the main loads that were used for optimisation are from the SPDT set-up.  
  \item Define constraints for the optimisation. The resulting geometry will be controlled by the chosen constraints. The most influential constraint is the volume fraction, which limits how much material can be removed from the original study space. Further constraints might include planar symmetry and minimum feature size.
  \item Convert the result into a solid volume mesh and perform FEA. Different combinations of study spaces, objectives and constraints were used, so the results needed to be compared and down selected. 
\end{enumerate}

\begin{figure}[h]
    \centering
    \includegraphics[width=0.5\linewidth]{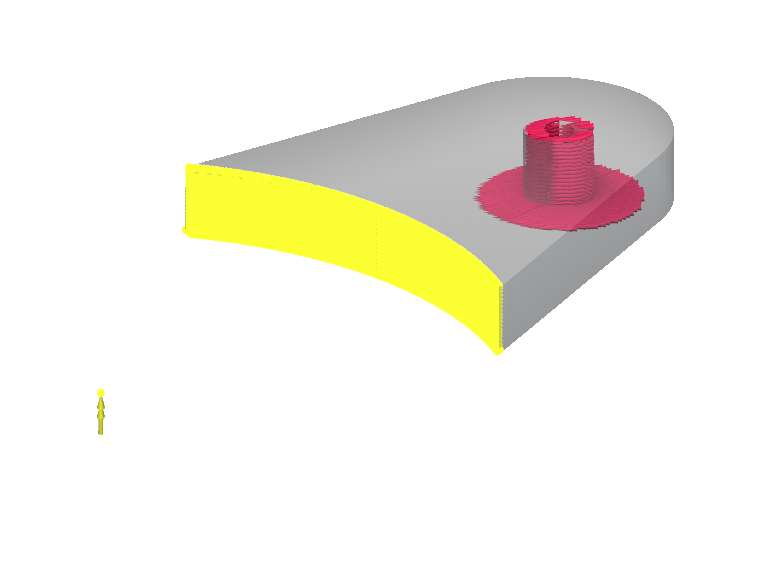}
    \caption{Point moment and displacement restraints applied to a TO study space}
    \label{fig:TopOptBCs}
\end{figure}

To reduce the computation time, the mounts were optimised individually. The mirror is symmetrical around one plane, so one of the optimised results could be mirrored in this plane, this further simplified the design process. A `passive' region around the bolt hole was set in the design space, which ensures no material is removed from this area. A displacement restraint was set in the bolt hole and on the bottom face of the screw pad - the solid cylinder that the bolt hole was drilled in, and a moment was applied to the concave curved face that would be touching the side wall of the mirror, Figure \ref{fig:TopOptBCs} shows these boundary conditions implemented in the study space. For a structural compliance objective, the optimised geometry is independent of the value used for the turning moment, so an arbitrary number was chosen, this was beneficial since the moment induced by SPDT is difficult to approximate. 

\begin{figure}[h]
    \centering
    \includegraphics[width=1\linewidth]{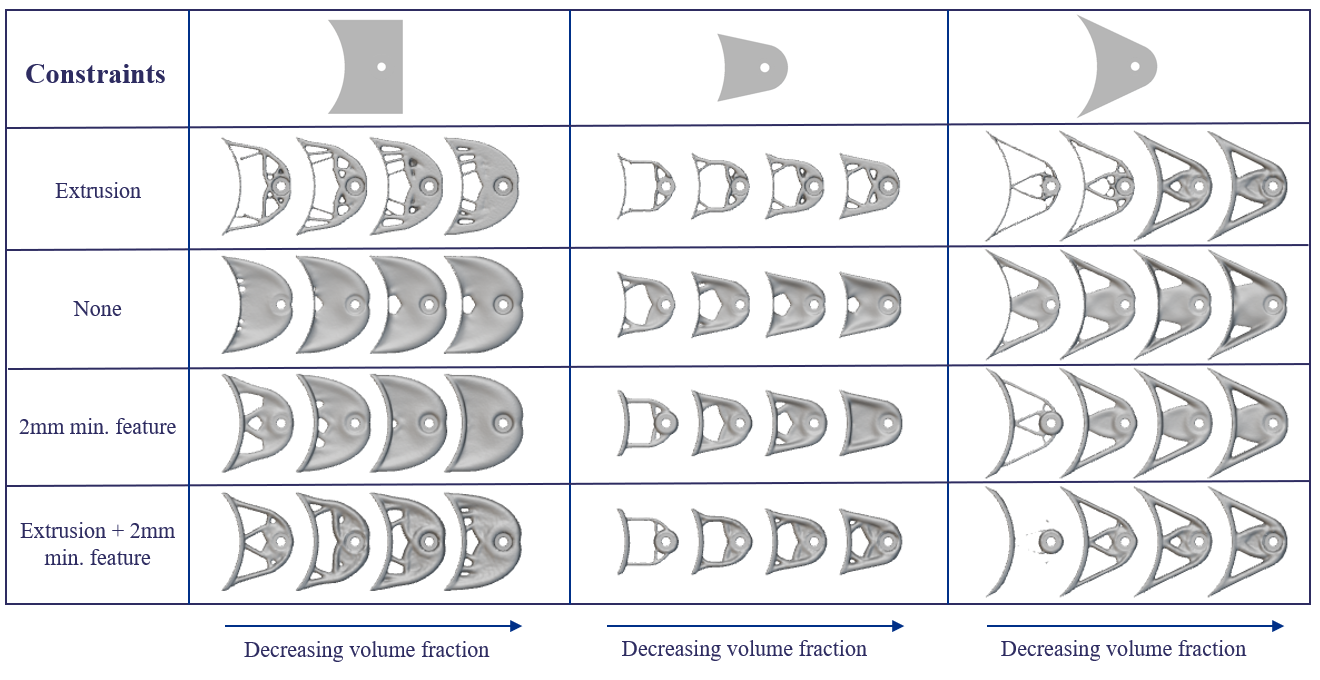}
    \caption{Initial TO study. Four combinations of constraints applied to three study spaces with varying volume fractions}
    \label{fig:TopOptStudy}
\end{figure}

Figure \ref{fig:TopOptStudy} shows TO results from three different design spaces, pictured at the top, each with different constraints, these results were structurally simulated to down select design spaces, and to see if there was any trend in performance with certain constraints. The initial study looked at different combinations of minimum feature and extrusion constraints at different volume fractions. An extrusion constraint forces the geometry to maintain a constant cross section along a specified direction. The minimum feature size constraint was used to ensure that optimised geometry was not less than what could be confidently printed, the mounts needed to be supported during printing, so breaking off these support structures could damage mounts that are too thin. It was found after simulating the results from the initial study, that extrusion constraints negatively affected the stiffness and that for the same mass, the smaller study space generally produced stiffer mounts.  

The final mounts were optimised to minimise structural compliance with a volume reduction of 60\% from the original design space. There was a \SI{3}{mm} minimum feature size constraint to limit the TO process from making features smaller than this and a planar symmetry constraint. A symmetry constraint was necessary to ensure that the mounts would be stiff regardless of which direction the SPDT lathe spins. It can be seen from Figure \ref{fig:Top Down Mirrors} that the three holes are not equally spaced, nor are they on the same radius. The two holes separated by 90\degree are \SI{1}{mm} further away from the centre of the mirror, this caused the centre of gravity (CG) to be off-centre from the central axis of the mirror. As a result of this, the set-up will not be balanced on the lathe, this can affect the quality of the cut. To spare the effort of manually balancing the work piece, the volume fraction constraint was reduced for the one mount until the CG aligned with the centre of the mirror. 

\subsection{Design for Manufacture}
\label{sec:dfam}

\begin{figure}[!h]
\centering
    \begin{subfigure}[b]{0.8\textwidth}
      \includegraphics[width=\textwidth]{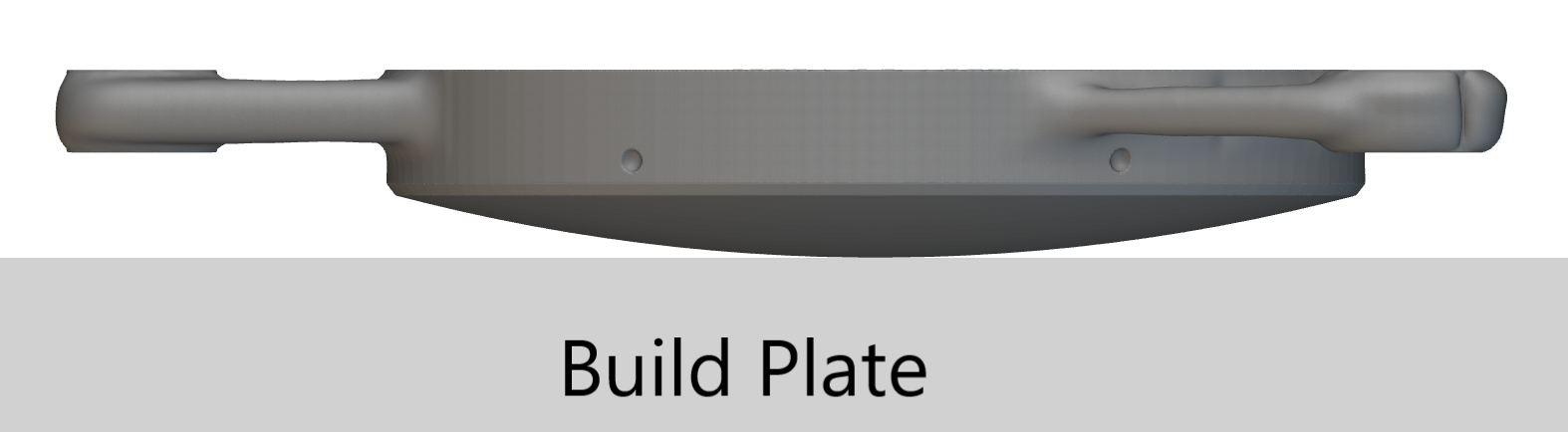} 
    \end{subfigure}
    \begin{subfigure}[b]{0.8\textwidth}
      \includegraphics[width=\textwidth]{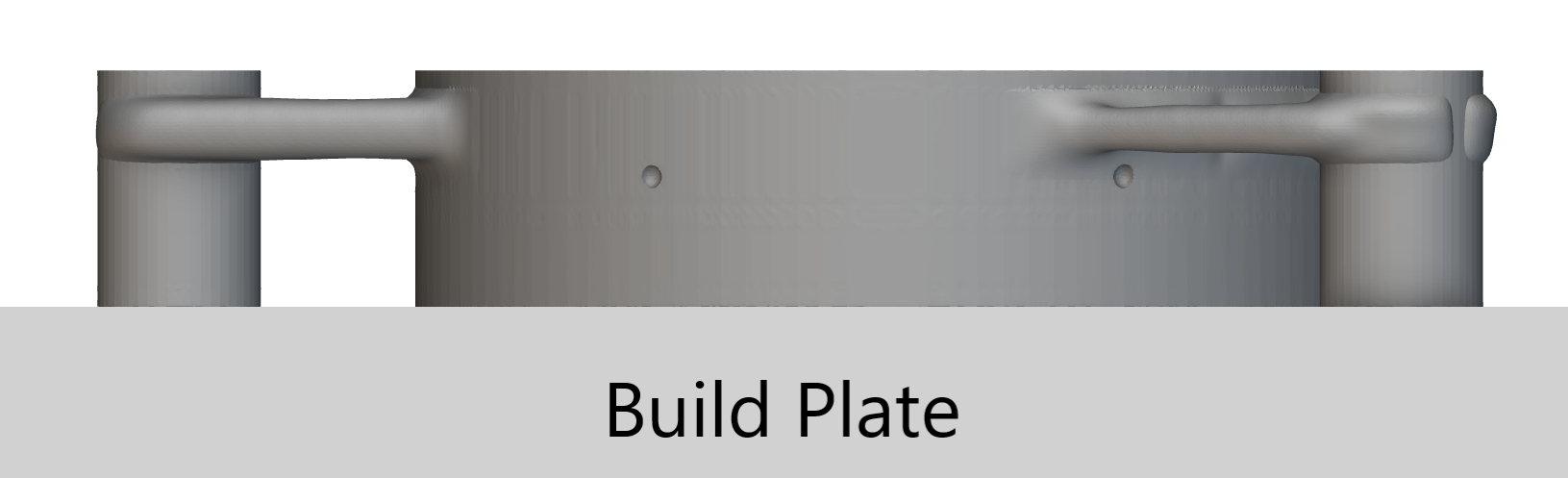}
    \end{subfigure}
    \caption{M2 mirror prototype with no modifications (top) and after changes (bottom)}
    \label{fig:DfAM}
  \end{figure}

The decision was made to print the mirror prototypes with the lattice face up and the convex surface touching the build plate, this was to ensure easy powder removal, but also to fully support the optical surface during the print. In its current design state, the mirror prototype is not well suited to printing; the convex surface forms a large, steep overhang with the build plate, which will need to be supported along with the mounts and screw pads (as shown in Figure \ref{fig:DfAM}). Since the convex surface will need to be rough machined prior to SPDT, adding sacrificial material there to extend it to the build plate will not add a further machining step, this will also help anchor the part to the build plate. Through consultation with the AM bureau, they requested that both the convex surface and the screw pads be made flat to the build plate. They also suggested that the M3 bolt holes should be printed solid and machined out to the correct size and position at a later stage. Extra material will need to be added to any surfaces that are going to be machined since printed parts exhibit relatively high surface roughness and would not be suitable for use as flat datums, this is also shown in Figure \ref{fig:DfAM}. 

The changes to the original design to meet manufacturing requirements highlights the shift in design mindset needed for additive manufacture. The layer-by-layer build process enforces rules on part geometry as demonstrated, but it also allows for the inclusion of features to assist manufacturing and post processing. All of the changes to the original design are summarised below:
\begin{itemize}
  \item The convex surface was made flat to anchor it to the build plate and extended by \SI{1}{mm} to allow for machining.
  \item The screw pads were extended to the build plate as well to secure them during the build and minimise any warping.
  \item \SI{1}{mm} was added to the top faces, excluding the lattice to allow for machining of a flat datum.
  \item The bolt clearance holes were printed solid to precisely machine them in after.
  \item Additional material was added to raise the part \SI{3}{mm} from the build plate to allow for WEDM.
\end{itemize}

\subsection{Final designs}

\begin{figure}[h]
    \begin{subfigure}[b]{0.5\textwidth}
      \includegraphics[width=\textwidth]{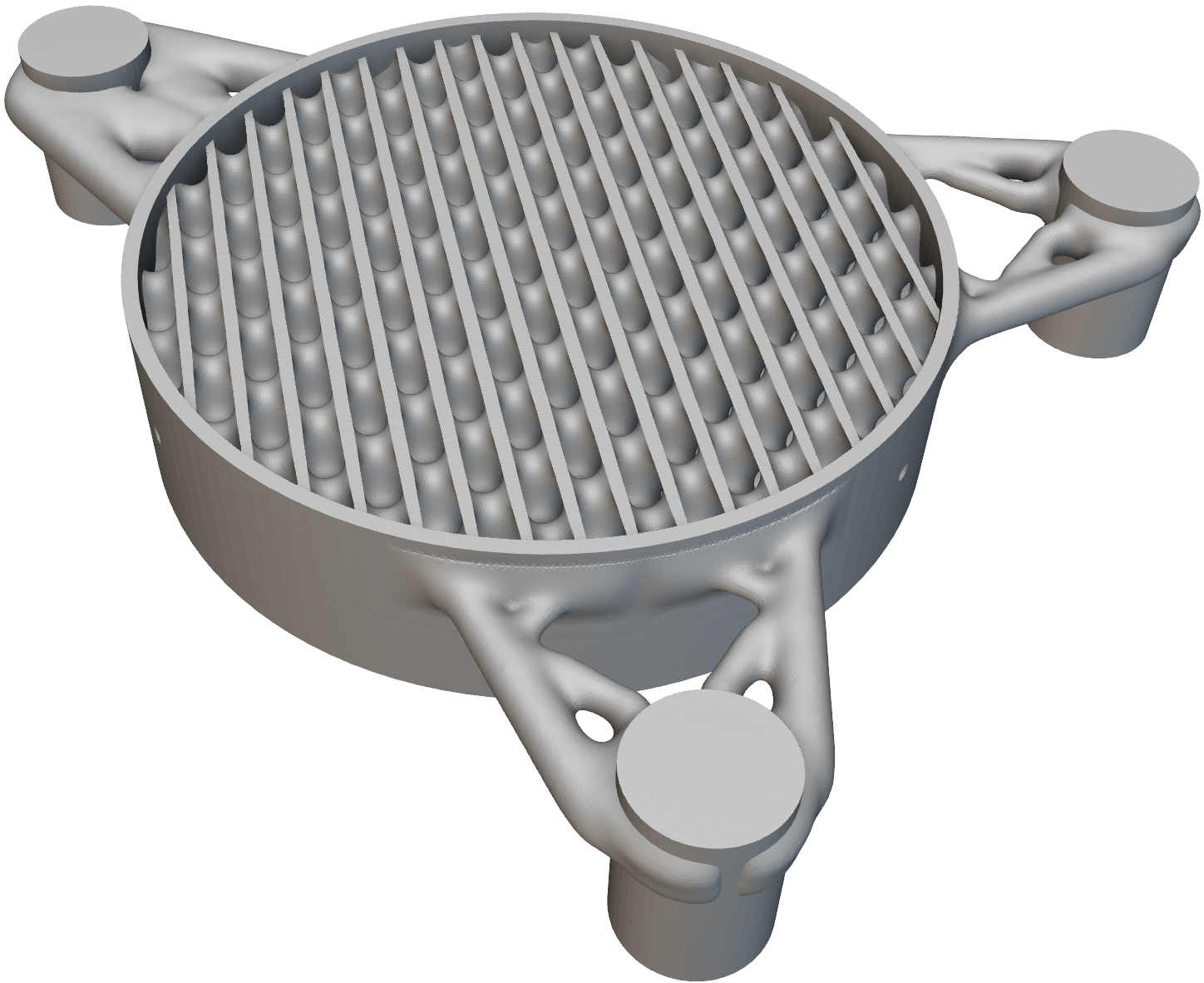}
      \caption{Diamond Cubic 50\% WR}
      \label{fig:20a}
    \end{subfigure}
    \begin{subfigure}[b]{0.5\textwidth}
      \includegraphics[width=\textwidth]{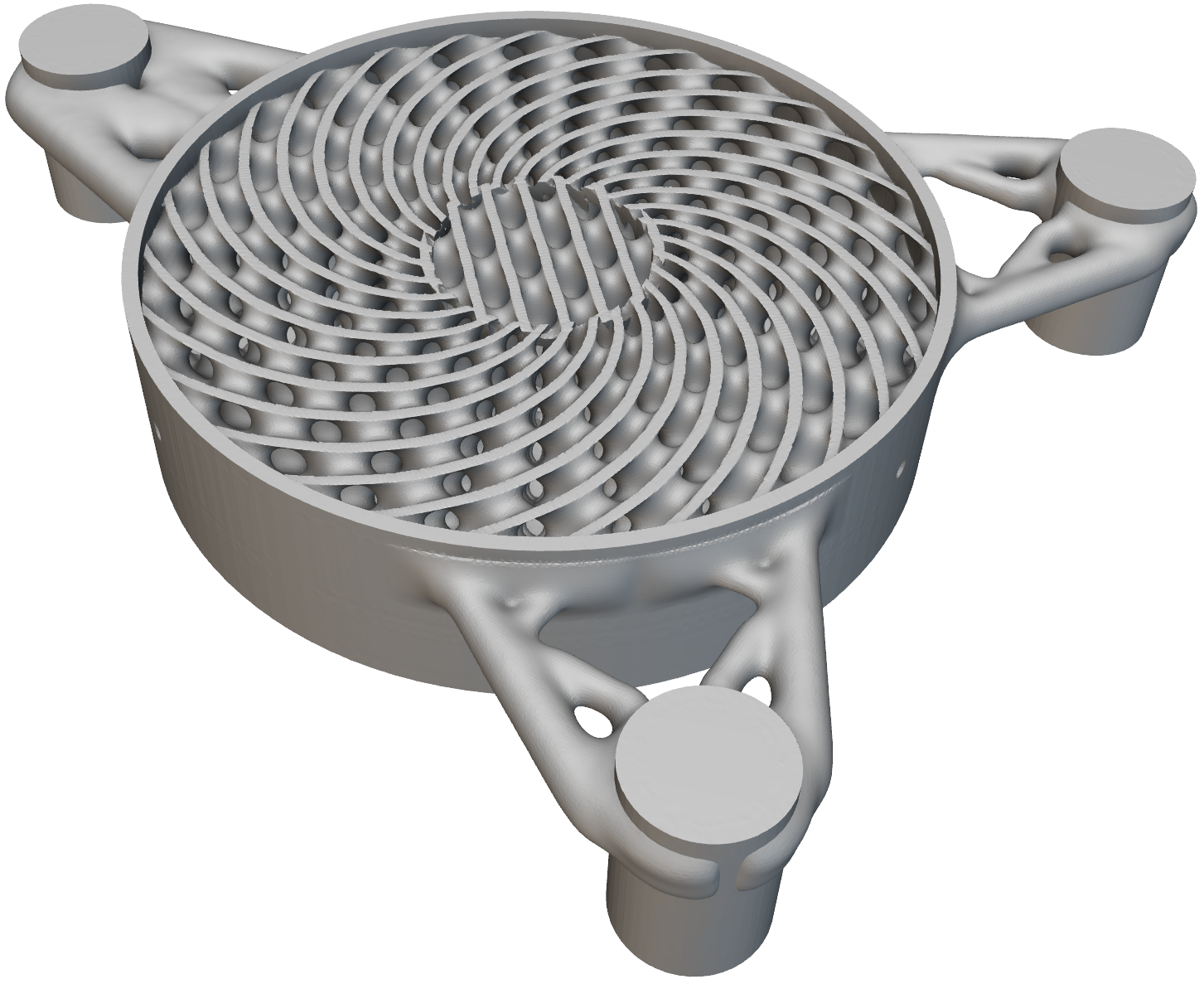}
      \caption{Diamond Conformal 50\% WR}
      \label{fig:20b}
    \end{subfigure}
    \begin{subfigure}[b]{0.5\textwidth}
      \includegraphics[width=\textwidth]{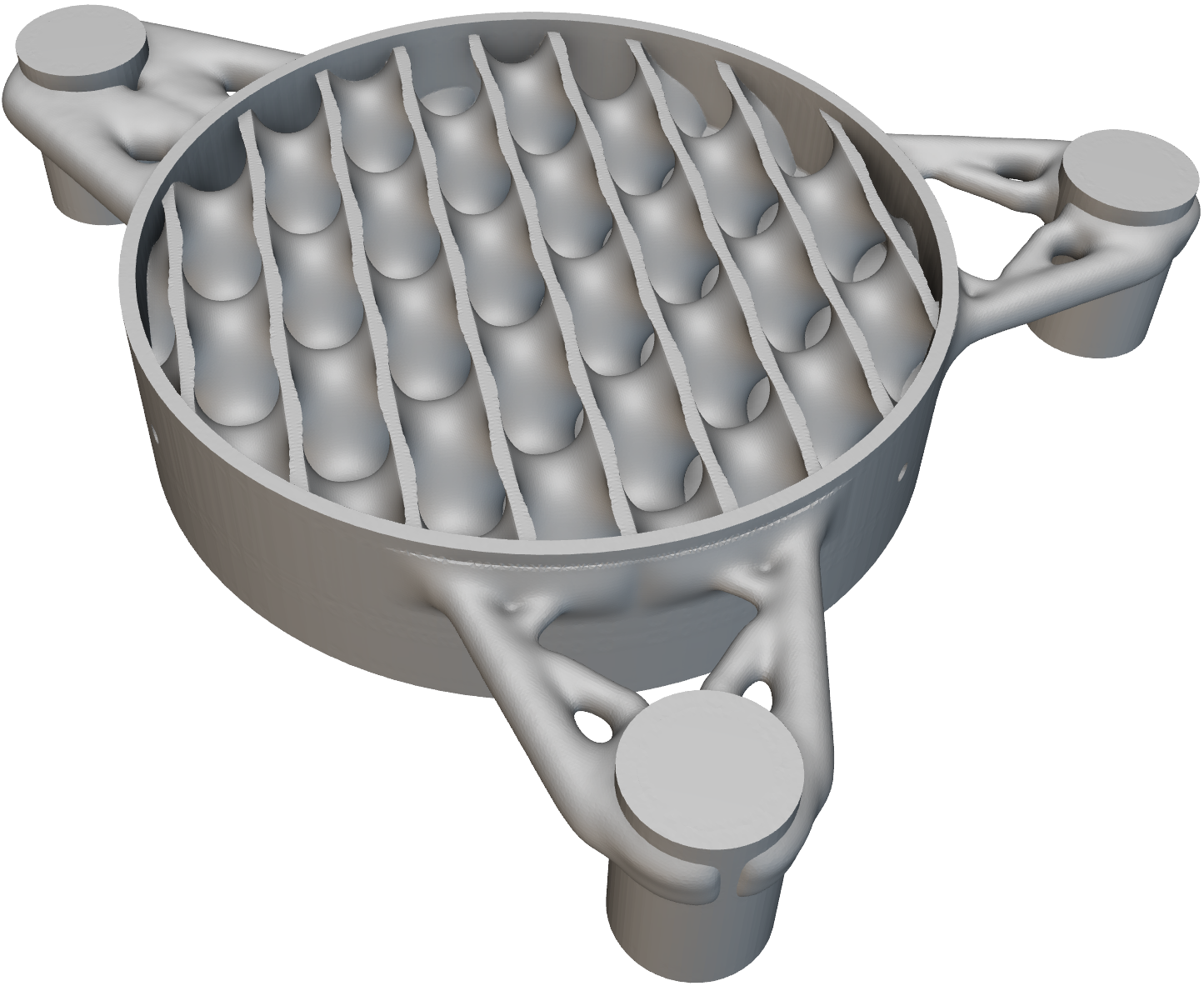}
      \caption{Diamond 70\% WR}
      \label{fig:20c}
    \end{subfigure}
    \begin{subfigure}[b]{0.5\textwidth}
      \includegraphics[width=\textwidth]{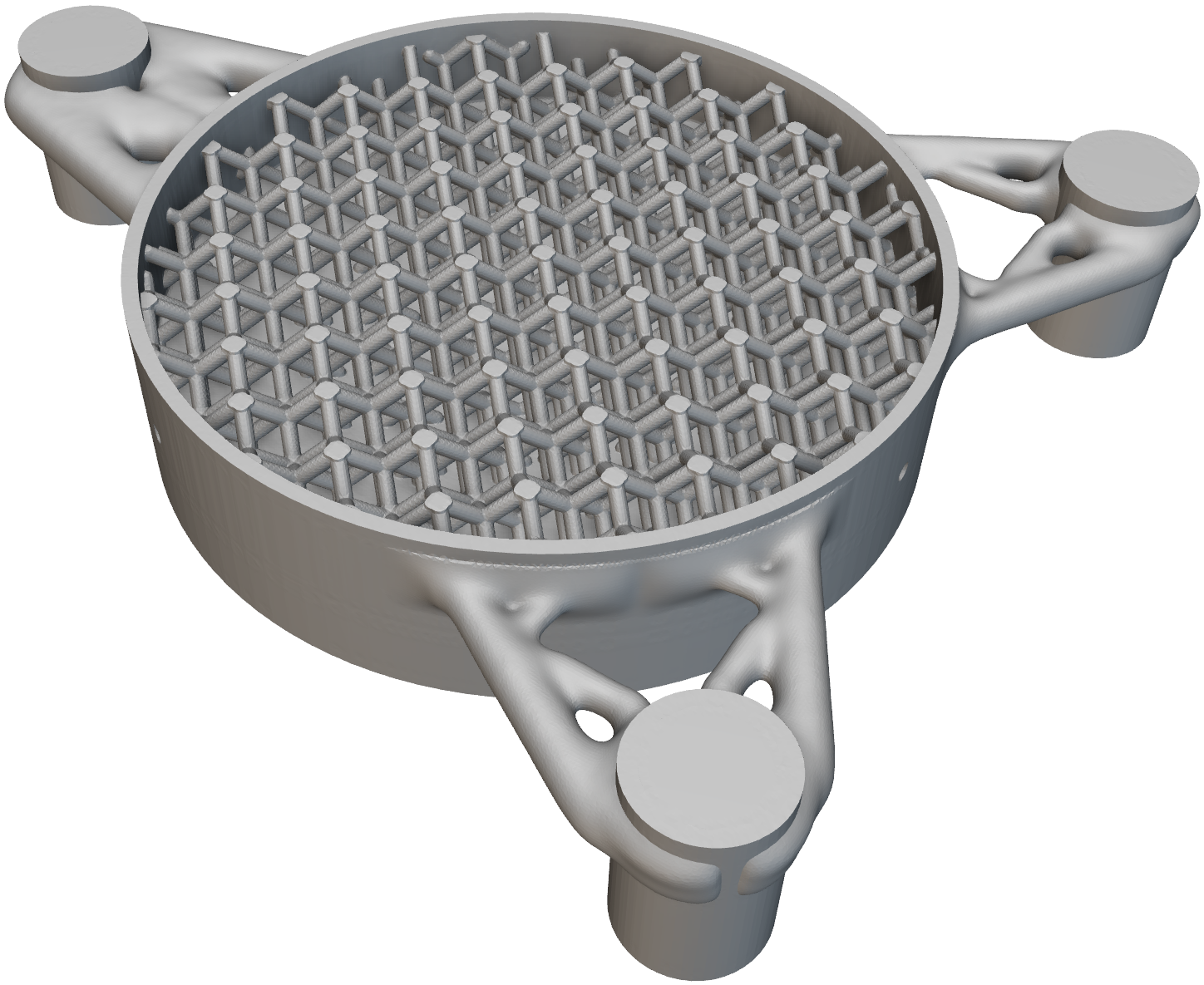}
      \caption{Fluorite 70\% WR}
      \label{fig:20d}
    \end{subfigure}
    \caption{Final Designs}
    \label{fig:finaldesigns}
  \end{figure}

As mentioned in Section \ref{sec:powderremovaltest}, it is difficult during the design phase to know if a given lattice design is printable; this was especially true for the two 50\% WR prototypes. In an attempt to decrease the uncertainty around printability for any future works, all of the lattice parameters of the printed designs are documented. CAD models of the final designs and lattices are shown in Figure \ref{fig:20a}-\ref{fig:20d}. Six equally spaced fiducial markers (\SI{0.5}{mm} spherical indents) were placed in the wall of the mirror.

\begin{itemize}
  \item Diamond cubic 50\% WR used 8 x 8 x 8 mm cells, the wall thickness of the lattice varies from \SI{0.6}{mm} to \SI{0.9}{mm} based on the surface distortion of the constant wall thickness lattice. 
  \item Diamond cubic 70\% WR used 16 x 16 x 16 mm cells, the wall thickness of the lattice is \SI{0.8}{mm}, the field driven wall thickness did not have as much of an effect with the 70\% weight removed lattices. 
  \item Diamond conformal 50\% WR has two different types of lattices. The inner cubic lattice used 7 x 7 x 7 mm cells  with a \SI{0.7}{mm} wall thickness. The outer conformal lattice is defined by the cell radius, height and arc count (number of segments in the circle), in this case they are 6, 6 and 14 respectively and the wall thickness varies from \SI{0.5}{mm} to \SI{0.8}{mm}. 
  \item Fluorite cubic 70\% WR used 7 x 7 x 7 mm cells with \SI{1}{mm} struts. The thickness does not vary because of the sparse nature of the 70\% mass reduced lattice. 
\end{itemize}

\section{Zernike Analysis}\label{sec:Zernikes}

\begin{figure}[!h]
  \centering
  \begin{subfigure}{0.3\textwidth}
    \includegraphics[width=\textwidth]{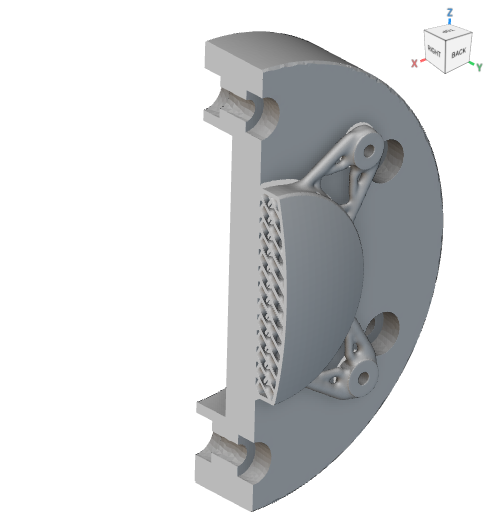}
    \caption{Section view of the CAD model}
    \label{fig:plate section}
  \end{subfigure}
  \hfill
  \begin{subfigure}{0.3\textwidth}
    \includegraphics[width=\textwidth]{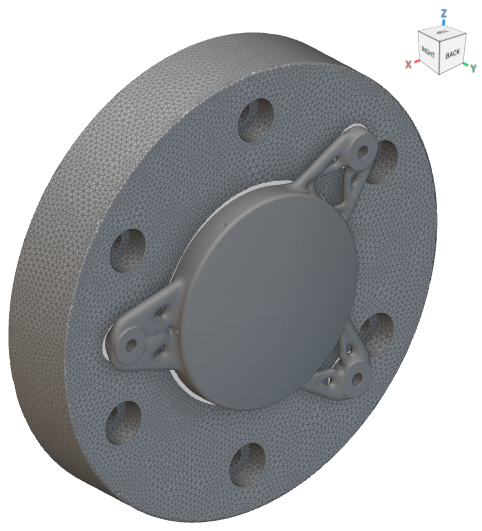}
    \caption{Finite element mesh}
    \label{fig:spdt FEA mesh}
  \end{subfigure}
   \hfill
  \begin{subfigure}{0.3\textwidth}
    \includegraphics[width=\textwidth]{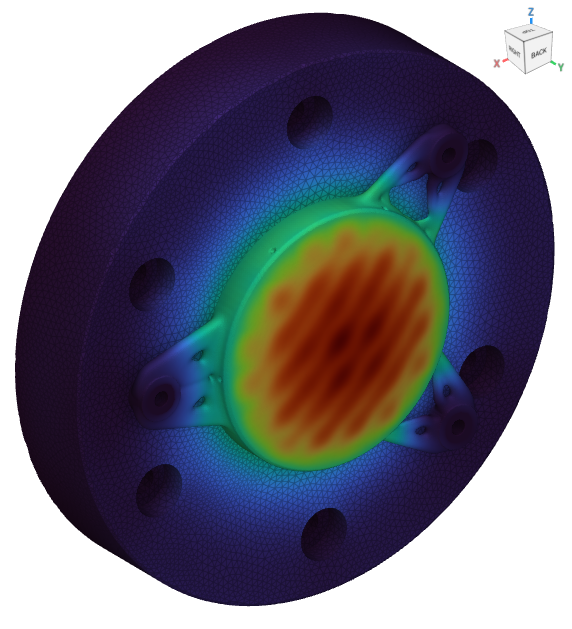}
    \caption{Displacement map}
    \label{fig:SPDT displ}
  \end{subfigure}
  \caption{Simulation model}
  \label{fig:Simulation model}
\end{figure}

Using the final designs, numerical simulations of the turning process were performed to assess the mirror surface deformation and stress concentrations in the mounts for each lattice. A fixture plate was needed to interface the mirror to the lathe, Figure \ref{fig:plate section} shows how the plate has a large cut out on one of the faces. The entire model, including lattice and mounts was meshed, alongside the fixture plate, see Figure \ref{fig:spdt FEA mesh}. The finite element model on average contained five million elements. The three screw holes were constrained in all directions and the bottom face of the mirror was set as a contact point on the fixture plate, which was also fixed. These simulations were performed for all of the lattices using the uniform \SI{3500}{Pa} pressure. The displacement map is shown in Figure \ref{fig:SPDT displ}. 

The maximum Von Mises stress for each design averages roughly \SI{0.1}{MPa} and is located in the contact point between the mirror and fixture plate meshes. This highlights the how the constraints on the model are not representative of reality. The connection of the two meshes is fixed in all degrees of freedom, whereas in reality, the mirror would be free to move in \textit{x} and \textit{z}, using the coordinate system of Figure \ref{fig:Simulation model}. This could potentially cause incorrect surface distortions as a result of the overconstrained model. The simulation tool in the AM design software did not have the functionality for this and other software packages could not be tested in the given time frame.

As with the FEA data from the lattice downselection stage, the nodal displacements were filtered to just the reflective surface, these were exported to a Zernike fitting Matlab script. The numerical values from the FEA were not considered as they are almost certainly not representative of the true loads imparted on the mirror. For one, only an approximation for the SPDT load was considered, no effects from any of the rough machining operations were investigated. The objective of the Zernike analysis was not to see the absolute displacement after Zernike fitting, rather the relative deformation between the different lattices. 

\begin{figure}[h]
    \begin{subfigure}[b]{0.5\textwidth}
      \includegraphics[width=\textwidth]{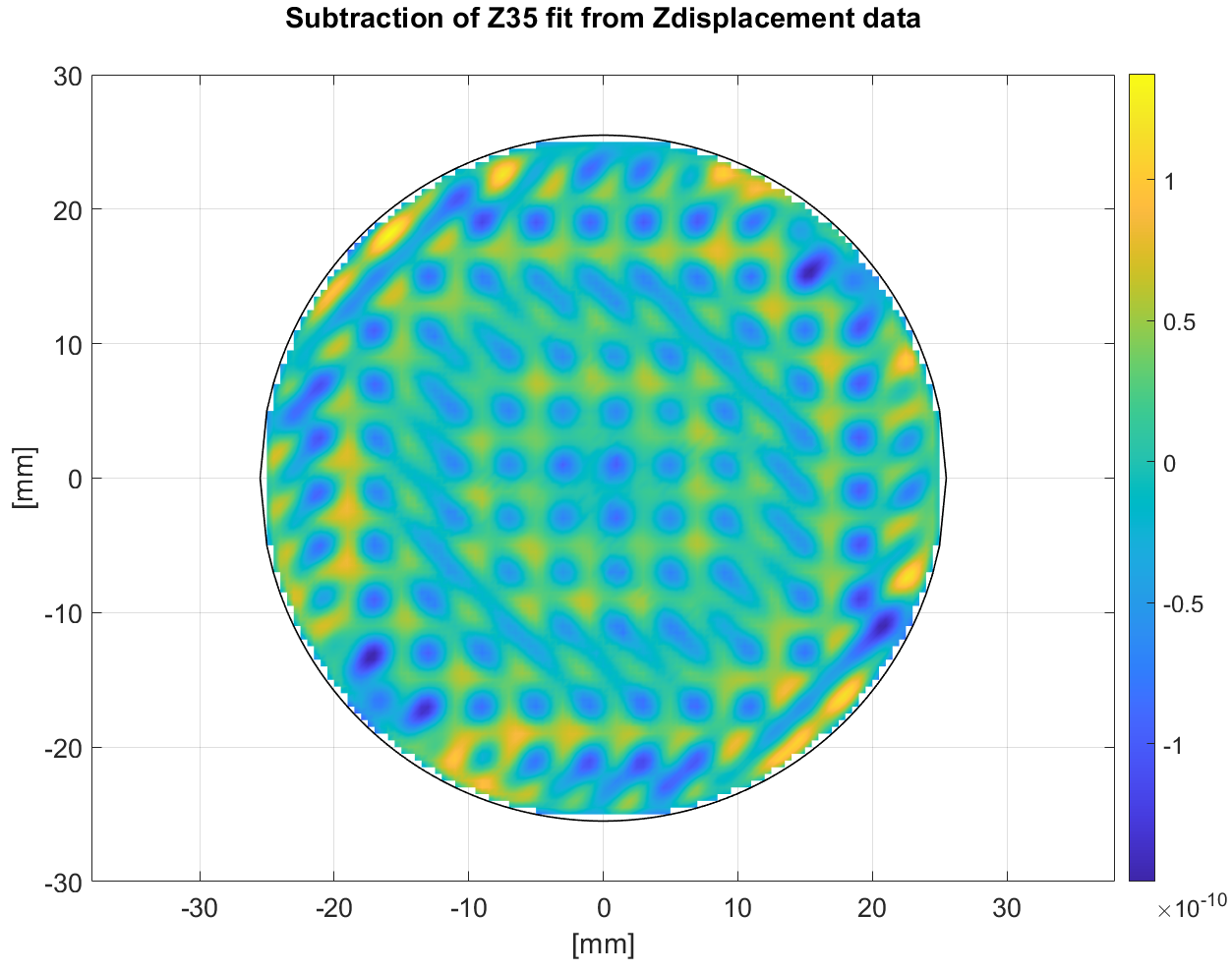}
      \caption{Diamond Cubic 50\% WR}
      \label{fig:Z35a}
    \end{subfigure}
    \begin{subfigure}[b]{0.5\textwidth}
      \includegraphics[width=\textwidth]{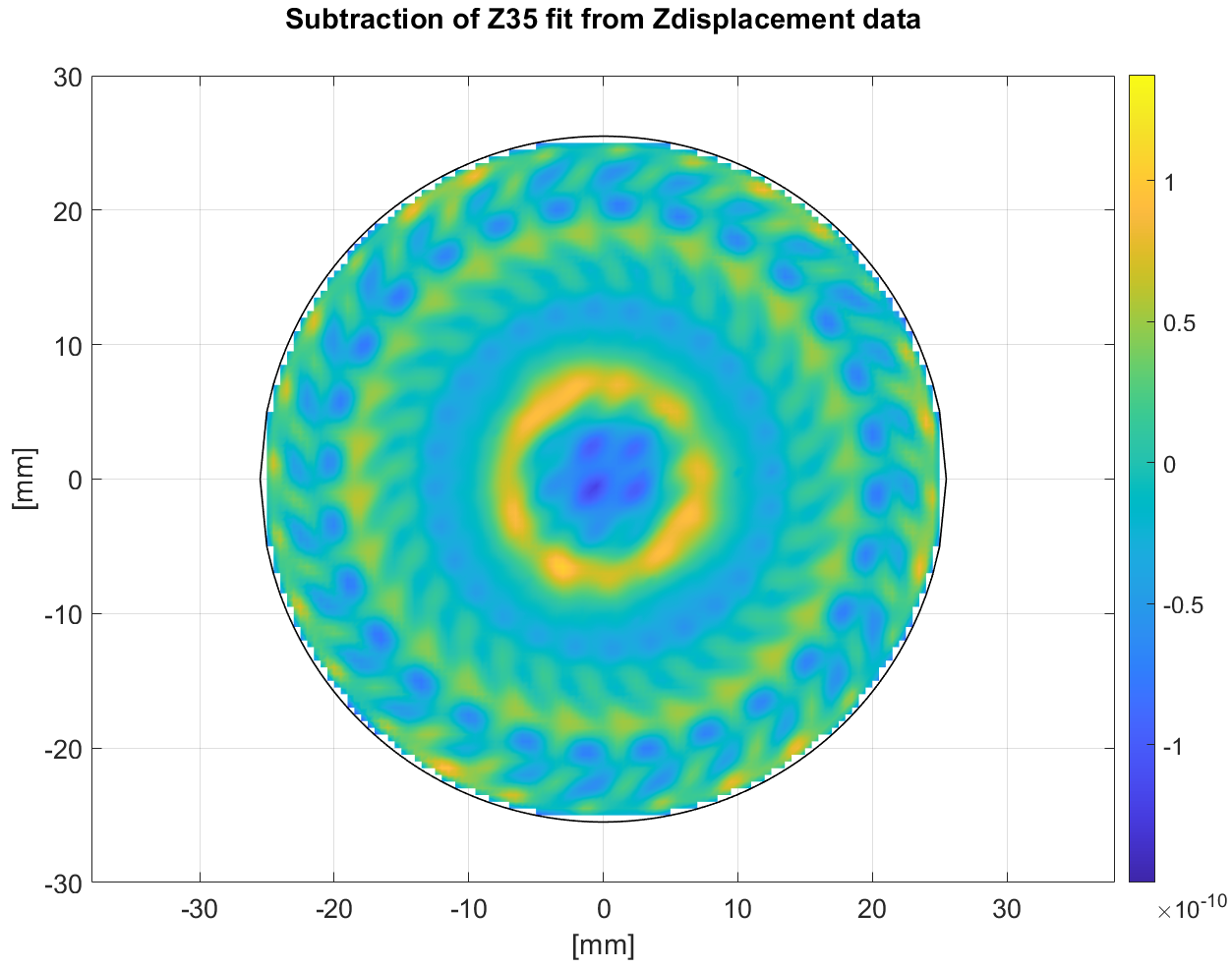}
      \caption{Diamond Conformal 50\% WR}
      \label{fig:Z35b}
    \end{subfigure}
    \begin{subfigure}[b]{0.5\textwidth}
      \includegraphics[width=\textwidth]{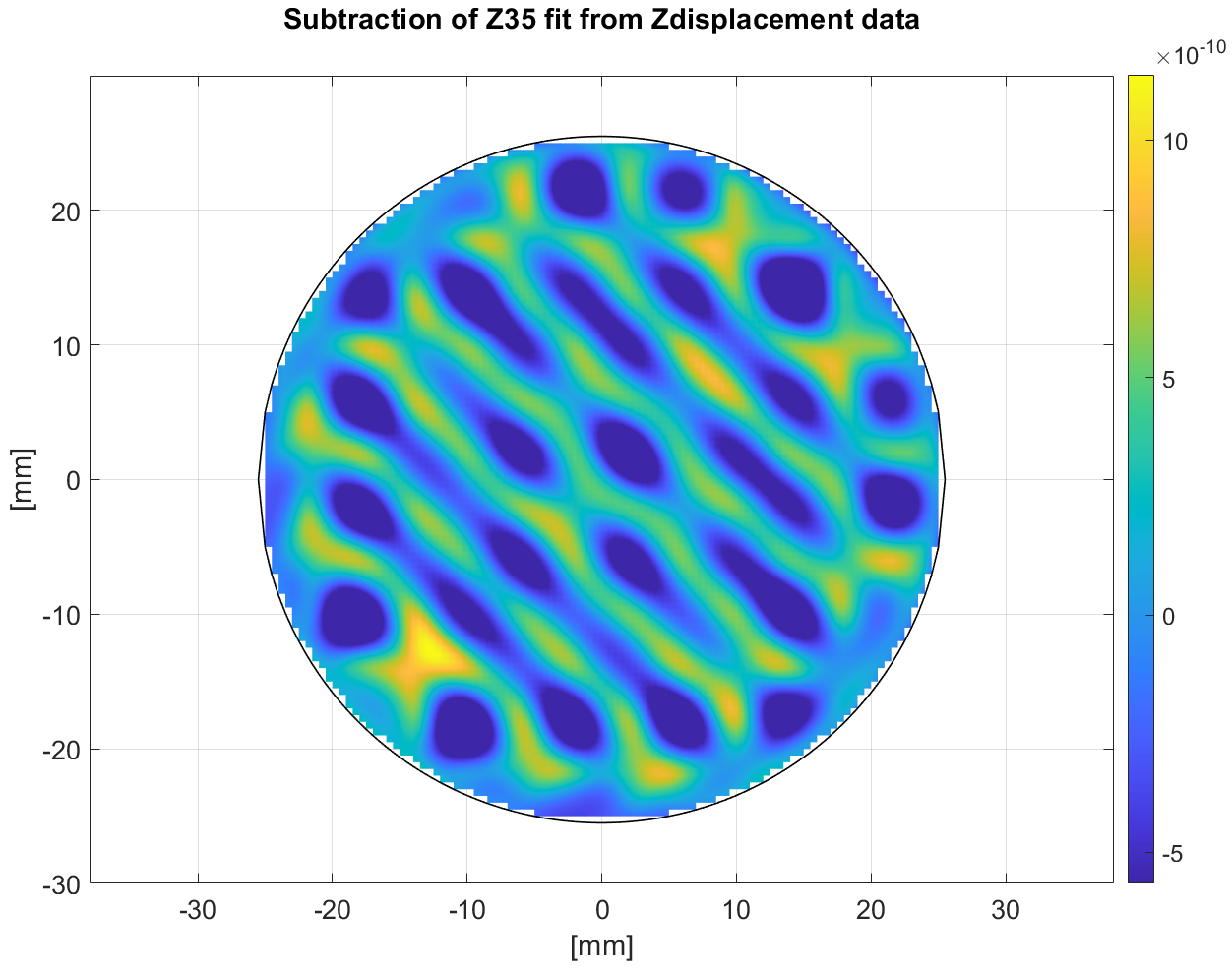}
      \caption{Diamond Cubic 70\% WR}
      \label{fig:Z35c}
    \end{subfigure}
    \begin{subfigure}[b]{0.5\textwidth}
      \includegraphics[width=\textwidth]{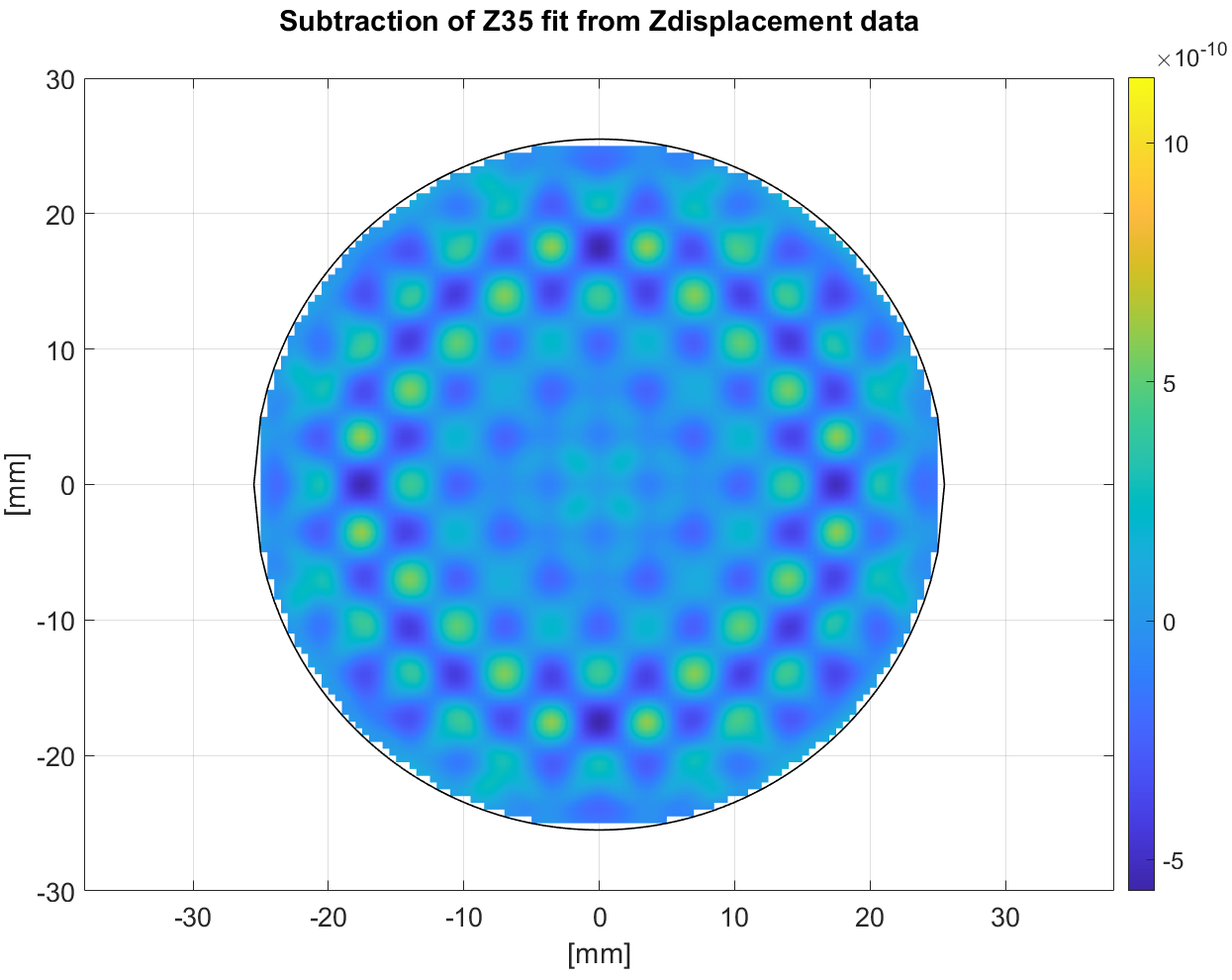}
      \caption{Fluorite Cubic 70\% WR}
      \label{fig:Z35d}
    \end{subfigure}
    \caption{Subtraction of Z35 fit}
  \end{figure}

\begin{figure}[h]
    \begin{subfigure}[b]{0.5\textwidth}
      \includegraphics[width=\textwidth]{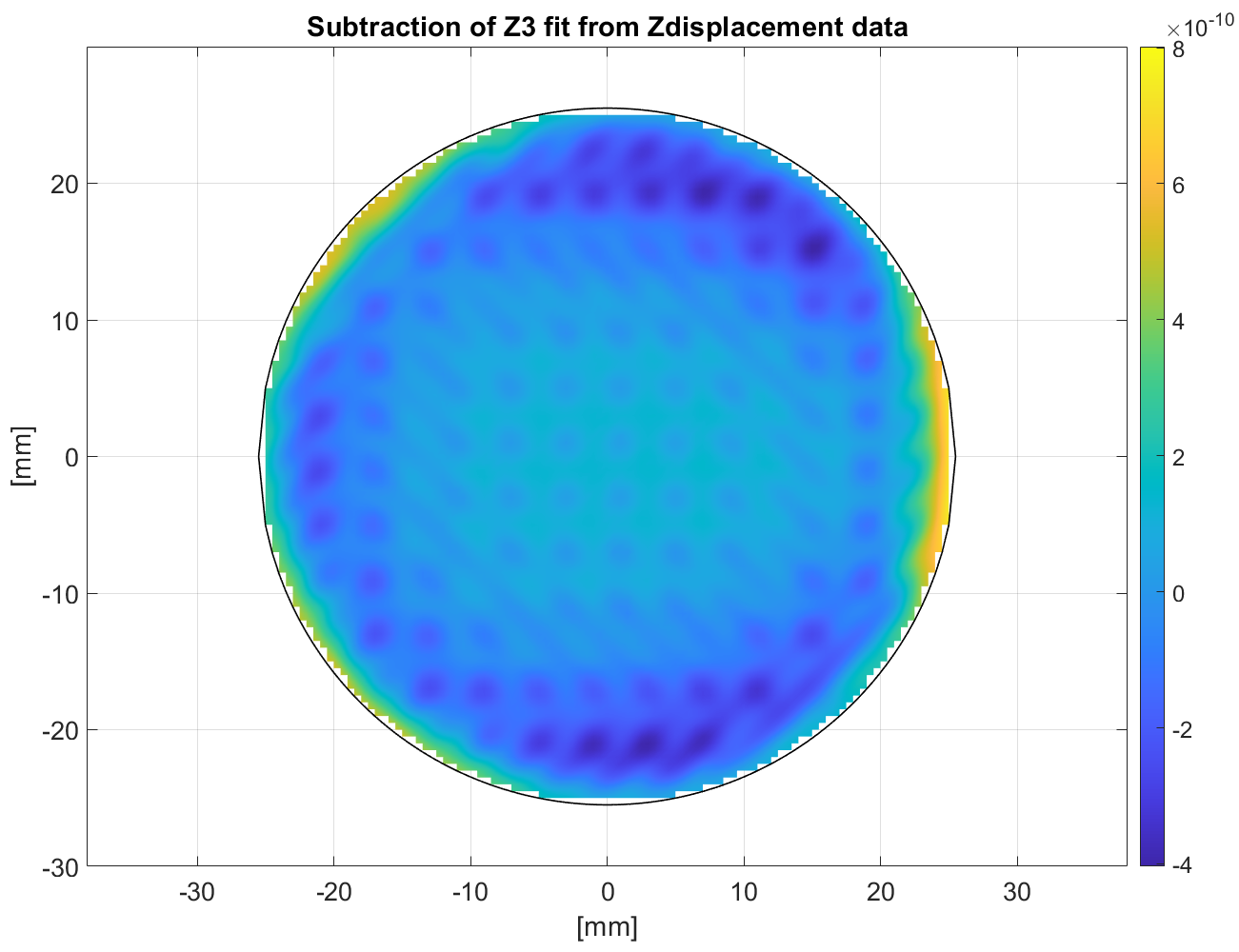}
      \caption{Diamond Cubic 50\% WR}
      \label{fig:Z3a}
    \end{subfigure}
    \begin{subfigure}[b]{0.5\textwidth}
      \includegraphics[width=\textwidth]{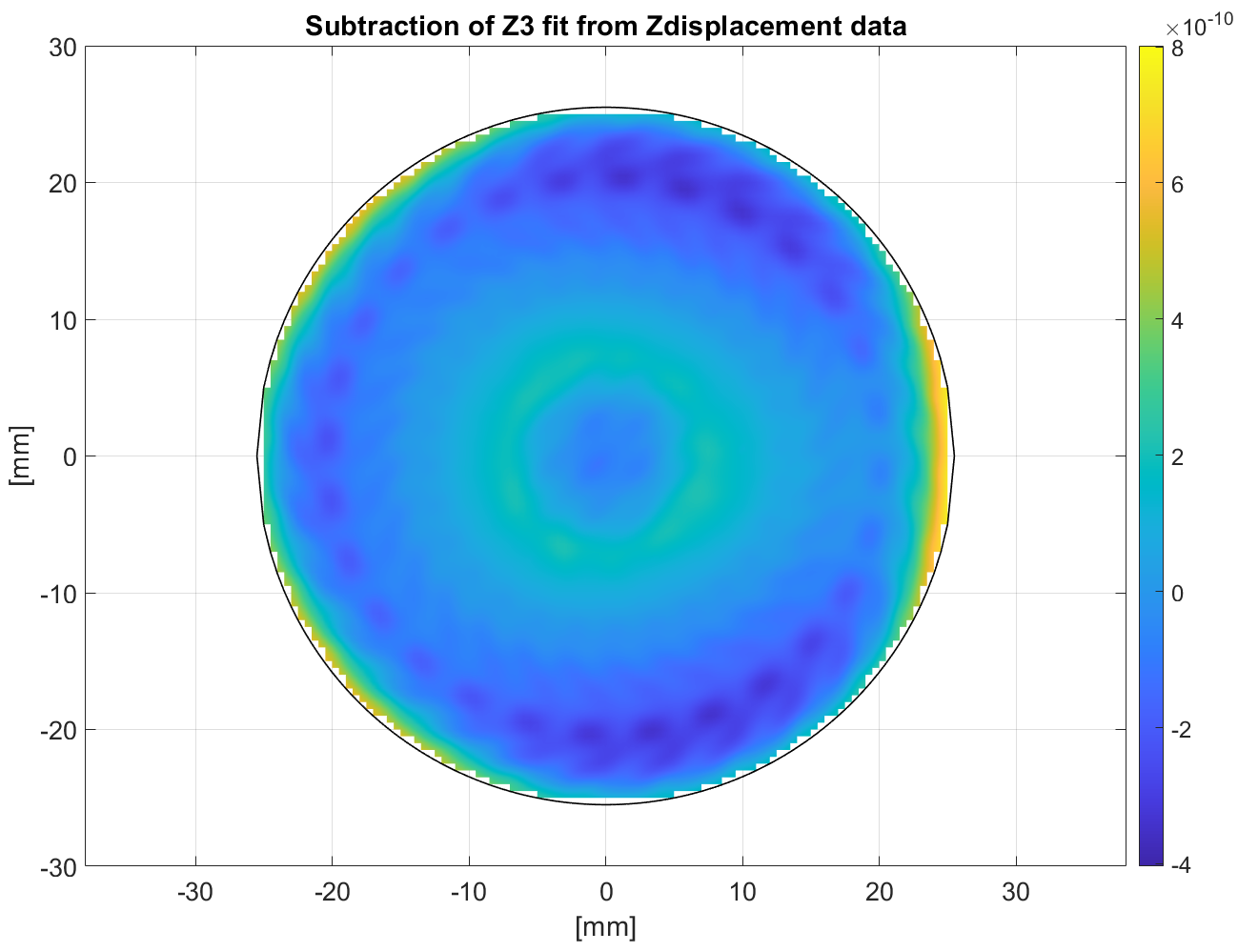}
      \caption{Diamond Conformal 50\% WR}
      \label{fig:Z3b}
    \end{subfigure}
    \begin{subfigure}[b]{0.5\textwidth}
      \includegraphics[width=\textwidth]{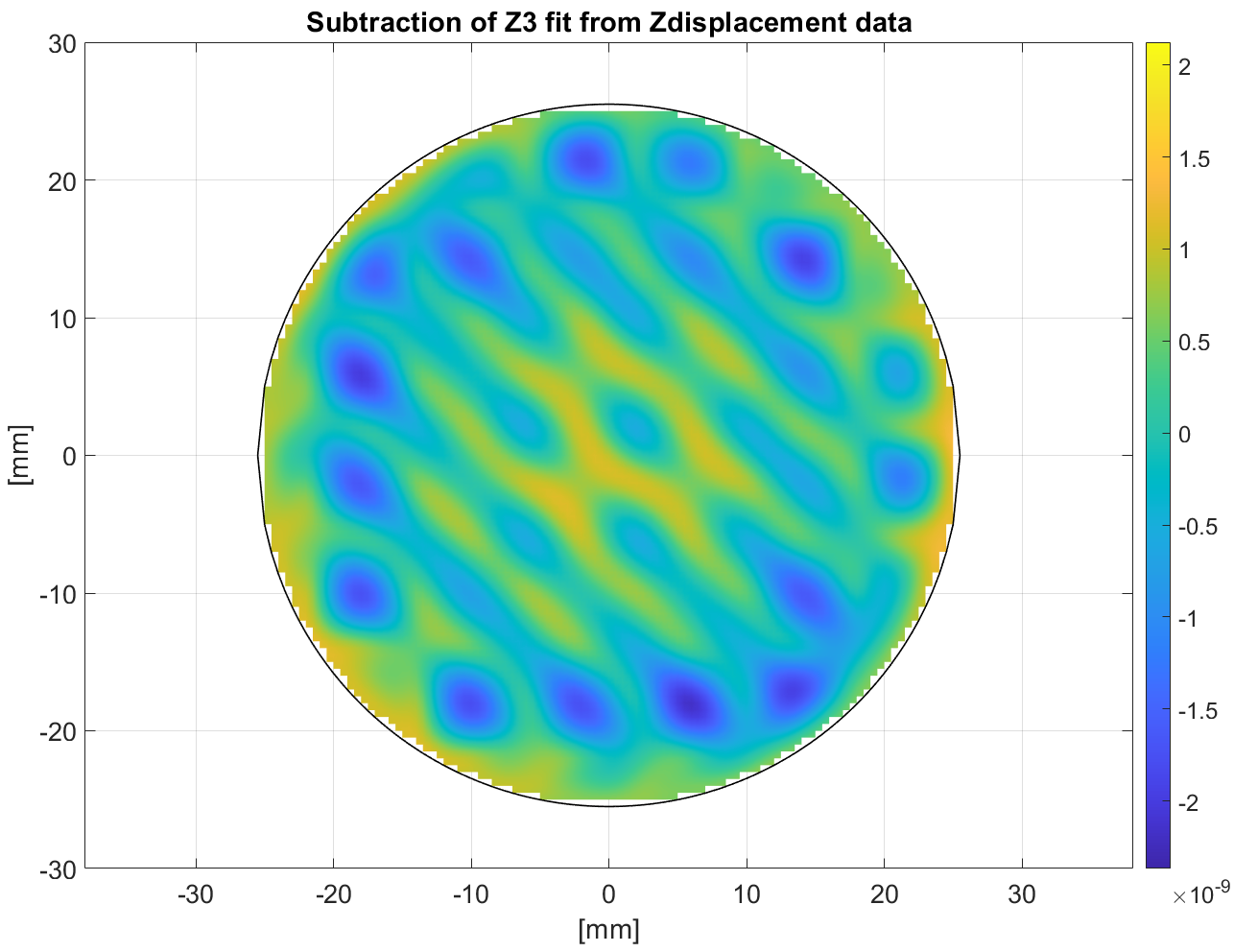}
      \caption{Diamond Cubic 70\% WR}
      \label{fig:Z3c}
    \end{subfigure}
    \begin{subfigure}[b]{0.5\textwidth}
      \includegraphics[width=\textwidth]{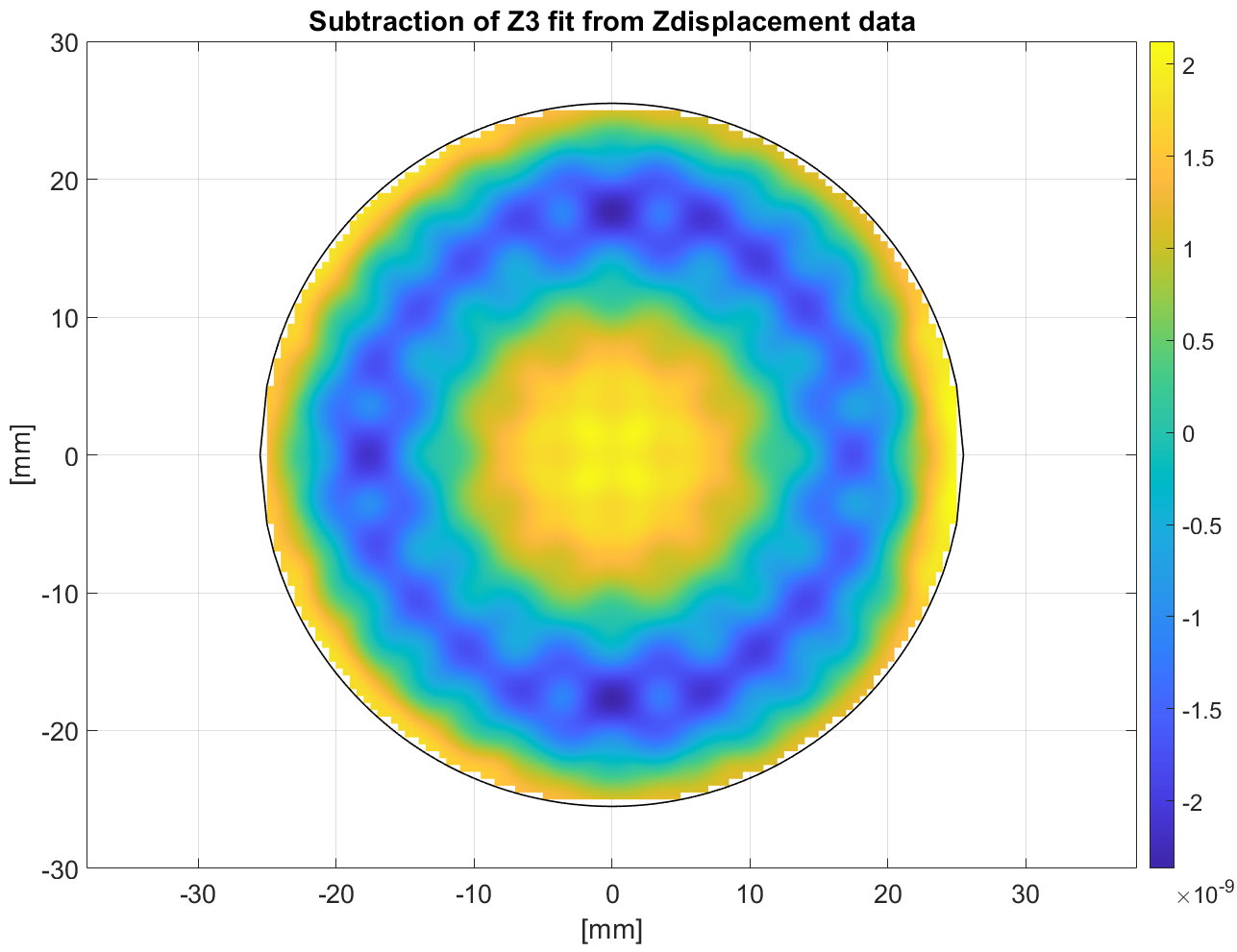}
      \caption{Fluorite Cubic 70\% WR}
      \label{fig:Z3d}
    \end{subfigure}
    \caption{Subtraction of Z3 fit}
  \end{figure}

The first 36 Zernike terms were fit to the input FEA data via a least squares fit method and the Zernike fit was subtracted from the input data to highlight the lattice print through. The subtracted data is presented in Figures \ref{fig:Z35a}-\ref{fig:Z35d} for each lattice, the print through pattern is clear. Additionally, plots for the subtraction of the first three Zernike terms are shown in Figures \ref{fig:Z3a}-\ref{fig:Z3d}.  

\begin{figure}[h]
    \centering
    \includegraphics[width=1\linewidth]{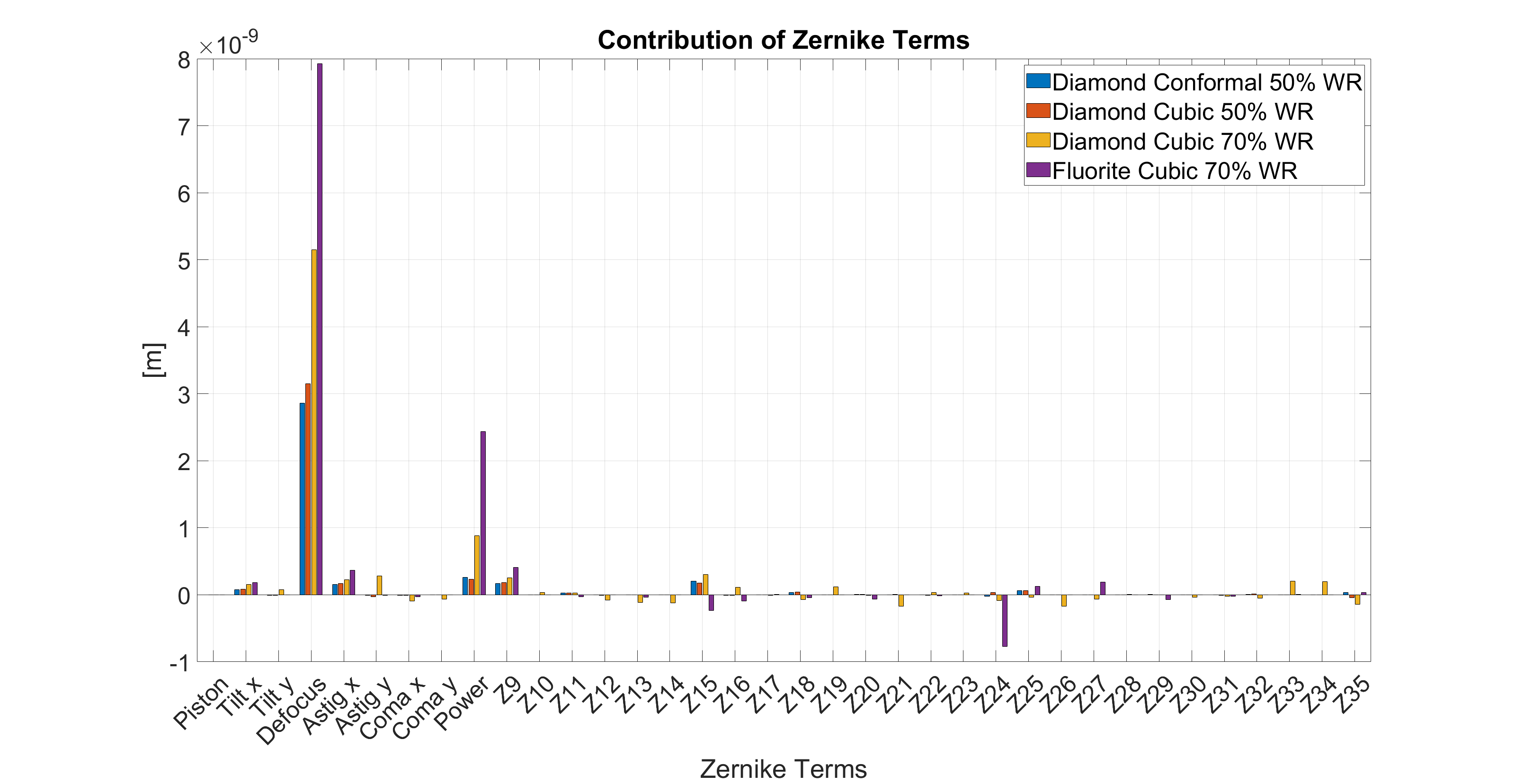}
    \caption{Contribution of Zernike terms for each lattice}
    \label{fig:contribution of z terms}
\end{figure}

\begin{figure}
    \centering
    \includegraphics[width=1\linewidth]{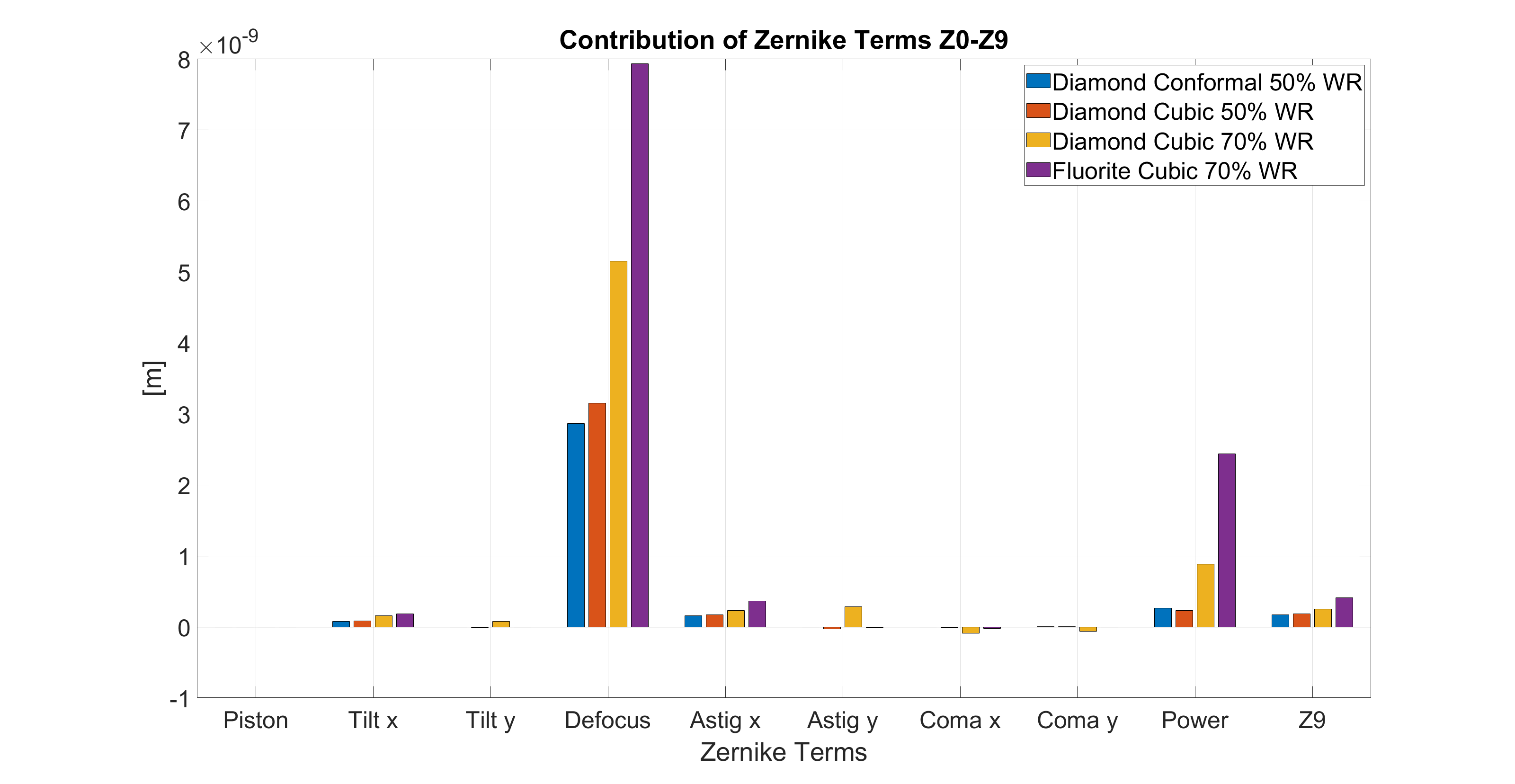}
    \caption{Contribution of just Z0-Z9 terms}
    \label{fig:contribution of z9 terms}
\end{figure}

Figure \ref{fig:contribution of z terms} shows the contribution of Zernike terms 0-35 for each lattice, all display large amounts of defocus, other terms with noticeable contribution are Astigmatism in \textit{x}, Power and Trefoil (Z9). Defocus can be explained by the solid wall around the lattice deflecting much less under pressure than the lattice. The trefoil is expected since the mirror is supported by three mounts at close to even separation and these regions will have slightly different properties due to the increased stiffness from the mounts. Since the higher order Zernike terms are relatively small compared to the lower order ones, they are omitted in Figure \ref{fig:contribution of z9 terms} to more clearly see the difference between each lattice. The Fluorite lattice can be seen to perform the worst, especially in Defocus, which is expected since strut unit cells are less stiff and material efficient\cite{TPMSSuperior,UnitCellComparison}. 


\section{Manufacture}\label{sec:manufacturing}
\subsection{Printing}

In total, nine prototypes were printed in the aluminium alloy AlSi10Mg, a duplicate set of the final four designs for redundancy, destructive testing, X-Ray computed tomography (XCT) and hot isostatic pressing (HIP) and one high resolution print for comparison. The high resolution part was printed on a Concept Laser MLabs, which has a build volume of 90 x 90 x 90 mm and a quoted layer height of 15-25\(\mu\)m. The other eight parts were built using an EOS M290, with a build volume of 250 x 250 x 305 mm and quoted layer height of 30-60\(\mu\)m. The parts were removed from the build plate using wire electrical discharge machining (WEDM), support material was manually removed and the areas where they contacted the part were filed smooth. Due to the high residual stresses introduced during the build process, parts are stress relieved. They were also tumble finished to smoothen the outer surfaces. 

\begin{figure}[h]
    \centering
    \includegraphics[width=1\linewidth]{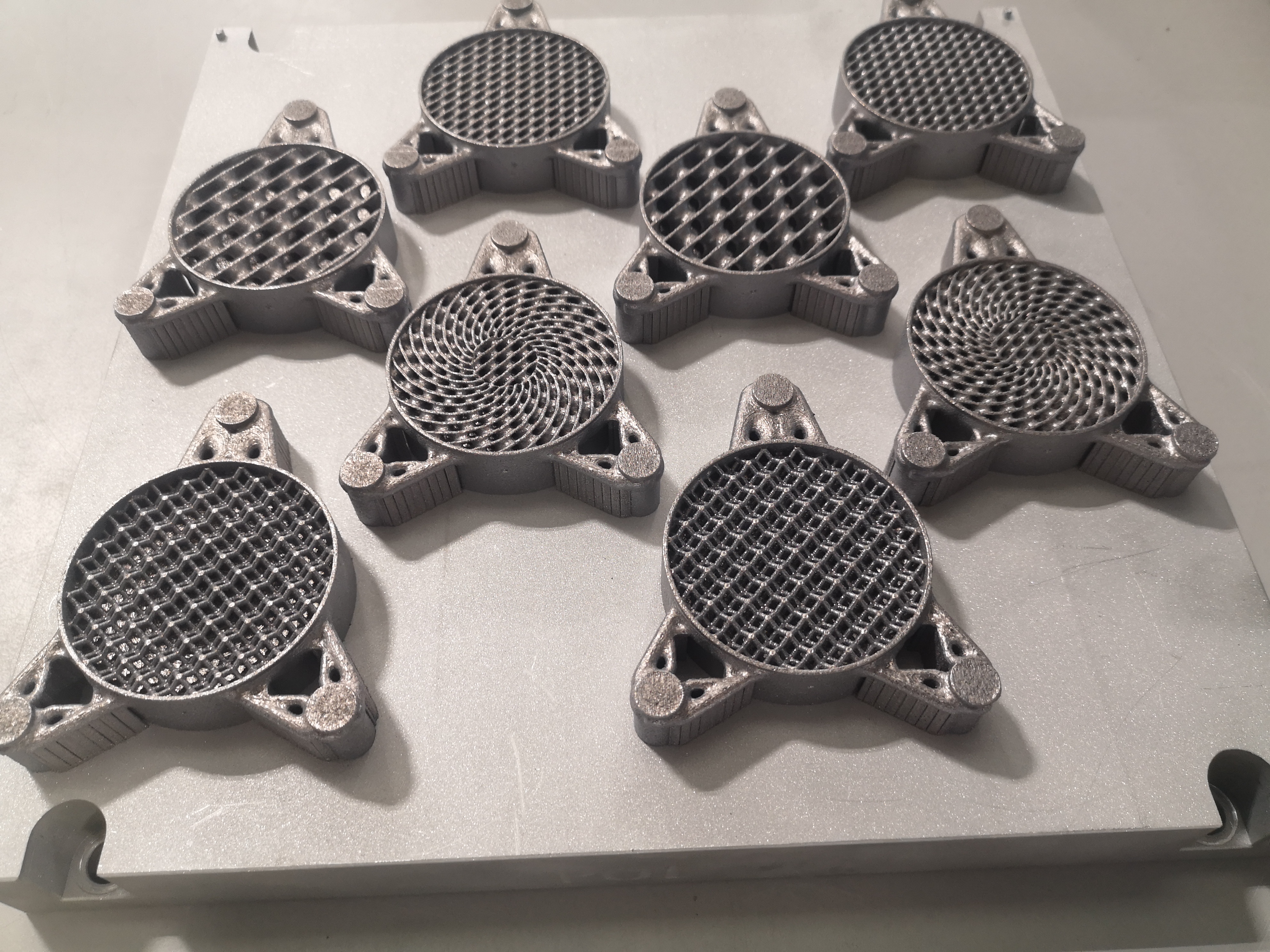}
    \caption{Eight normal resolution parts still on the build plate, support structures can be seen on the mounts. Image credit: Alan Magowan, LPE}
    \label{fig:buildplate}
\end{figure}

\begin{figure}[h]
    \centering
    \includegraphics[width=1\linewidth]{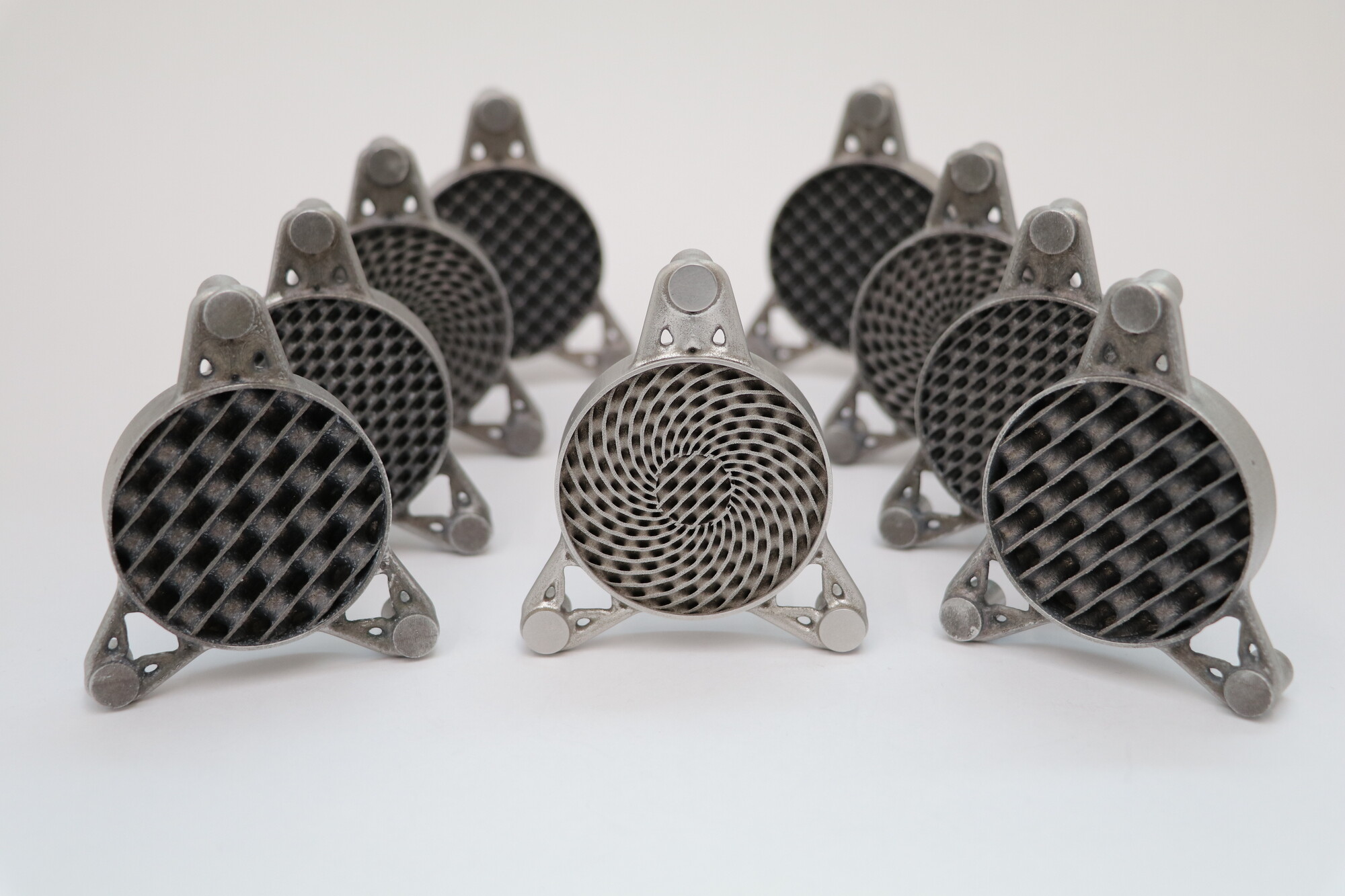}
    \caption{All aluminium mirror prototypes. Image credit: Jason Cowan}
    \label{fig:mirrorsjason}
\end{figure}

Figure \ref{fig:buildplate} shows the eight normal resolution parts still attached to the build plate, the support structures on the mounts are clearly visible. All lattices printed successfully, even the resolution of the separating wall in the conformal lattices, discussed in Section \ref{sec:conformalatdesign}, was very good. The two 70\% WR lattices, which are much less dense, also printed well with no signs of sagging or crumbling across the larger overhangs. Figure \ref{fig:mirrorsjason} pictures the nine prototypes, with the high resolution part in the middle. The high resolution process has a smaller layer height and can print finer features, it was decided that the conformal lattice would benefit most from this increased resolution of all the lattice choices because it has the most intricate geometry. This comes at the cost of a more expensive print with a smaller build volume, one mirror only just fits in the printer's area.  

\subsection{Post processing}

\subsubsection{Machining}\label{sec:machining}
\begin{figure}[h]
    \centering
    \includegraphics[width=1\linewidth]{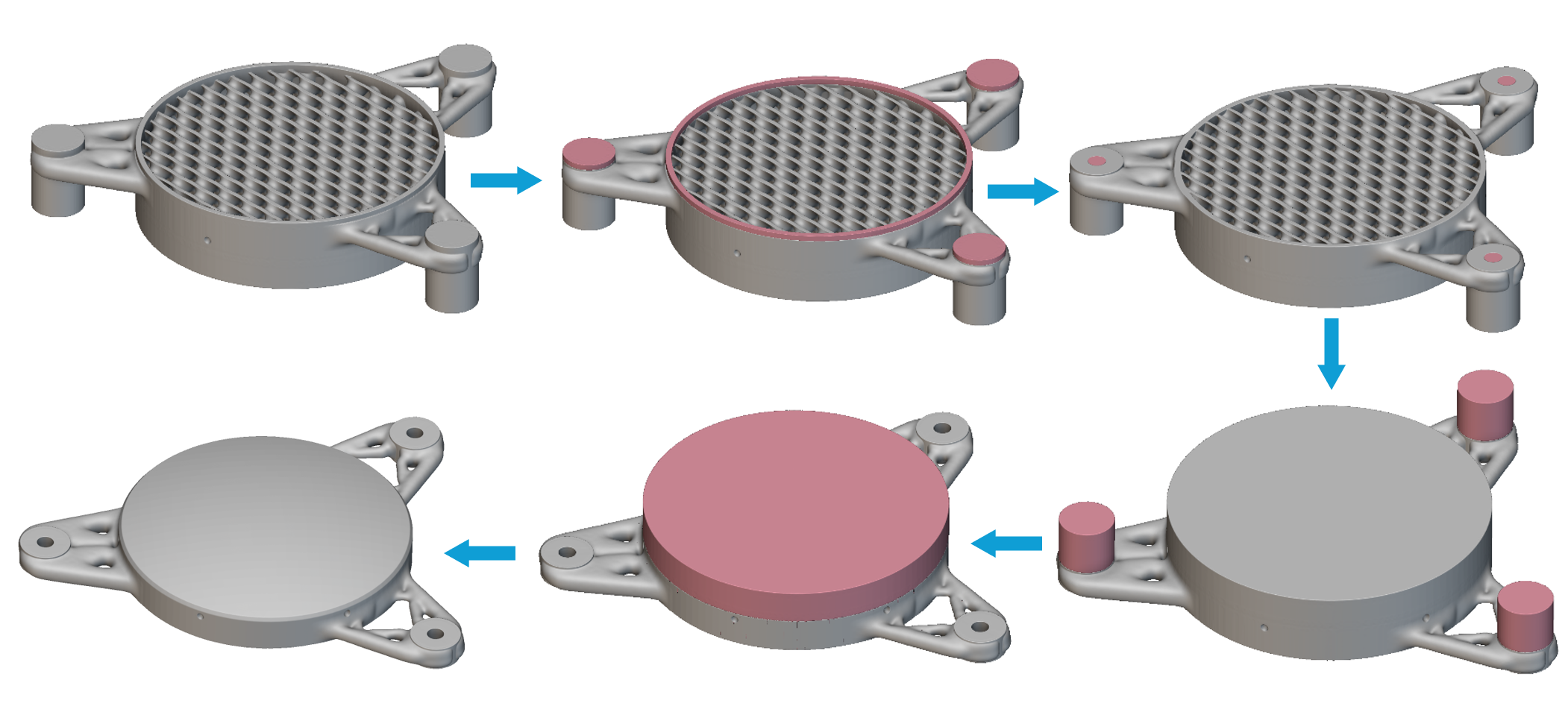}
    \caption{Machining steps}
    \label{fig:Machining steps}
\end{figure}

\begin{figure}
  \centering
  \begin{subfigure}{0.48\textwidth}
    \includegraphics[width=\textwidth]{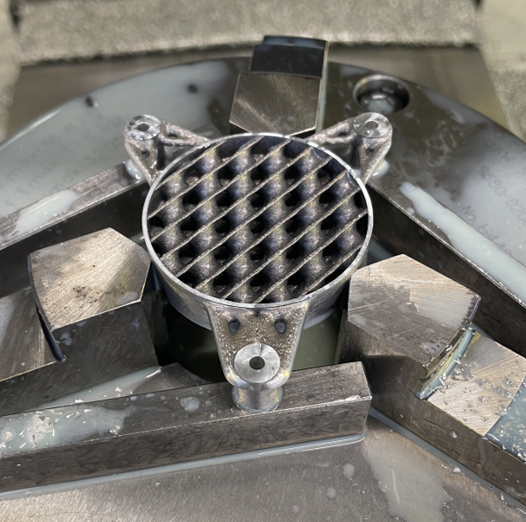}
    \caption{Operation 1: Skim top and drill holes}
    \label{fig:Op1}
  \end{subfigure}
  \hfill
  \begin{subfigure}{0.48\textwidth}
    \includegraphics[width=\textwidth]{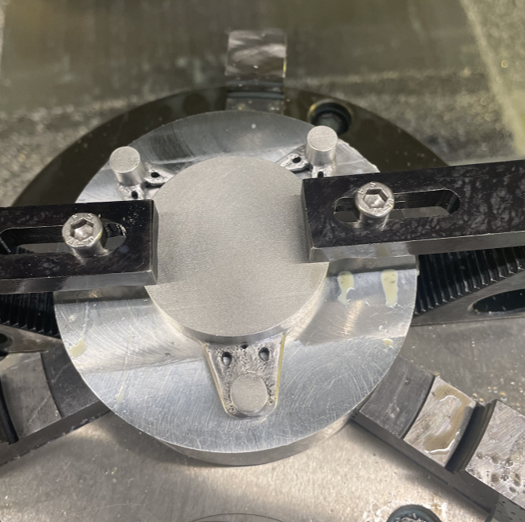}
    \caption{Operation 2: Part located with pins}
    \label{fig:Op2 setup}
  \end{subfigure}
  \hfill
  \begin{subfigure}{0.48\textwidth}
    \includegraphics[width=\textwidth]{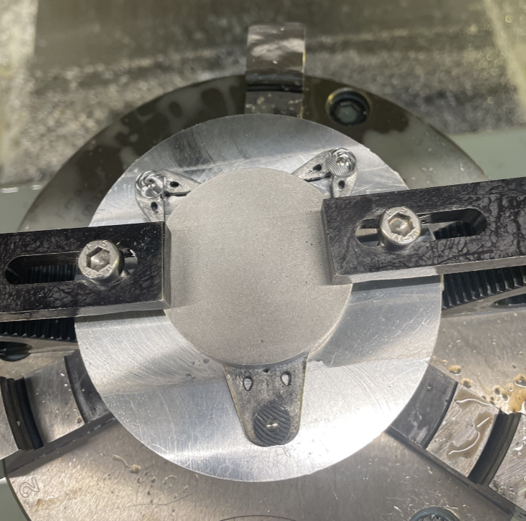}
    \caption{Operation 2: Remove excess material on screw pads}
    \label{fig:Op2}
  \end{subfigure}
  \hfill
  \begin{subfigure}{0.48\textwidth}
    \includegraphics[width=\textwidth]{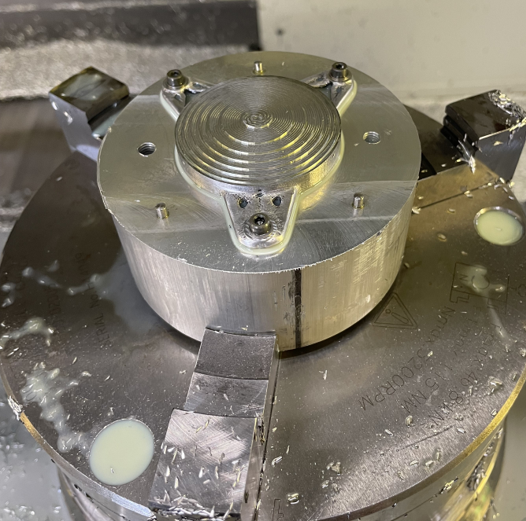}
    \caption{Operation 3: Milling convex radius of curvature}
    \label{fig:Op3}
  \end{subfigure}
  \caption{Subtractive Machining. Image credits: Richard Kotlewski} 
  \label{fig:Workholding}
\end{figure}

Machining was needed to prepare the printed part for SPDT. Several operations were needed to remove the additional features added in Section \ref{sec:dfam}, Figure \ref{fig:Machining steps} shows the order of machining operations and which parts of the printed model were removed at each stage. Figure \ref{fig:Op1}-\ref{fig:Op3} shows how the part was held and machined in house at the UKATC. In Figure \ref{fig:Op2 setup}, the mirror is located using pins in the holes drilled in Operation 1. The convex surface was machined to a spherical radius of \SI{100.1}{mm} to allow for \SI{100}{\micro\meter} of extra material to be removed during SPDT to reach the target ROC. 

A suspected risk during the print was that the mounts would warp, extending the screw pads to the build plate was done to mitigate this risk by more securely attaching them. However, it is clear after machining that there was still a small amount of warping as the faces of the three screw pads did not lie in the same plane. Process simulation could have been used to assess the deformation due to thermal stresses, but given the time frame, this was not possible. 

\subsubsection{Single Point Diamond Turning}

\begin{figure}
 \centering
  \begin{minipage}{0.9\textwidth}
    \includegraphics[width=\textwidth]{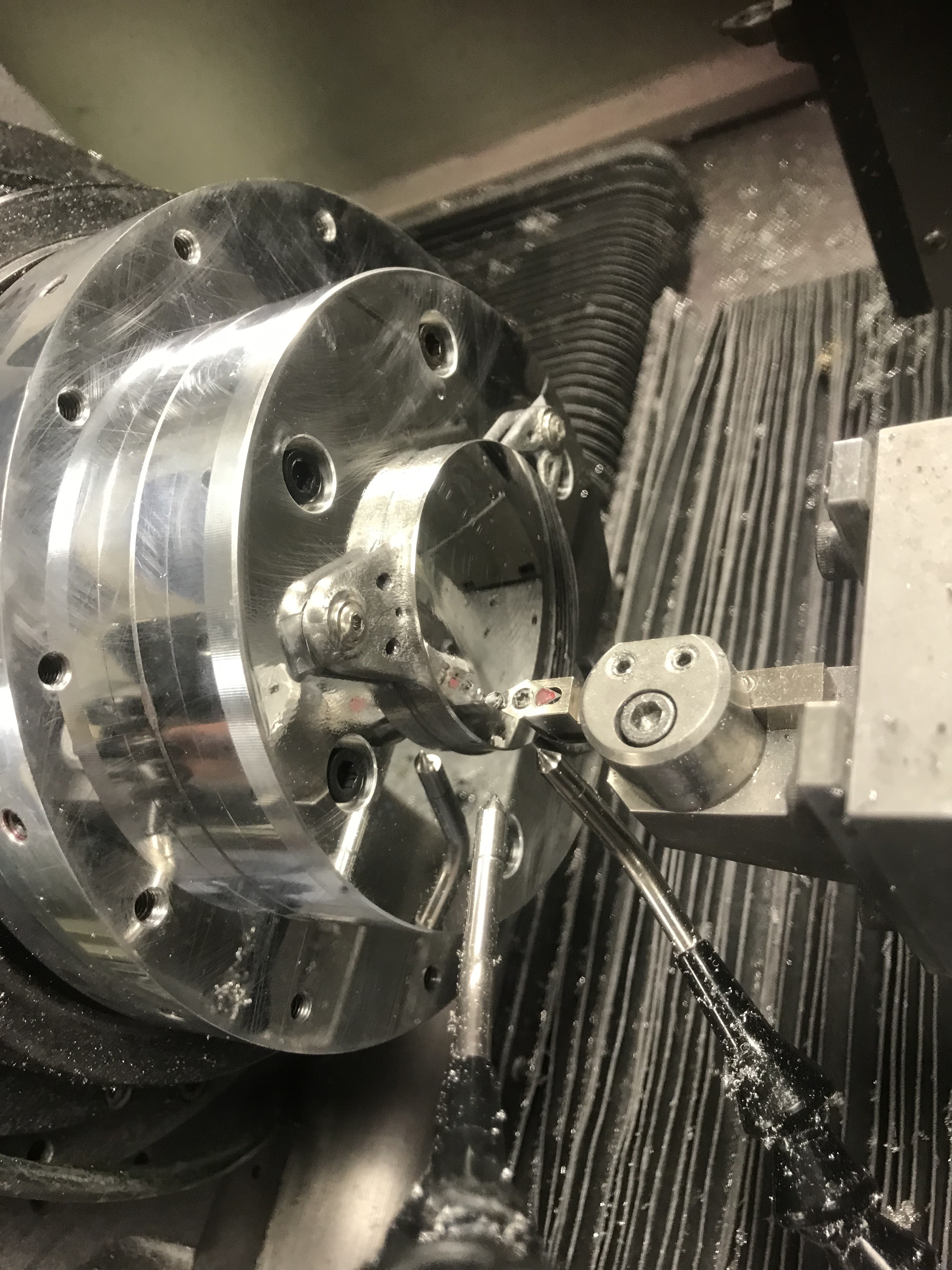}
    \caption{Mirror on diamond turning lathe. Image credit: Ken Parkin}
    \label{fig:MirroronMachine} 
  \end{minipage}
\end{figure}

SPDT was performed at the Centre for Advanced Instrumentation, Durham University on a Moore Nanotech 250 UPL. Three prototypes were cut, the high and normal resolution conformal lattices, and a 70\% WR Diamond cubic part. Two more were intended to have hot isostatic pressing (HIP) done before SPDT, to assess the difference in surface roughness. However, due to circumstances outside of our control, the HIP process was delayed and these parts could not be included. The parts had several rough cuts made at \SI{10}{\micro\meter} depth of cut to clean the surface, then a single \SI{2}{\micro\meter} deep finishing cut. Figure \ref{fig:MirroronMachine} shows the mirror during turning.

\section{Metrology}\label{sec:metrology}

\subsection{Mass}
The mirrors were weighed after machining and compared to the predicted values from the CAD models. Table \ref{tab:masses} presents the masses and errors, the actual masses don't deviate significantly from the predicted values. There can be several sources of mass error across the workflow, STL meshes may not accurately capture all geometry, density values in datasheets are approximate, tumble finishing, which smoothens the surfaces will remove material and more.   

\begin{table}[h]
\caption{Mass measurements} 
\label{tab:masses}
\begin{center}       
\begin{tabular}{ |p{4.8cm}|p{2.5cm}|p{2.5cm}|p{2.5cm}| }
\hline
\textbf{Design} & \textbf{Measured (g)} & \textbf{Predicted (g)} & \textbf{Error (\%)} \\
\hline
\rule[-1ex]{0pt}{3.5ex}Diamond cubic 50\% WR & 27.7 & 27.9 & -0.7 \\
\hline
\rule[-1ex]{0pt}{3.5ex}Diamond cubic 70\% WR & 20.6 & 21.1 & -2.3 \\
\hline
\rule[-1ex]{0pt}{3.5ex}Fluorite cubic 70\% WR & 20.3 & 20.5 & -1.0   \\
\hline
\rule[-1ex]{0pt}{3.5ex}Diamond conformal 50\% WR & 29.9 & 29.6 & 1.0 \\
\hline
\rule[-1ex]{0pt}{3.5ex}Diamond conformal 50\% WR* & 27.5 & 29.6 & -7.1 \\
\hline
\end{tabular}
 \begin{tablenotes}
    \footnotesize
        \item[a] \hspace{1cm} \parbox[t]{4cm}{*High resolution}
    \end{tablenotes}
\end{center}
\end{table}

\subsection{Surface Roughness}\label{sec:surfaceroughness}

\begin{figure}[!h]
  \centering
  \begin{subfigure}{0.32\textwidth}
    \includegraphics[angle=-90, width=\textwidth]{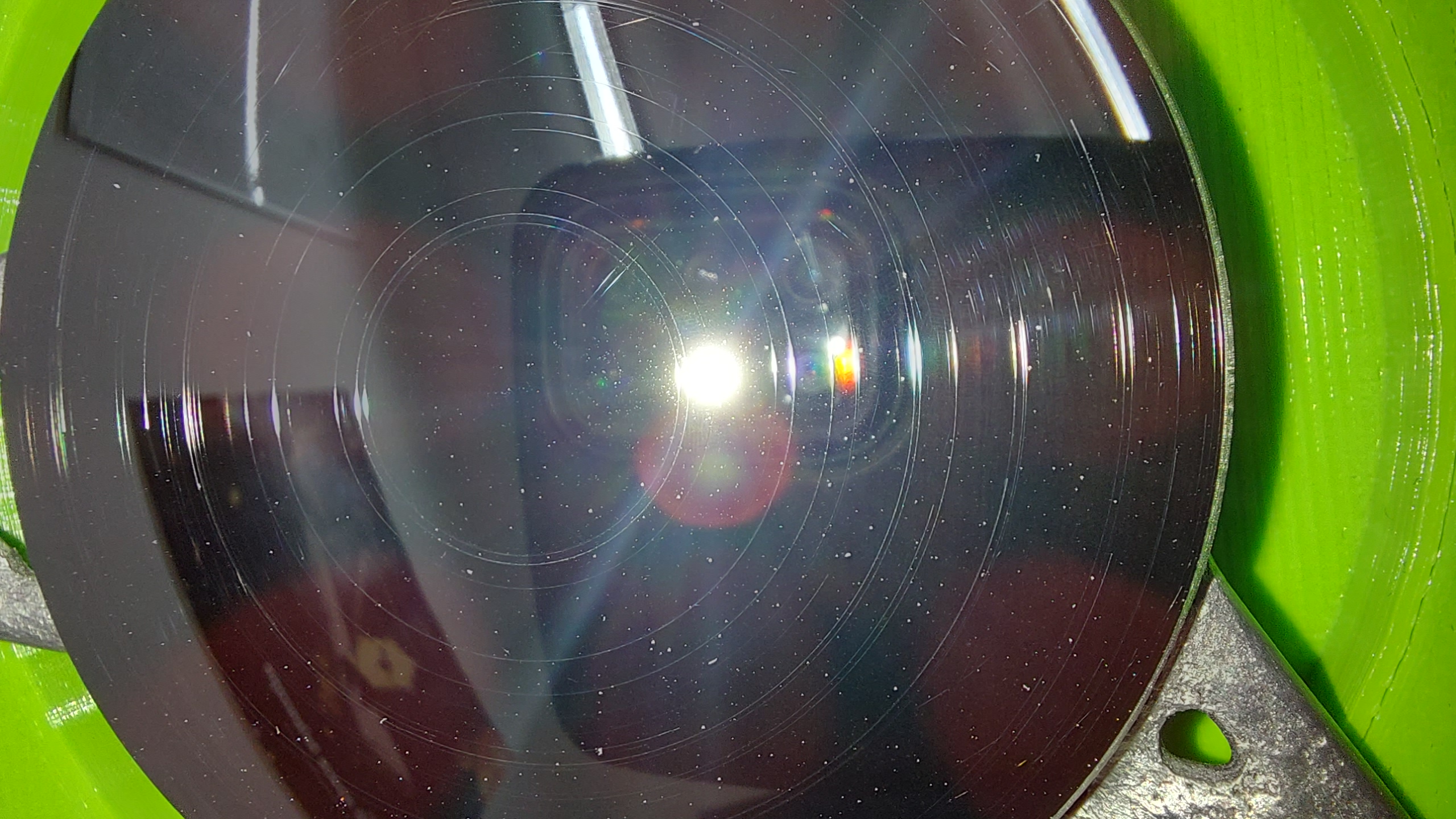}
    \caption{Normal resolution Diamond 70\% WR}
    \label{fig:Part3a}
  \end{subfigure}
  \hfill
  \begin{subfigure}{0.32\textwidth}
    \includegraphics[angle=-90, width=\textwidth]{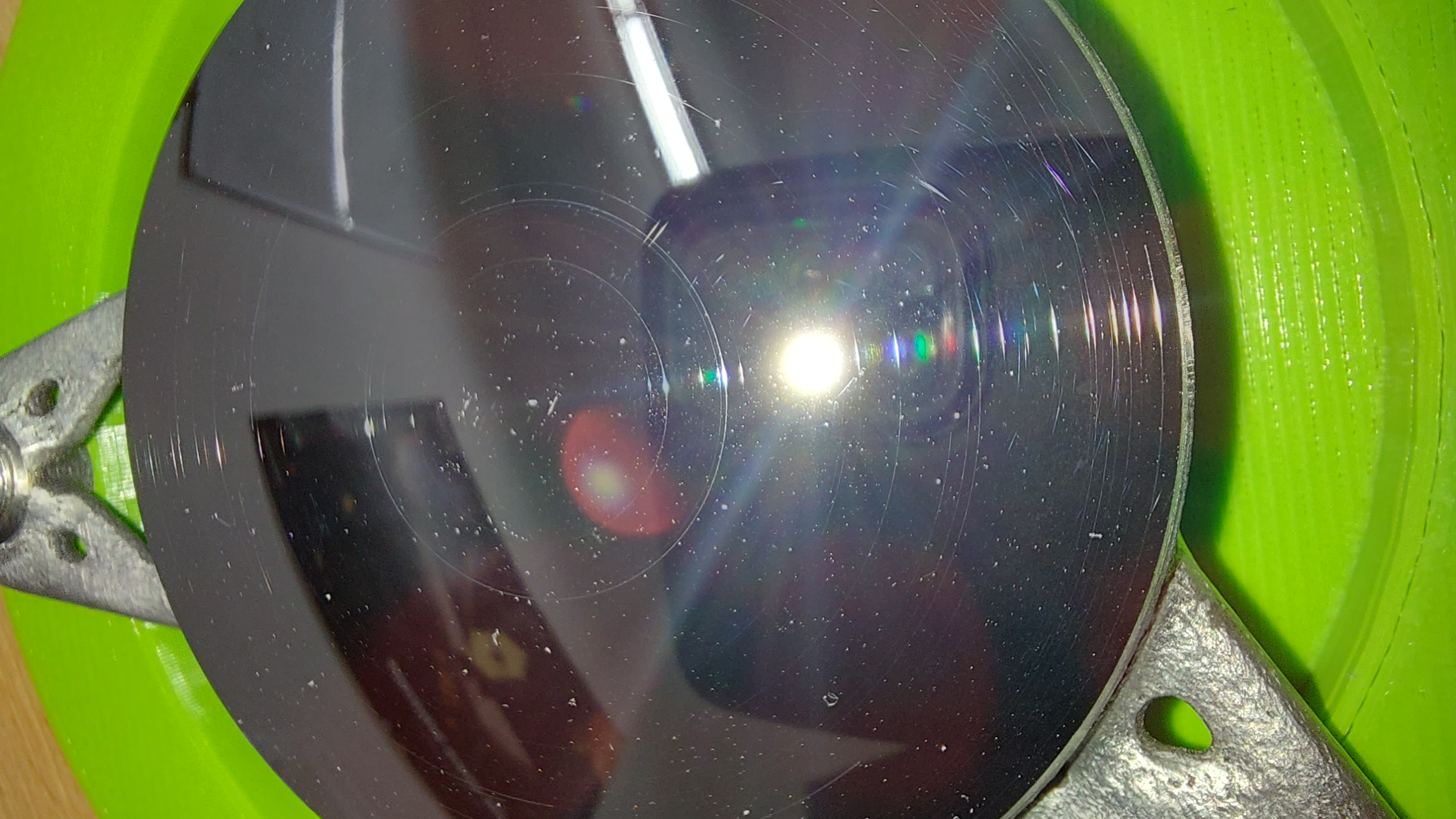}
    \caption{Normal resolution Diamond conformal 50\% WR}
    \label{fig:Part2a}
  \end{subfigure}
  \hfill
  \begin{subfigure}{0.32\textwidth}
    \includegraphics[angle=-90, width=\textwidth]{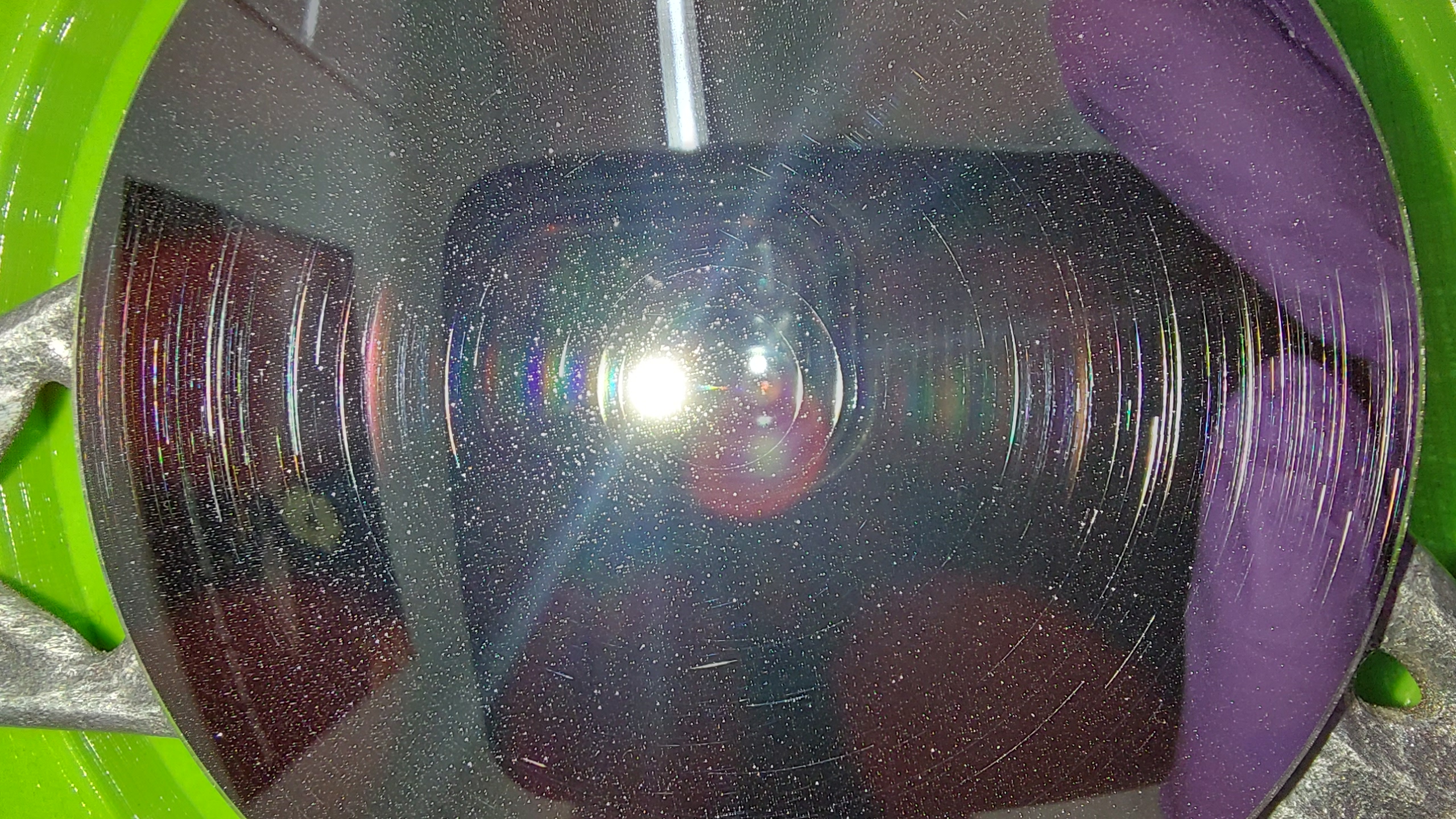}
    \caption{High resolution Diamond conformal 50\% WR}
    \label{fig:Part1a}
  \end{subfigure}
  \hfill
  \caption{Pictures of mirror surfaces}
  \label{fig:pictures of mirror surfaces}
\end{figure}

Figure \ref{fig:pictures of mirror surfaces} shows the three reflective surfaces after SPDT, all exhibit concentric scratches and porosity. Qualitatively, the high resolution print has more diffuse reflection, this is likely due to different machine settings. 

Surface roughness measurements were taken at the Centre for Advanced Instrumentation, Durham University using a Zygo Zegage Plus, X20 magnification, with 8\textsuperscript{th} order polynomial removed from the data. Measurements were made in four locations on the mirror surface: at the centre, off centre, midway along the radius and towards the outer edge. Table \ref{tab:Surfaceroughness} presents the average (Sa), RMS (Sq) and PV (Sz) values in each location for the three designs. Part 1 off centre is missing its dataset and Part 2 centre sampled a defect, so was excluded. The parts are numbered as follows: 

\begin{enumerate}
    \item Diamond Conformal 50\% WR High resolution
    \item Diamond Conformal 50\% WR Normal resolution
    \item Diamond Cubic 70\% WR Normal resolution
\end{enumerate}

\begin{table}[h]
\caption{Surface Roughness Results}
\label{tab:Surfaceroughness}
\centering
\resizebox{\textwidth}{!}{%
\begin{tabular}{|l|*{12}{p{0.8cm}|}}
\hline
 & \multicolumn{12}{c|}{\textbf{Measurement / nm}} \\ \hline
 & \multicolumn{3}{c|}{\textbf{Centre}} & \multicolumn{3}{c|}{\textbf{Off Centre}} & \multicolumn{3}{c|}{\textbf{Midway}} & \multicolumn{3}{c|}{\textbf{Outer}} \\ \hline
\textbf{Part No.} & \textbf{Sa}  & \textbf{Sq}  & \textbf{Sz}  & \textbf{Sa}  & \textbf{Sq}  & \textbf{Sz}  & \textbf{Sa}  & \textbf{Sq}  & \textbf{Sz}  & \textbf{Sa}  & \textbf{Sq}  & \textbf{Sz}  \\ \hline
1 & 8.30 & 13.2 & 726 & N/A & N/A & N/A & 11.8 & 34.2 & 3150 & 34.3 & 315 & 25500 \\ \hline
2 &  &  &  & 4.60 & 5.90 & 126 & 5.70 & 8.40 & 9.40 & 4.90 & 15.2 & 38.5 \\ \hline
3 & 8.00 & 11.5 & 407 & 4.50 & 5.80 & 219 & 5.70 & 9.40 & 988 & 8.80 & 38.5 & 9160 \\ \hline
\end{tabular}%
}
\end{table}

\begin{figure}
    \centering
    \includegraphics[width=1\linewidth]{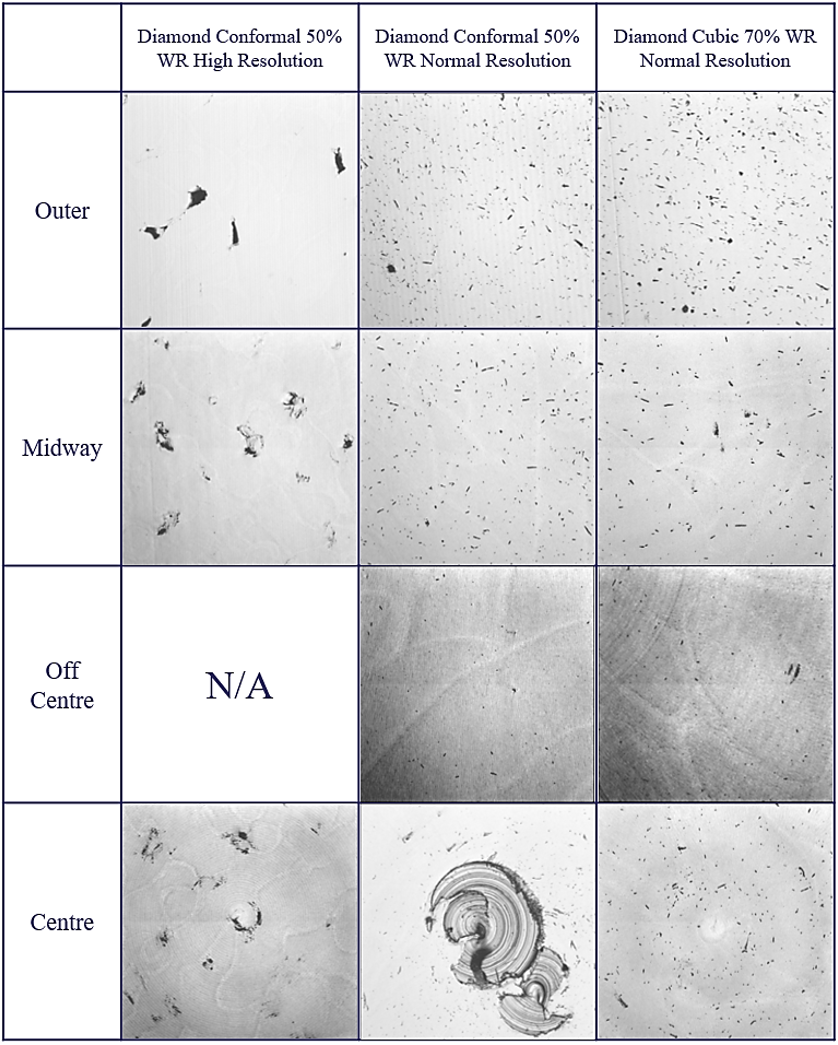}
    \caption{Microscope images of each prototype}
    \label{fig:Microscopic}
\end{figure}

\begin{figure}
    \centering
    \includegraphics[width=1\linewidth]{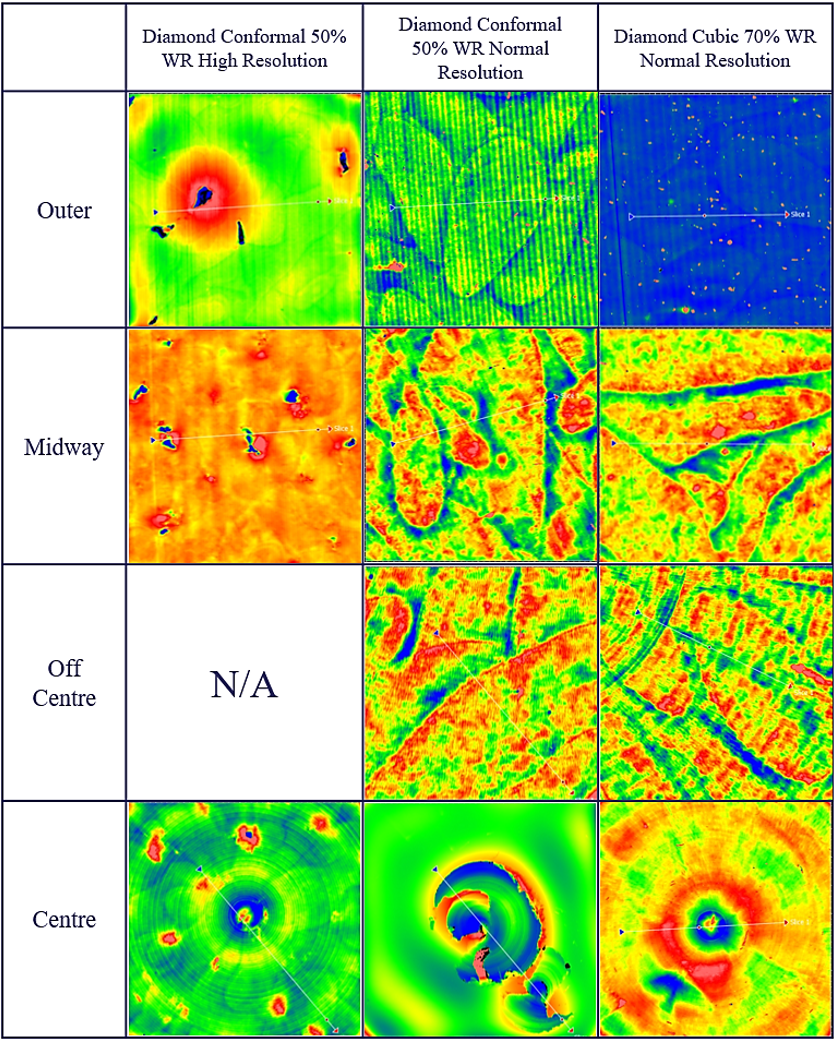}
    \caption{Topographic images of each prototype}
    \label{fig:Topographic}
\end{figure}

Figures \ref{fig:Microscopic} and \ref{fig:Topographic} show both microscopic and topographic images respectively for each measurement location and design. Each square represents a roughly \SI{420}{\micro\meter} x \SI{420}{\micro\meter} area of the reflective surface. Pores are visible in every measurement. Upon a first look from a limited set of samples, it seems as though the high resolution print has fewer, but larger pores compared to the normal resolution part. However, most of the defects seen in the normal resolution microscope images are probably not porosity, there are a number of long, thin black marks which are not characteristic of pores. When comparing the topographic images to the microscope images, only very few of these needle-like structures can be identified, but they do not present as holes in the surface, rather protrusions, it is theorised these are silicon particles. The relatively larger, irregular pores in the high resolution part are indicative of lack of fusion in the layer. The difference between the porosity between the two resolution parts likely comes down to the machine setting. Orientation is not a factor as all parts were printed in the same direction, and the chemical composition of the two powders is the same. A roughly \SI{200}{\micro\meter} defect is present in the reflective surface at the centre of the Diamond conformal normal resolution part which results in a roughness RMS of over \SI{1000}{nm}, it was suggested that this could be caused by hard material being dragged into the centre during the cut. A roughness RMS of \SI{5.8}{nm} was achieved off centre of the normal resolution Diamond cubic part.
 
Eliminating the defects in the reflective surface is crucial before AM mirrors can be taken further for use in final products, however, this was not the goal for the prototypes. On-going work into optimising print parameters\cite{RSnellDefects} and investigating the effect that HIP has on surface roughness aims to resolve the key challenges with AM mirrors\cite{CAtkinsHIP}.


\section{Fused Silica Prototypes}\label{sec:Fused silica}

\begin{figure}[h]
    \centering
    \includegraphics[width=1\linewidth]{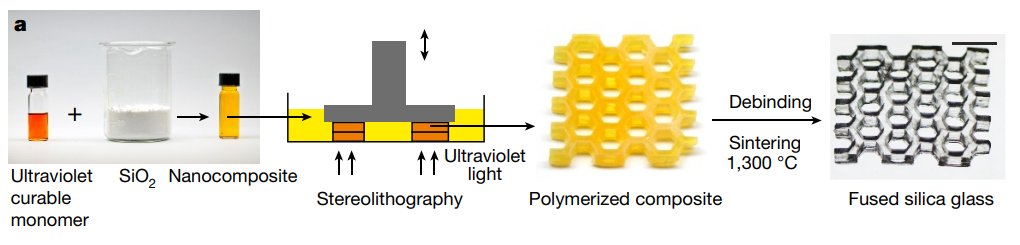}
    \caption{Fused silica printing and post processing. Image credit: \textit{Kotz et al.(2017)\cite{FusedSilica}}}
    \label{fig:fusedsilicaprocess}
\end{figure}

Part of this study was to investigate M2 prototypes in a 3D printed ceramic material, with the motivation to target shorter wavelength applications. Ceramics are typically much more mechanical hard materials than metals, therefore the achievable surface roughness is lower. The ceramic material chosen for the prototypes was fused silica, \textit{Kotz et al.(2017)\cite{FusedSilica}} presented a novel, high-resolution method for 3D printing fused silica using resin printing, or stereolithography (SLA), resulting in transparent parts suitable for optical components. A fine silica (silicon dioxide) powder is held in a polymer binder, which makes up the resin. Following the conventional SLA process, thin layers (tens of microns) of resin are cured using an ultraviolet light source ($\lambda$=\SI{385}{nm} for the specific printer) to build a part. Green parts (newly printed parts), undergo thermal debinding to remove the polymer binder, then sintering to fuse the remaining silica into a fully dense part, this workflow is shown in Figure \ref{fig:fusedsilicaprocess}. 

\subsection{Fused silica prototype design}

As 3D printed fused silica mirrors are less mature than metal mirrors, the scope of the ceramic investigation was reduced to focus on printability of a lightweight mirror; material defects will be investigated at later stages. Parallel investigations into 3D printed ceramics present polishing results of solid \SI{50}{mm} diameter by \SI{5}{mm} deep, flat fused silica disks\cite{CAtkinsCeramics}, the potential for short wavelength applications was demonstrated, a few defects were observed, but not yet fully identified. Pursuing a curved surface mirror will be valuable for future mirror projects to understand which print orientations are suitable. 

The design of the fused silica prototype was adapted from the aluminium prototype. While the aluminium topology optimised mounts did not consider launch conditions (shock loading and vibrations), they were included to demonstrate the ability of AM to consolidate parts and were a convenient workholding solution for machining/SPDT. The polishing setup will be different to the SPDT one, for the ease of holding the mirrors, the mounts were omitted from the design. Fused silica prototypes will aim for only 50\% mass reduction with one lattice design, to minimise uncertainty around lattice printability. 

\begin{figure}[h]
    \centering
    \includegraphics[width=0.75\linewidth]{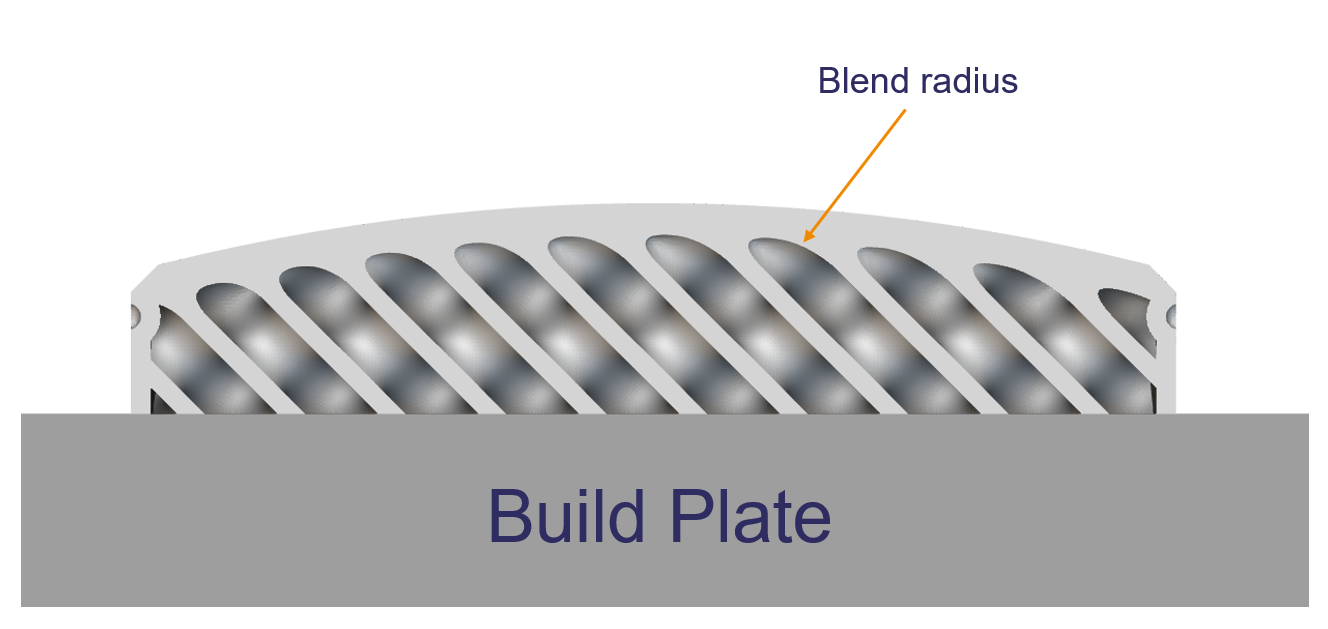}
    \caption{Section view of a fused silica design}
    \label{fig:fusedsilicaonbuildplate}
\end{figure}

Diamond (TPMS) cubic was chosen as the lattice. Unlike the aluminium prototype, the fused silica part will not have any post machining performed (aside from polishing), this meant that the mirror could not be printed with the curved surface on the build plate, instead, it was printed with the curved surface pointing up, as shown in Figure \ref{fig:fusedsilicaonbuildplate}. In this orientation, the optical surface did not have full support from the lattice underneath, as mentioned in Section \ref{sec:amdesigncons}, it is not acceptable to have support structures in the lattice. To overcome this challenge, a blend radius was implemented to adequately support the surface during the print, this can be seen in Figure \ref{fig:fusedsilicaonbuildplate}. Additional modifications included extending the optical surface by \SI{0.5}{mm} and increasing the size of the chamfer.

It was also decided to print a prototype with a flat surface to compare to commercially available fused silica blanks, both mirrors use the same lattice parameters, which were \SI{1}{mm} wall thickness and 10 x 10 x 10 mm cells. Table \ref{tab:fusedsilicaspecs} presents the mechanical and optical specifications of each mirror. 

\begin{table}[]
\caption{Specifications of fused silica mirrors}
\label{tab:fusedsilicaspecs}
\centering
\begin{tabular}{|l|ll|}
\hline
                     & \multicolumn{2}{c|}{\textbf{Fused Silica Mirrors}}                   \\ \hline
\textbf{Parameter}   & \multicolumn{1}{l|}{\textbf{Curved Surface}} & \textbf{Flat Surface} \\ \hline
Mechanical aperture  & \multicolumn{2}{c|}{\SI{52}{mm}\diameter}                                              \\ \hline
Clear aperture       & \multicolumn{2}{c|}{\SI{50}{mm}\diameter}                                              \\ \hline
Height (at peak)     & \multicolumn{1}{l|}{\SI{10.5}{mm}}                    & \SI{9}{mm}                     \\ \hline
Optical prescription & \multicolumn{1}{l|}{Convex, spherical}       & Flat                  \\ \hline
ROC                  & \multicolumn{1}{l|}{\SI{100.5}{mm}}                   & -                     \\ \hline
Surface thickness    & \multicolumn{2}{c|}{\SI{1.5}{mm}}                                             \\ \hline
Edge chamfer         & \multicolumn{2}{l|}{\SI{1}{mm} x 45\degree}                                          \\ \hline
Extra feature        & \multicolumn{2}{l|}{6 x fiducial markers}                            \\ \hline
\end{tabular}
\end{table}

\subsection{Printing}

\begin{figure}[!h]
  \centering
  \begin{subfigure}[t]{0.48\textwidth}
    \includegraphics[width=\textwidth]{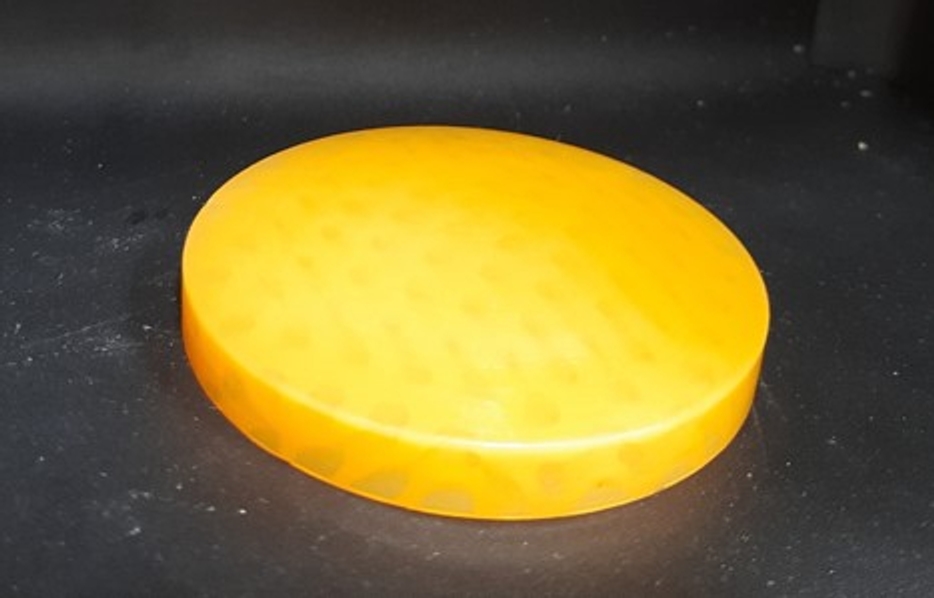}
    \caption{Curved surface part directly after printing}
  \end{subfigure}
  \hfill
  \begin{subfigure}[t]{0.48\textwidth}
    \includegraphics[width=\textwidth]{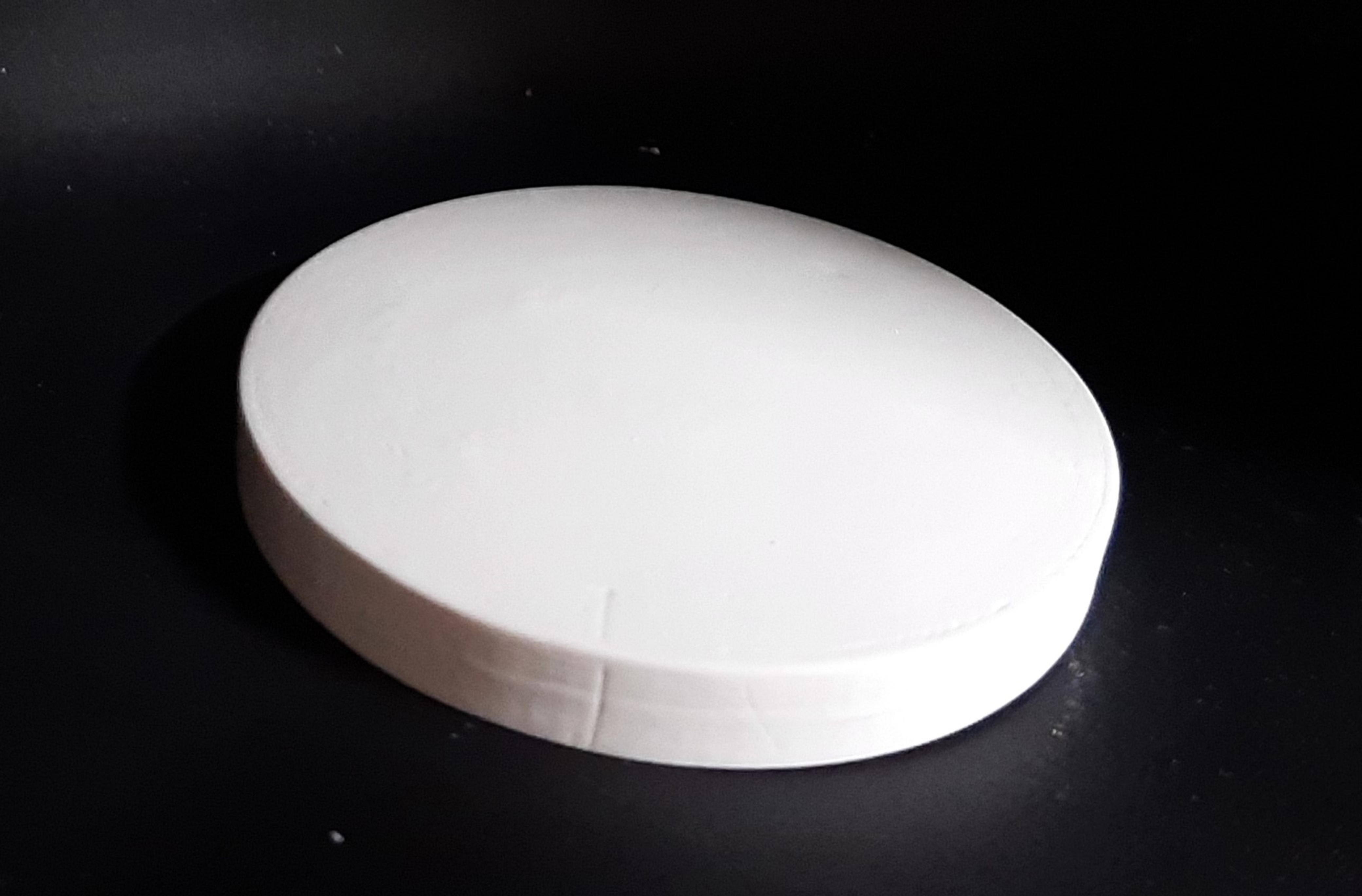} 
    \caption{Curved surface part after 600\textdegree{C} debinding}
  \end{subfigure}
  \caption{Fused silica parts in production. Image credits: Patrick Risch}
  \label{fig:fspartsproduction}
\end{figure}

In total, eight parts were printed, three curved surface parts, three flat surface parts and two of the solid polishing test sample disks from \textit{Atkins et al.(2024)\cite{CAtkinsHIP}}. They were printed on an Asiga Pro at a layer height of \SI{50}{\micro\meter}. Figure \ref{fig:fspartsproduction} shows one of the curved surface parts in production, the orange part is just after printing and it becomes white after thermal debinding, it will eventually become transparent after sintering.

\section{Conclusion and Future Work}

To conclude, this study has presented additively manufactured aluminium and fused silica mirror prototypes that approximate the secondary mirror from the 6U ADOT CubeSat project. The aluminium mirrors were mass reduced by roughly 50\% and 70\% compared to a solid mirror using different lattices, the mounts for the mirror in the ADOT assembly were topology optimised for stiffness and consolidated into the part, they were also used for work holding during machining and SPDT operations. Final lattice designs were downselected using FEA, then prototype validation. Eight normal resolution parts were printed in AlSi10Mg through an external bureau, a mixture of strut (Fluorite) and TPMS (Diamond) unit cells were implemented into cubic (Diamond cubic 50\% WR, Diamond cubic 70\% WR and Fluorite cubic 70\% WR) and conformal (Diamond conformal 50\% WR) lattices. A high resolution Diamond conformal design was also printed to investigate any differences in porosity and roughness between normal and high resolution printers. Relative surface distortions and print through effects from SPDT were shown using Zernike analysis. Machining was conducted at the UKATC to prepare the parts for SPDT which was carried out at the Centre for Advanced Instrumentation, along with the metrology. Fused silica prototypes were adapted from the aluminium designs and modified to suit a different print orientation. In addition to printing parts with a convex surface, flat mirrors were also printed to compare to commercial fused silica blanks. 

Future work for the aluminium prototypes include diamond turning two parts that have undergone HIP, then comparing surface roughness data to the dataset presented in Section \ref{sec:surfaceroughness}. Two more parts, that are yet to be turned, could undergo bead blasting to assess the effect on porosity and roughness. X-Ray Computer Tomography (XCT) will be used to analyse the effectiveness of HIP with regards to porosity reduction, in addition to quantifying and verifying the presence, location and shape of porosity. Fused silica mirrors will undergo surface profilometry of the as-printed parts, to observe any print-through and stair stepping effects. Polishing will be carried out at INAF Astronomical Observatory of Brera.

\section*{Acknowledgements}
The authors acknowledge the UKRI Future Leaders Fellowship `Printing the future of space telescopes' under grant \# MR/T042230/1, in addition to Patrick Risch of Glassomer, and Alan Magowan of Laser Prototypes Europe for their support and advice during design and manufacture of prototypes.

\bibliography{report} 

\begin{thebibliography}{10}

\bibitem{Hipreplacement}
Krebs, V., ``{3D-Printed Implant Reconstructs Hip in Patient with No Proximal Femur or Pelvis}.'' \url{https://consultqd.clevelandclinic.org/3d-printed-implant-reconstructs-hip-in-patient-with-no-proximal-femur-or-pelvis/} (2022).
\newblock (Accessed: 11 January 2024).

\bibitem{Porschepistons}
``{3D printing technology optimises pistons for the powerful 911 GT2 RS}.'' \url{https://media.porsche.com/mediakit/porsche-innovationen/en/porsche-innovationen/3d-printed-pistons}.
\newblock (Accessed: 11 January 2024).

\bibitem{CookBook}
Atkins, C., van~de Vorst, L. G. T.~B., Conley, A., Farkas, S., Hugot, E., Mező, G., Morris, K., Roulet, M., Snell, R.~M., Tenegi-Sangin{\'e}s, F., Todd, I., Vega-Moreno, A., and Schnetler, H., ``{The OPTICON A2IM Cookbook: an introduction to additive manufacture for astronomy},'' in [{\em Advances in Optical and Mechanical Technologies for Telescopes and Instrumentation V}{\nolinebreak\hspace{0.1em}]},  Navarro, R. and Geyl, R., eds.,  {\bf 12188},  121880W, International Society for Optics and Photonics, SPIE (2022).

\bibitem{XRayOpticsprintthru}
{Atkins}, C., {Feldman}, C., {Brooks}, D., {Watson}, S., {Cochrane}, W., {Roulet}, M., {Doel}, P., {Willingale}, R., and {Hugot}, E., ``{Additive manufactured x-ray optics for astronomy},'' in [{\em Society of Photo-Optical Instrumentation Engineers (SPIE) Conference Series}{\nolinebreak\hspace{0.1em}]},  {O'Dell}, S.~L. and {Pareschi}, G., eds., {\em Society of Photo-Optical Instrumentation Engineers (SPIE) Conference Series} {\bf 10399},  103991G (Aug. 2017).

\bibitem{Woodard}
{Woodard}, K.~S. and {Myrick}, B.~H., ``{Progress on high-performance rapid prototype aluminum mirrors},'' in [{\em Society of Photo-Optical Instrumentation Engineers (SPIE) Conference Series}{\nolinebreak\hspace{0.1em}]},  {Vizgaitis}, J.~N., {Andresen}, B.~F., {Marasco}, P.~L., {Sanghera}, J.~S., and {Snyder}, M.~P., eds., {\em Society of Photo-Optical Instrumentation Engineers (SPIE) Conference Series} {\bf 10181},  101810T (May 2017).

\bibitem{RSnellDefects}
Snell, R., Atkins, C., Schnetler, H., Chahid, Y., Beardsley, M., Harris, M., Zhang, C., Pears, R., Thomas, B., Saunders, H., Sloane, A., Maddison, G., and Todd, I., ``{Towards understanding and eliminating defects in additively manufactured CubeSat mirrors},'' in [{\em Advances in Optical and Mechanical Technologies for Telescopes and Instrumentation V}{\nolinebreak\hspace{0.1em}]},  Navarro, R. and Geyl, R., eds.,  {\bf 12188},  121880V, International Society for Optics and Photonics, SPIE (2022).

\bibitem{CAtkinsHIP}
Atkins, C., Chahid, Y., Lister, G., Tuck, R., Kotlewski, R., Snell, R.~M., Livera, E.~R., Fauor, M., Todd, I., Deffley, R., Shipley, J., Walsh, T., Gardstam, J., Bourgenot, C., White, P., Davies, S., and Tammas-Williams, S., ``{Targeting low micro-roughness for 3D printed aluminium mirrors using a hot isostatic press},'' in [{\em Advances in Optical and Mechanical Technologies for Telescopes and Instrumentation VI}{\nolinebreak\hspace{0.1em}]},  Navarro, R. and Jedamzik, R., eds.,  {\bf 13100},  13100--141, International Society for Optics and Photonics, SPIE (2024).

\bibitem{PorosityMech}
Harkin, R., Wu, H., Nikam, S., Yin, S., Lupoi, R., Walls, P., McKay, W., and McFadden, S., ``Evaluation of the role of hatch-spacing variation in a lack-of-fusion defect prediction criterion for laserbased powder bed fusion processes,'' {\em The International Journal of Advanced Manufacturing Technology}~{\bf 126},  659--673 (March 2023).

\bibitem{PorosityonMechandFatigue}
Al-Maharma, A.~Y., Patil, S.~P., and Markert, B., ``Effects of porosity on the mechanical properties of additively manufactured components: a critical review,'' {\em Materials Research Express}~{\bf 7},  122001 (dec 2020).

\bibitem{SurfRoughFatigueLife}
Xiao, W., Chen, H., and Yin, Y., ``Effects of surface roughness on the fatigue life of alloy steel,'' {\em Key Engineering Materials}~{\bf 525-526},  417--420 (11 2012).

\bibitem{CBreenOutGassing}
Breen, C., Walpole, J., Atkins, C., McPhee, S., Cliffe, M., Moffat, J., Edwards-Mowforth, M., Lister, I., Reynolds, L., Conley, A., Allum, S., Snell, R.~M., Tammas-Williams, S., and Watson, S., ``{Outgassing properties of additively manufactured aluminium},'' in [{\em Advances in Optical and Mechanical Technologies for Telescopes and Instrumentation V}{\nolinebreak\hspace{0.1em}]},  Navarro, R. and Geyl, R., eds.,  {\bf 12188},  121882I, International Society for Optics and Photonics, SPIE (2022).

\bibitem{Cubesatpaper}
{Schwartz}, N., {Brzozowski}, W., {Ali}, Z., {Milanova}, M., {Morris}, K., {Bond}, C., {Keogh}, J., {Harvey}, D., {Bissell}, L., {Sauvage}, J.-F., {Dumont}, M., {Correia}, C., {Rees}, P., and {Bruce}, H., ``{6U CubeSat deployable telescope for optical Earth observation and astronomical optical imaging},'' in [{\em Space Telescopes and Instrumentation 2022: Optical, Infrared, and Millimeter Wave}{\nolinebreak\hspace{0.1em}]},  {Coyle}, L.~E., {Matsuura}, S., and {Perrin}, M.~D., eds., {\em Society of Photo-Optical Instrumentation Engineers (SPIE) Conference Series} {\bf 12180},  1218031 (Aug. 2022).

\bibitem{Westik}
Westsik, M., Wells, J.~T., Chahid, Y., Morris, K., Milanova, M., Beardsley, M., Harris, M., Ward, L., Alcock, S.~G., Nistea, I.-T., Cottarelli, S., Tammas-Williams, S., and Atkins, C., ``{From design to evaluation of an additively manufactured, lightweight, deployable mirror for Earth observation},'' in [{\em Astronomical Optics: Design, Manufacture, and Test of Space and Ground Systems IV}{\nolinebreak\hspace{0.1em}]},  Hull, T.~B., Kim, D., and Hallibert, P., eds.,  {\bf 12677},  1267704, International Society for Optics and Photonics, SPIE (2023).

\bibitem{ImplicitModelling}
Courter, B., ``{Implicit modeling for engineering design}.'' \url{https://www.ntop.com/resources/blog/implicit-modeling-for-mechanical-design/} (2019).
\newblock (Accessed: 11 January 2024).

\bibitem{FieldDrivenDesign}
Allen, G., ``White paper: Field-driven design,'' tech. rep., nTopology.

\bibitem{BonePaper}
Kunio, I., ``Bone substitute fabrication based on dissolution-precipitation reactions,'' {\em Materials}~{\bf 3} (02 2010).

\bibitem{UnitCellComparison}
Alketan, O., Rowshan, R., and Abu Al-Rub, R., ``Topology-mechanical property relationship of 3d printed strut, skeletal, and sheet based periodic metallic cellular materials,'' {\em Additive Manufacturing}~{\bf 19},  167–183 (01 2018).

\bibitem{TPMSSuperior}
Hayashi, K., Kishida, R., Tsuchiya, A., and Ishikawa, K., ``Superiority of triply periodic minimal surface gyroid structure to {Strut-Based} grid structure in both strength and bone regeneration,'' {\em ACS Appl Mater Interfaces}~{\bf 15},  34570--34577 (July 2023).

\bibitem{ConformalTheory}
Atkins, C., Brzozowski, W., Dobson, N., Milanova, M., Todd, S., Pearson, D., Bourgenot, C., Brooks, D., Snell, R., Sun, W., Cooper, P., Alcock, S., and Nistea, I.-T., ``Lightweighting design optimisation for additively manufactured mirrors,''  42 (09 2019).

\bibitem{RobSconformal}
Snell, R., Atkins, C., Schnetler, H., Todd, I., Hern{\'a}ndez-Nava, E., Lyle, A.~R., Maddison, G., Morris, K., Miller, C., Roulet, M., Hugot, E., Sangin{\'e}s, F.~T., Vega-Moreno, A., van~de Vorst, L. T. G.~B., Dufils, J., Brouwers, L., Farkas, S., Mező, G., Beardsley, M., and Harris, M., ``{An additive manufactured CubeSat mirror incorporating a novel circular lattice},'' in [{\em Advances in Optical and Mechanical Technologies for Telescopes and Instrumentation IV}{\nolinebreak\hspace{0.1em}]},  Navarro, R. and Geyl, R., eds.,  {\bf 11451},  114510C, International Society for Optics and Photonics, SPIE (2020).

\bibitem{ConformalActual}
Wu, J., Wang, W., and Gao, X., ``Design and optimization of conforming lattice structures,'' {\em IEEE Transactions on Visualization and Computer Graphics}~{\bf PP},  1--1 (09 2019).

\bibitem{Conformalntop}
Horvath, N. and Davies, M., ``Advancing lightweight mirror design: a paradigm shift in mirror preforms by utilizing design for additive manufacturing,'' {\em Appl. Opt.}~{\bf 60},  681--696 (Jan 2021).

\bibitem{overhangunitcellexample}
Liu, X., Sekizawa, K., Suzuki, A., Takata, N., Kobashi, M., and Yamada, T., ``Compressive properties of al-si alloy lattice structures with three different unit cells fabricated via laser powder bed fusion,'' {\em Materials}~{\bf 13}(13) (2020).

\bibitem{polishingpressure}
Atkins, C., Brzozowski, W., Dobson, N., Milanova, M., Todd, S., Pearson, D., Bourgenot, C., Brooks, D., Snell, R., Sun, W., Cooper, P., Alcock, S.~G., and Nistea, I.-T., ``{Lightweighting design optimisation for additively manufactured mirrors},'' in [{\em Astronomical Optics: Design, Manufacture, and Test of Space and Ground Systems II}{\nolinebreak\hspace{0.1em}]},  Hull, T.~B., Kim, D.~W., and Hallibert, P., eds.,  {\bf 11116},  1111617, International Society for Optics and Photonics, SPIE (2019).

\bibitem{FusedSilica}
{Kotz, F.}, {Arnold, K.}, {Bauer, W.}, {Schild, D.}, {Keller, N.}, {Sachsenheimer, K.}, {Nargang, T}, {Richter, C.}, {Helmer, D}, and {Rapp, B}, ``Three-dimensional printing of transparent fused silica glass,'' {\em Nature}~{\bf 544},  337--339 (April 2017).

\bibitem{CAtkinsCeramics}
Atkins, C., Chahid, Y., Lister, G., Tuck, R., Isherwood, D., Civitani, M., Vecchi, G., Pareschi, G., Rongyan, S., Yamamura, K., I.~Noto, Y.~N., Alcock, S.~G., and Ioana-Theodora~Nistea, M. B. D.~S., ``{Additive manufacturing in ceramics: targeting lightweight mirror applications in the visible, UV \& X-ray},'' in [{\em Advances in Optical and Mechanical Technologies for Telescopes and Instrumentation VI}{\nolinebreak\hspace{0.1em}]},  Navarro, R. and Jedamzik, R., eds.,  {\bf 13100},  13100--123, International Society for Optics and Photonics, SPIE (2024).

\end{thebibliography}
\bibliographystyle{spiebib} 

\end{document}